\documentclass[11pt,a4paper]{article}
\pdfoutput=1
\usepackage{jheppub}

\usepackage{multirow, graphicx,amssymb,url,mathrsfs,amsmath}
\usepackage{wrapfig,boxedminipage,setspace,subfigure,epsfig}
\usepackage{amsxtra,amstext,latexsym,dsfont,amsfonts}
\usepackage{color,eucal}
\usepackage[dvipsnames]{xcolor}
\usepackage{float}
\usepackage{slashed,comment}
\usepackage{kotex}
\usepackage{ulem}
\usepackage{tikz}
\usetikzlibrary{calc,patterns,angles,quotes}
\usetikzlibrary{decorations.pathreplacing,decorations.markings,snakes}

\usepackage{tabularx, array}
\newcolumntype{L}[1]{>{\raggedright\arraybackslash}p{#1}}
\newcolumntype{C}[1]{>{\centering\arraybackslash}p{#1}}
\newcolumntype{R}[1]{>{\raggedleft\arraybackslash}p{#1}}

\newcommand{\dd}{\mathrm{d}}

\newcommand{\grad}{\nabla}

\title{Collective dynamics and the Anderson-Higgs
mechanism in a bona fide holographic superconductor}
\author[a,b,c,d]{Hyun-Sik Jeong,}
\author[e,f]{Matteo Baggioli,}
\author[g,h]{Keun-Young Kim,}
\author[c,d]{and Ya-Wen Sun}

\emailAdd{hyunsik.jeong@uam.es}
\emailAdd{b.matteo@sjtu.edu.cn}
\emailAdd{fortoe@gist.ac.kr}
\emailAdd{yawen.sun@ucas.ac.cn}

\affiliation[a]{Instituto de F\'isica Te\'orica UAM/CSIC, Calle Nicol\'as Cabrera 13-15, 28049 Madrid, Spain}
\affiliation[b]{Departamento de F\'isica Te\'orica, Universidad Aut{\'o}noma de Madrid, Campus de Cantoblanco, 28049 Madrid, Spain}
\affiliation[c]{School of physics $\&$ CAS Center for Excellence in Topological Quantum Computation, University of Chinese Academy of Sciences, Zhongguancun east road 80, Beijing 100049, China}
\affiliation[d]{Kavli Institute for Theoretical Sciences, University of Chinese Academy of Sciences, \\ Zhongguancun east road 80, Beijing 100049, China}
\affiliation[e]{Wilczek Quantum Center, School of Physics and Astronomy, Shanghai Jiao Tong University, Shanghai 200240, China.}
\affiliation[f]{Shanghai Research Center for Quantum Sciences, Shanghai 201315.}
\affiliation[g]{Department of Physics and Photon Science, Gwangju Institute of Science and Technology, \\
123 Cheomdan-gwagiro, Gwangju 61005, Korea}
\affiliation[h]{Research Center for Photon Science Technology, Gwangju Institute of Science and Technology, \\
123 Cheomdan-gwagiro, Gwangju 61005, Korea}

\abstract{The holographic superconductor is one of the most popular models in the context of applied holography. Despite what its name suggests, it does not describe a superconductor. On the contrary, the low temperature phase of its dual field theory is a superfluid with a spontaneously broken U(1) global symmetry. As already observed in the previous literature, a \textit{bona fide} holographic superconductor can be constructed using mixed boundary conditions for the bulk gauge field. By exploiting this prescription, we study the near-equilibrium collective dynamics in the Higgs phase and reveal the characteristic features of the Anderson-Higgs mechanism. We show that second sound disappears from the spectrum and the gauge field acquires a finite energy gap of the order of the plasma frequency. We observe an overdamped to underdamped crossover for the Higgs mode which acquires a finite energy gap below $\approx T_c/2$, with $T_c$ the superconducting critical temperature. Interestingly, the energy gap of the Higgs mode at low temperature is significantly smaller than $2\Delta$, with $\Delta$ the superconducting energy gap. Finally, we interpret our results using Ginzburg-Landau theory and we confirm the validity of previously derived perturbative analytic expressions.
}

\begin{document}
\maketitle

%%%%%%%%%%%%%%%%%%%%%%%%%%%%%%%%
%    
%%%%%%%%%%%%%%%%%%%%%%%%%%%%%%%%
\section{Introduction}
In the last decade, the holographic correspondence, or gauge-gravity duality, has become an invaluable complementary tool to investigate the many-body dynamics of strongly correlated materials and strongly coupled condensed matter systems \cite{Hartnoll:2016apf,Zaanen:2015oix,Baggioli:2019rrs,Natsuume:2014sfa}, with a particular emphasis on the problem of strange metals and high-$T_c$ superconductors \cite{Baggioli:2022pyb,Davison:2013txa,Zaanen:2018edk,Hartnoll:2021qyl}.

The so-called holographic superconductor, or HHH model, introduced by Hartnoll, Herzog and Horowitz \cite{Hartnoll:2008vx,Hartnoll:2008kx}, is one of the most popular models in the context of holography applied to condensed matter and it has received an enormous amount of attention in the last years (see \cite{Cai:2015cya,Herzog:2009xv} for reviews on the topic). Nevertheless, it presents a ``small'' problem: it does not describe a superconductor. On the contrary, since the U(1) symmetry of the dual field theory is global rather than local, it describes a superfluid. One could argue that for some questions, e.g. the electric conductivity, the difference between the two is not important and hence one could still consider the holographic superfluid model as a weakly gauged holographic superconductor. Unfortunately, for many other features (e.g., the nature and dynamics of vortices, the collective low energy modes, etc.), a superfluid is profoundly different from a superconductor. 

In order to investigate these different aspects, it is imperative to construct a \textit{bona fide} holographic superconductor model. As a matter of fact, that has already been considered by many authors in the past \cite{Montull:2009fe,Domenech:2010nf,Maeda:2010br,Silva:2011zzc,Rozali:2012ry,Gao:2012yw,Salvio:2012at,Salvio:2013ja,Salvio:2013jia,Dias:2013bwa,Zeng:2019yhi,delCampo:2021rak,Li:2021dwp,Natsuume:2022kic,Keranen:2009re,Albash:2009iq,Maeda:2009vf}. The ``trick" to transform a holographic superfluid into a holographic superconductor consists in modifying the boundary conditions for the bulk gauge field from Dirichlet to mixed boundary conditions, as introduced in the seminal works by Witten \cite{Witten:1998qj,Witten:2003ya} (see also \cite{Klebanov:1999tb,Leigh:2003ez,Yee:2004ju,Breitenlohner:1982jf}), and described in detail by Marolf and Ross \cite{Marolf:2006nd}.\footnote{See \cite{Cottrell:2017gkb} for some subtleties about the two different approaches.} This procedure, which is equivalent to a Legendre transform of the dual field theory generating functional together with the introduction of a boundary Maxwell kinetic term, allows to gauge the boundary U(1) symmetry and bring in dynamical electromagnetism in the dual description.

The same type of boundary conditions have resulted to be important in several other holographic applications including the study of plasmons \cite{Gran:2017jht,Gran:2018iie,Gran:2018vdn,Gran:2018jnt,Baggioli:2019aqf,Gran:2019djz,Baggioli:2019sio,Baggioli:2021ujk,Romero-Bermudez:2019lzz,Mauri:2018pzq,Romero-Bermudez:2018etn}, Friedel oscillations \cite{Faulkner:2012gt}, anyons \cite{Jokela:2013hta,Brattan:2013wya,Brattan:2014moa} and magnetohydrodynamics \cite{Ahn:2022azl}.\footnote{In this context, it has been proved in \cite{DeWolfe:2020uzb} that the mixed boundary conditions are equivalent to the action of an electromagnetic duality in the bulk and the usage of higher-form bulk fields therein \cite{Grozdanov:2017kyl}.} An analogous procedure can also be used to make the boundary metric dynamical and obtain semiclassical Einstein equations in the boundary dynamics \cite{Compere:2008us,Ecker:2021cvz,Ishibashi:2023luz}.

One fundamental difference between superfluids and superconductors is the spectrum of collective low-energy excitations. Superfluids are characterized by the appearance of an additional sound mode \cite{doi:10.1063/1.3248499}, known as \textit{second sound}\footnote{Or fourth sound if translational invariance is explicitly broken.}. This new excitation is a direct manifestation of the emergent Goldstone mode of the spontaneously broken U(1) global symmetry. The latter coincides with the fluctuations of the phase of the order parameter which cost no energy. On the contrary, the fluctuations of the amplitude of the order parameter, collectively labelled as the Higgs mode, are not hydrodynamic\footnote{Below the critical point, $T<T_c$, the frequency $\omega(k)$ of the Higgs mode does not go to zero as $k\rightarrow 0$.}, and they are overdamped close to the critical temperature. At low temperature, the Higgs mode is expected to develop a real energy gap which is proportional to the superconducting gap $\Delta$. This whole dynamics can be directly derived using a phenomenological time-dependent Ginzburg Landau (GL) description \cite{larkin2009theory,kopnin2001theory}. At the same time, the late time and long distance dynamics of a superfluid in the broken phase can be consistently described using relativistic superfluid hydrodynamics \cite{doi:10.1063/1.1703944,Schmitt:2014eka,1974anh.....3.....P,Nicolis:2011cs,Son:2002zn,Bhattacharya:2011tra,Herzog:2011ec}, as a formal extension of the two-fluid Tisza-Landau model \cite{tisza1938transport,landau1941theory}. 

\begin{figure}
    \centering
    \includegraphics[width=0.4
    \linewidth]{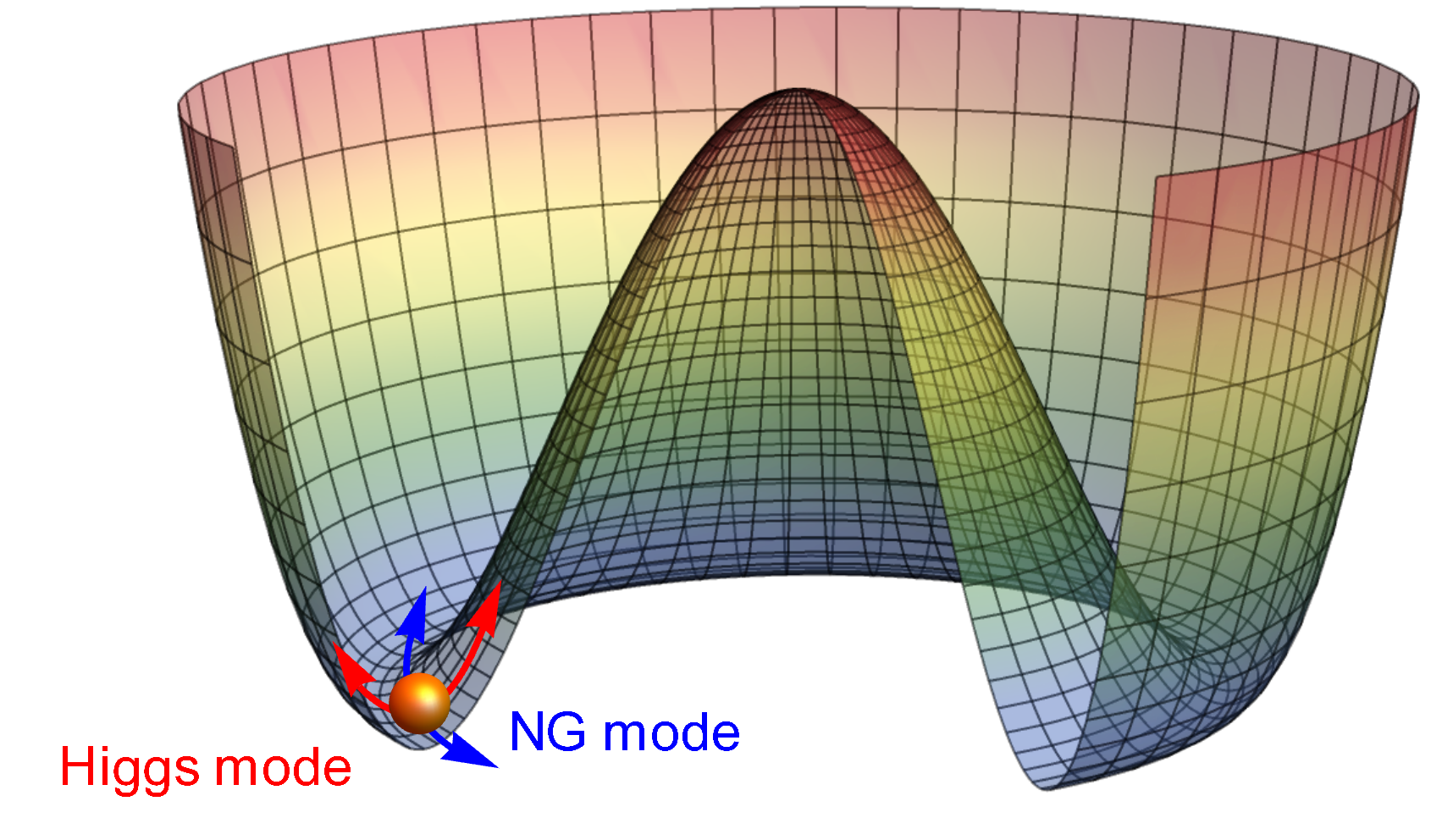}
    \qquad
    \includegraphics[width=0.4\linewidth]{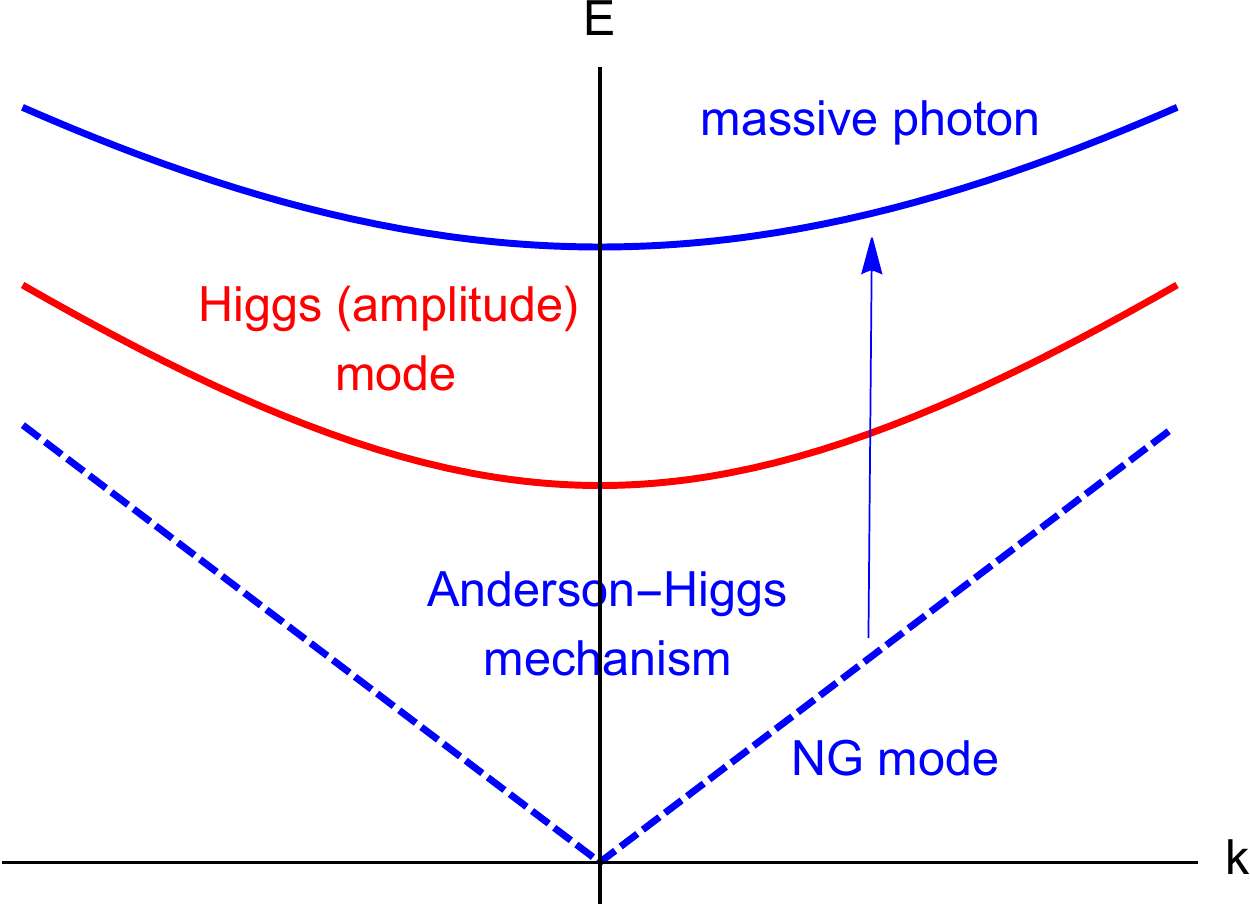}
    \caption{\textbf{Left: }The typical Mexican-hat potential in the GL phenomenological description of 2nd order phase transition. In blue, the fluctuations of the phase of the order parameter, the Nambu Goldstone mode. In red, the fluctuations of the amplitude of the order parameter, the Higgs mode. \textbf{Right: } The spectrum of low-energy collective modes in a superconductor. The Higgs mode is expected to have an energy gap of the size of the superconducting gap. The NG mode is ``eaten" by the gauge field and the photon becomes massive with an energy gap of order of the plasma frequency.}
    \label{fig:0}
\end{figure}

In a superconductor, the major difference with what just described is due to the famous Anderson-Higgs mechanism~\cite{anderson1958coherent,anderson1958random,PhysRevLett.13.508,PhysRev.130.439}. The massless Nambu-Goldstone mode, which in superfluids corresponds to the fluctuations of the phase of the order parameter, gets ``eaten'' by the dynamical gauge field and the corresponding photon becomes massive (see cartoon in Fig. \ref{fig:0}). The mass of the photon is expected to be order of the plasma frequency $\omega_p$ and it is a direct effect of the presence of dynamical electromagnetism. In other words, apart from the presence of first sound\footnote{In the rest of this manuscript, we will not consider the fluctuations of energy and momentum, therefore we will not discuss the dynamics of first sound arising from those.}, in a superconductor, and differently from a superfluid, we do not expect any other gapless excitation.

The low-energy spectrum of holographic superfluids has been investigate numerically by computing the quasinormal modes at finite frequency and wave-vector. In the probe limit, this task has been originally achieved in \cite{Amado:2009ts,Amado:2013aea,Amado:2013xya}.\footnote{See also \cite{Zhao:2022jvs} for the extension to more exotic superfluid phase transitions and \cite{Ammon:2021pyz} for the generalization in presence of a small explicit breaking of the global U(1) symmetry.} More completely, in \cite{Arean:2021tks}, a fully backreacted analysis has been done and matched $1$-to-$1$ with the expectations from relativistic superfluid hydrodynamics. Perturbative computations near the critical point were originally performed in \cite{Herzog:2010vz}. More recently, using a more advanced method based on the concept of symplectic current, extended analytical results have been presented \cite{Donos:2021pkk,Donos:2022www}. Those studies investigated the dynamics of the overdamped order parameter fluctuations \cite{Donos:2022xfd} (see also \cite{Plantz:2015pem,She:2011cm} for earlier studies) and provided a concrete comparison near $T_c$ between the holographic superfluid model, time-dependent Ginzburg Landau theory and model F in Hoenberg-Halperin classification \cite{Donos:2022qao}. To the best of our knowledge, an underdamped Higgs mode with a real energy gap, obeying the standard effective theory expectations \cite{doi:10.1146/annurev-conmatphys-031214-014350,doi:10.1146/annurev-conmatphys-031119-050813}, has never been observed in holographic superfluids. On the contrary, in \cite{Bhaseen:2012gg}, the authors observed the emergence of a pair of underdamped complex valued modes at low temperature arising from microscopic degrees of freedom and not related to the dynamics of the order parameter.\footnote{We thank Aristomenis Donos to point this out to us.}

At the same time, we are not aware of any computation of the low energy collective modes in a \textit{bona fide} holographic superconductor model. In \cite{Gao:2012yw}, the authors made an attempt in this direction by considering purely alternative (i.e., Neumann) boundary conditions for the bulk gauge field. As explained in \cite{Ahn:2022azl}, and re-iterated below, those boundary conditions simply perform a Legendre transform of the boundary action but do not introduce any kinetic term for the boundary gauge field. In other words, those boundary conditions correspond to the limit of infinite boundary gauge coupling and miss most of the relevant physics.

The scope of this work is to fill this gap and study in detail the collective dynamics of a \textit{bona fide} holographic superconductor model at finite frequency and wave-vector. The manuscript is organized as follows. In section \ref{secGL}, following the work of \cite{Grigorishin2021}, we present a phenomenological time dependent Ginzburg Landau description of the collective dynamics; in section \ref{secmod}, we present the holographic setup and all the details related to it; in section \ref{phase}, we describe the main thermodynamic and transport properties in the Higgs phase; in section \ref{sectran}, we present the results for the transverse modes; in section \ref{seclong}, we present the results for the longitudinal modes and evidence for the Anderson-Higgs mechanism; finally, in section \ref{secfin}, we conclude with some final remarks and observations for the future.

%%%%%%%%%%%%%%%%%%%%%%%%%

\section{Ginzburg-Landau phenomenological approach: a review}
\label{secGL}
In this section, we present a brief review of the phenomenological Ginzburg-Landau theory \cite{Ginzburg:1950sr} in its different incarnations. Our task is not to construct a complete Ginzburg-Landau description for strongly coupled superconductors nor to exactly match the results from holography to the effective description. On the contrary, we will use the results presented here as a guidance for the interpretation and discussion of the holographic results. For simplicity, we will follow closely the presentation of Ref.\cite{Grigorishin2021} (see also \cite{HOHENBERG20151,Schmitt:2014eka}). In order to avoid clutter, the speed of light and the Planck constant $\hbar$ will be set to unity in the rest of the manuscript.

\subsection{Ginzburg-Landau theory}

Let us start from the most known form of Ginzburg-Landau theory which is a valid description for a superfluid transition close to the critical point. The starting point is the  free energy density $\mathcal{F}\left[\Psi\right]$ which is expressed as a function of a complex order parameter field $\Psi$,
\begin{equation}\label{GL1}
    F\left[\Psi\right] = F_n\left(T\right)+\int \dd^3r \, \mathcal{F}\left[\Psi\right] =F_n\left(T\right)+\int \dd^3r \left[ a \,|\nabla \Psi|^2 + b\,|\Psi|^2 + \frac{c}{2}\,|\Psi|^4\right] \,,
\end{equation}
where $a,b,c$ are phenomenological parameters. For vanishing order parameter, $\Psi=0$, the free energy 
 $F$ coincides with the normal phase free energy, $F_n\left(T\right)$. 
The complex scalar field can be conveniently parameterized as $\Psi = |\Psi(r)| \, e^{i \theta(r)}$, where $|\Psi(r)|$ is its modulus and $\theta(r)$ its phase.
In order to implement the spontaneous symmetry breaking of the global U(1) symmetry and the transition to a superfluid phase at small temperature, one phenomenologically assumes that $b=\beta \left( T - T_c \right)$. In this way, for $T<T_c$, the quadratic term in the free energy density becomes negative and the minima of the latter are shifted to a finite value of $\Psi$. This is the familiar dynamics of the Mexican-hat potential (see Fig.\ref{fig:0}).
 
Minimizing the functional in Eq.\eqref{GL1}, we obtain the classical equation of motion
\begin{equation}\label{} 
a \nabla^2 \Psi - b \Psi - c |\Psi|^2 \Psi = 0 \,,
\end{equation}
which, for homogeneous solutions, gives rise to an equilibrium value $\Psi_0$:
\begin{equation}\label{EQVAL}
    |\Psi| \,=\, \sqrt{\frac{|b|}{c}} \,\,=:\, \Psi_0 \,\sim\, \sqrt{T_c-T} \,.
\end{equation}
At the critical temperature, the susceptibility $\chi$ diverges, $\chi^{-1} \propto b$, and the heat capacity displays a jump \cite{HOHENBERG20151}.
By construction, the order parameter obeys the mean-field scaling behavior with critical exponent $1/2$.
For later use, we also define the superfluid density $n_s$ and the normal density $n_n$ as
\begin{align}\label{DENSITIES}
\begin{split}
n_s :=2 |\Psi|^2\,,\qquad n_n = n - n_s = n - 2 \Psi_0^2  \,.
\end{split}
\end{align}
Here, $n$ indicates the total density and the factor of $2$ comes from the comparison with the microscopic theory in which the condensate is formed by a pair of electrons \cite{tinkham2004introduction}. 
%\color{blue}Another way to derive this is by analyzing the kinetic term for the phase $\theta$ which arises from the free energy in Eq.\eqref{GL1}. In a non-relativistic system, that term, $a |\Psi|^2 \left(\nabla \theta\right)^2$, should be equal to:
%\begin{equation}
 %   \frac{n_s}{2 m_e} \left(\nabla \theta\right)^2 \quad \text{or equivalently}\quad \frac{m_e \,n_s}{\hbar^2}\,v_s^2 \quad \text{with}\quad v_s := \frac{\hbar}{m_e}\,\nabla \theta\,,
%\end{equation}
%where $v_s$ is the superfluid velocity, $m_e$ the effective mass of the electrons. By setting $a=\hbar^2/(4m_e)$ \cite{Grigorishin2021}, one immediately recovers that $n_s=2 |\Psi|^2$, as already advertised.
%\kyr{Where does Eq.(2.5) come from? maybe there is some typo for the power of $\nabla \theta$?}\color{black}\MB{there is a typo but maybe just better to remove it all. its not very helpful to go talking about the relativistic limit.}

In order to promote this picture out of equilibrium, different routes can be followed. At first, we will ignore dissipative terms, and just insist on a field theory approach based on a Lagrangian formalism. Later, we will discuss in detail the shortcomings of this picture. The idea is to promote the Ginzburg-Landau functional to an action $S$ defined in Minkowski space with coordinates $\{v t, \vec{r}\}$:
\begin{equation}\label{SFUN}
    S = \int \dd t \,\dd^3 r  \, \mathcal{L} \,,
\end{equation}
where $v$ is an emergent lightcone velocity which does not depend a priori on temperature.
A simple way to build the Lagrangian $\mathcal{L}$ in Eq.\eqref{SFUN} is to recast the free energy density $\mathcal{F}$ in Eq.\eqref{GL1} in a relativistic-invariant form using the following substitution
\begin{equation}\label{}
\nabla \Psi \,\,\rightarrow\,\,  \partial_\mu \Psi \,, \qquad \partial_\mu := \left(\frac{\partial_t}{v},\nabla\right)  \,.
\end{equation}
The corresponding Lagrangian can be then written down as
\begin{align}\label{LAGEQ}
\begin{split}
\mathcal{L} = a \left(\partial_\mu \Psi\right)\left(\partial^\mu \Psi^{*}\right) - b\,|\Psi|^2 - \frac{c}{2}\,|\Psi|^4  \,=\,  \frac{a}{v^2} \, \frac{\partial \Psi}{\partial t} \frac{\partial \Psi^{*}}{\partial t} - \mathcal{F} \,.
\end{split}
\end{align}
For stationary solutions, i.e., equilibrium configurations, the dynamics obtained from the action principle in Eq.\eqref{LAGEQ} reduces to the standard  Ginzburg-Landau theory in Eq.\eqref{GL1}.

Decomposing the complex scalar order parameter into its modulus and phase, the Lagrangian in Eq.\eqref{LAGEQ} can be further expressed 
\begin{align}\label{LAGEQ2}
\begin{split}
\mathcal{L} = a \,\partial_\mu |\Psi| \,\, \partial^\mu |\Psi^{*}| + a \, |\Psi|^2 \,\partial_\mu \theta \,\, \partial^\mu \theta - b\,|\Psi|^2 - \frac{c}{2}\,|\Psi|^4   \,.
\end{split}
\end{align}
In order to study the dynamics out of equilibrium, let us consider a small deviation of the modulus from its equilibrium value
\begin{align}\label{OUEQ}
\begin{split}
|\Psi| = \Psi_0 + \phi   \,,\quad  (\phi \ll \Psi_0) \,,
\end{split}
\end{align}
where $\Psi_0$ is given in Eq.\eqref{EQVAL} and it is real valued.
Then, the Lagrangian in Eq.\eqref{LAGEQ2} reduces to
\begin{align}\label{LAGEQ3}
\begin{split}
\mathcal{L} = a \,\partial_\mu \phi \,\, \partial^\mu \phi  - 2 |b|\,\phi^2  + a \, \Psi_0^2 \,\partial_\mu \theta \,\, \partial^\mu \theta + \frac{b^2}{2c}   \,,
\end{split}
\end{align}
which consequently yields to the two dynamical equations
\begin{align}\label{LAGEQ4}
\begin{split}
 a \left(\frac{1}{v^2} \frac{\partial^2 \phi}{\partial t^2} - \nabla^2 \phi \right) +2 |b| \phi = 0 \,, \quad  \frac{1}{v^2} \frac{\partial^2 \theta}{\partial t^2} - \nabla^2 \theta = 0 \,.
\end{split}
\end{align}
The former is the equation for the modulus of the complex order parameter, often indicated as the Higgs/amplitude mode, while the latter is that for the phase, which is identified with the Goldstone mode. 

By going to Fourier space, and solving the above equations, we obtain two different low-energy excitations which are described by
\begin{align}\label{HGMO}
\begin{split}
\text{Higgs mode:} \quad  \omega^2 = \frac{2 |b| v^2}{a} + v^2 k^2 \,, \qquad
\text{Goldstone mode:} \quad \omega^2 = v^2 k^2 \,.
\end{split}
\end{align}
As expected, the Goldstone mode shows a gapless dispersion relation with velocity $v$. On the contrary, the Higgs mode presents an energy gap 
\begin{equation}
 \omega_H := \sqrt{\frac{2 |b| v^2}{a}}\,, 
\end{equation}
which vanishes at the critical temperature as $\sim \sqrt{T_c-T}$.
In addition, the Higgs mass in Eq.\eqref{HGMO} can be obtained using the relativistic formula for the energy, $\omega_H^2 = m_H^2 v^4 + P^2 v^2$, as
\begin{align}\label{HIGM}
\begin{split}
m_H := \frac{\omega_H}{v^2} = \sqrt{\frac{2 |b|}{a v^2}} \,\sim\, \sqrt{T_c-T} \,, 
\end{split}
\end{align}
In what follows, we use the word ``mass" interchangeably with the term ``energy gap".\\
Before continuing, let us emphasize the (many) shortcomings of this first simple approach. (I) All dissipative effects are neglected. The latter would have several effects on the dispersion relation of the modes discussed. First, they would introduce attenuation in the dispersion of the Goldstone mode. Second, they would make the Higgs mode overdamped close to the critical temperature. (II) The dynamics considered so far is restricted to the order parameter $\Psi$ and ignores completely its coupling to other conserved quantities as charge density, momentum and energy. Moreover, we ignored the coupling to a potential external gauge fields, parameterizing for example an external chemical potential or superfluid velocity. (III) The Lagrangian construction is completely phenomenological and poorly motivated. In particular, it is not able to reproduce the well-known fact that the speed of propagation of the Goldstone mode vanishes at the critical temperature $T=T_c$. This is simply because the velocity $v$ is introduced by hand using the emergent light-cone structure and it is not related to the superfluid density as it should. Within the standard GL picture, in order to obtain propagating modes, as second or fourth sound, one needs to include reactive couplings to other conserved quantities such as charge density (see for example \cite{HOHENBERG20151}). Proceeding in this section, we will describe some of the more advanced alternatives to this method and discuss the possibility to have a complete description of the dissipative dynamics.

\subsection{Anderson-Higgs mechanism}

So far, we have considered a system with a global U(1) symmetry and in particular the transition between a normal fluid to a superfluid state. Now, we want to promote the Ginzburg-Landau description to the case of superconductors where the U(1) symmetry is gauged. In order to do that, we perform the following transformation
\begin{align}\label{}
\begin{split}
\partial_\mu \Psi  \,\,\rightarrow\,\, D_\mu \Psi := \left( \partial_\mu + i \tilde q A_\mu \right)\Psi   \,,
\end{split}
\end{align}
where we have defined for convenience $\tilde q 
:= q/v$. Moreover, we add in the Lagrangian a coupling to an external current $J^\mu_{\text{ext}}$ and a kinetic term for the dynamical gauge field:
\begin{equation}
    \mathcal{L} \,\,\rightarrow\,\, \mathcal{L} -A_\mu J^\mu_{\text{ext}}-\frac{1}{4 \lambda} F^2 \,.
\end{equation}
Here, $\lambda$, parameterizes the strength of the gauge coupling or, in other words, the strength of the electromagnetic interactions.

By setting the external sources to zero, $J^\mu_{\text{ext}}=0$, we obtain:
\begin{align}\label{}
\begin{split}
\mathcal{L} = a \left( \partial_\mu + i \tilde q A_\mu \right)\Psi \, \left( \partial^\mu - i \tilde q A^\mu \right)\Psi^{*}  - b\,|\Psi|^2 - \frac{c}{2}\,|\Psi|^4  -\frac{1}{4 \lambda} F^2 \,,
\end{split}
\end{align}
which, following the same steps as before, can be expressed near equilibrium 
\begin{align}\label{LAGAMU}
\begin{split}
\mathcal{L} = a \,\partial_\mu \phi \,\, \partial^\mu \phi  - 2 |b|\,\phi^2  + \frac{b^2}{2c} + a \, \tilde q^2 \, \Psi_0^2 \,A_\mu A^\mu  -\frac{1}{4 \lambda} F^2  \,,
\end{split}
\end{align}
where we neglected mixed terms $\sim \Psi_0 \phi A_\mu A^\mu$ (see \cite{Grigorishin2021} for details regarding this approximation). In addition, the phase degree of freedom $\theta$ has disappeared from the Lagrangian as it can be simply reabsorbed into a gauge transformation.
Comparing the Lagrangian in Eq.\eqref{LAGAMU} with that in Eq.\eqref{LAGEQ3}, one can notice that the phase $\theta$ (the Goldstone mode) is absorbed into the gauge field $A_\mu$ which has now acquired a finite mass $\propto \Psi_0^2$. This is the famous Anderson-Higgs mechanism \cite{anderson1958coherent,anderson1958random,Higgs:1964ia,PhysRevLett.13.508,PhysRev.130.439}.

Using Eq.\eqref{LAGAMU}, the equations of motion for the gauge field can be derived as
\begin{equation}\label{LAMBDA2}
    \partial_\mu F^{\mu\nu} + \frac{1}{\lambda_{GL}^2} A^\nu = 0 \,, \qquad \lambda_{GL}^2 := \frac{1}{2 a \, \tilde{q}^2 \Psi_0^2 \, \lambda}\,,
\end{equation}
i.e., the famous London equation, where $\lambda_{GL}$ is the London penetration length. 

In Fourier space, the dispersion relation for the photon becomes
\begin{align}\label{DISGA}
\omega^2 = \omega_A^2 + v^2 k^2 \,, \qquad \omega_A := \frac{v}{\lambda_{GL}}= q \Psi_0 \sqrt{2 a \lambda} \,.
\end{align}
The gauge field mass in the relativistic form is then given by 
\begin{align}\label{GAGM}
\begin{split}
m_{A} := \frac{\omega_A}{v^2} = \frac{1}{v\,\lambda_{GL}}  \,\sim\, \sqrt{T_c-T} \,,
\end{split}
\end{align}
where we used the expression for $\lambda_{GL}$ in \eqref{LAMBDA2} and Eq.\eqref{EQVAL}.

As for the Higgs mode \eqref{HIGM}, mass of the gauge field \eqref{GAGM} vanishes at the critical temperature following the mean-field behavior $\left(T-T_c\right)^{1/2}$, but with a different multiplicative prefactor.
Taking the ratio between the two masses (or energy gaps), we get:
\begin{equation}\label{RATIOHA}
    \frac{m_H}{m_A} = \frac{\omega_H}{\omega_A} = \sqrt{2}\,\frac{\lambda_{GL}}{\xi_{GL}} =: \sqrt{2} \,\kappa_{GL} \,,
\end{equation}
where we have defined the GL parameter $\kappa_{GL}$, and the correlation length $\xi_{GL}$
\begin{align}\label{CLNEWDEF}
\begin{split}
\xi_{GL} := \sqrt{\frac{a}{|b|}} \,.
\end{split}
\end{align}
This shows that, depending on the type of superconductor, one mass could be larger or smaller than the other. Indeed, for type-I superconductors one has $\kappa_{GL}<1/\sqrt{2}$ while, for type-II superconductors, $\kappa_{GL}>1/\sqrt{2}$.
In our general scenario, in which the EM coupling is taken as arbitrary, this distinction depends on the value of $\lambda$ since $\kappa_{GL}\sim1/\sqrt{\lambda}$.

By projecting these expressions to zero temperature, one obtains an interesting result regarding the mass of the photon field in the zero temperature limit. Let us stress that this extrapolation is a priori not trustable since the GL framework is reliable only close to the critical point around which the value of the order parameter $\Psi$ is small, and the free energy can therefore be legitimately expanded in powers of it. On the contrary, going at low temperature, the order parameter grows and the GL treatment is not well grounded. Nevertheless, let us abuse of this approximation and see what we get. In the limit of zero wave-vector, $k=0$, and zero temperature, the dispersion relation in Eq.\eqref{DISGA} becomes
\begin{align}\label{}
\begin{split}
\omega_A(T=0) = \frac{v}{\lambda_{GL}(T=0)} =  \sqrt{2 a\, q^2 \Psi_0^2(T=0) \, \lambda} =  \sqrt{a\, q^2 \, n\, \lambda} \,,
\end{split}
\end{align}
where we have used the expression for the London penetration length $\lambda_{GL}$ in Eq.\eqref{LAMBDA2}.
In addition, in the last equality, $\Psi_0$ is replaced by the total density $n$ using \eqref{DENSITIES}:
\begin{align}\label{}
\begin{split}
n_n = n - 2 \Psi_0^2  \quad\xrightarrow[]{T=0}\quad  n = 2 \Psi_0^2 \,.
\end{split}
\end{align}
Here, we have assumed that the normal component $n_n$ is vanishing at $T=0$. Under this assumption, one can see that 
\begin{align}\label{PLAFOR22}
\begin{split}
\omega_A(T=0) =  \omega_p \,,\qquad \omega_p:= \sqrt{a \, q^2 \, n \, \lambda} \,,
\end{split}
\end{align}
where $\omega_p$ is the plasma frequency.\footnote{In the non-relativistic limit, one has:
\begin{equation}
    a = \frac{\hbar^2}{4m_e}\,,\qquad q = \frac{2e}{\hbar}\,,
\end{equation}
and then recovers the familiar expression for the plasma frequency~\cite{Grigorishin2021}:
\begin{equation}
    \omega_p^2  = \frac{\lambda\, e^2 n }{m_e}\,.
\end{equation}
For relativistic systems, the mass density $n m_e$ has to be substituted with the relativistic form $\epsilon+p$. More in general, what appears in the denominator is the momentum susceptibility $\chi_{\pi\pi}$. In relativistic systems, $\chi_{\pi\pi}$ takes the aforementioned form because of the equivalence between energy current and momentum density imposed by the Lorentz boosts Ward identity.}

Following this argument, we find that the ``mass" of the photon in the zero temperature limit is determined by the value of the plasma frequency.
This can also be thought as a consequence of the Anderson-Higgs mechanism. In other words, we do expect the sound mode to be pushed by Coulomb interactions to the plasma frequency value. Since, via the Anderson-Higgs mechanism, the Goldstone mode is absorbed into the gauge field, the ``mass" of the gauge field at small temperature is pushed to the plasma frequency value as well.

\subsection{Dissipative effects}
So far, we have completely ignored any dissipative effects coming from the conductivity, the viscosity, etc. For simplicity, we will first follow the treatment in \cite{Grigorishin2021} and then discuss possible improvements.

The first effect of dissipation comes from the fact that the material is a conductor, with a finite conductivity $\sigma$. Because of this reason, the electric permittivity cannot be assumed to be a constant. On the contrary, in the simplest scheme of approximations, it becomes a complex and frequency dependent quantity given by
\begin{equation}\label{sub}
    \epsilon_0 \rightarrow \epsilon(\omega)=\epsilon_0\left(1+i\,\lambda\, \frac{\sigma(\omega)}{\omega}\right) \,.
\end{equation}
This substitution arises naturally in the standard treatment of electromagnetism in conductors \cite{griffiths2014introduction}, and it can be easily derived from the Maxwell equation:
\begin{equation}
    \nabla \cdot D=\rho\,,\qquad D=\epsilon\,E\,,
\end{equation}
where $D$ is the displacement vector and $E$ the electric field. By using Ohm's law, $J=\sigma E$, together with the continuity equation $\partial_t \rho +\nabla \cdot J=0$, one obtains that:
\begin{eqnarray}
    \rho=\frac{k \cdot J}{\omega}=\sigma \,\frac{k \cdot E}{\omega} \,,
\end{eqnarray}
which, plugged into the Gauss law for the displacement vector, gives rise to the substitution in Eq.\eqref{sub}. Under this simple replacement, the dispersion relation of the massive gauge field in Eq.\eqref{DISGA} becomes \cite{Grigorishin2021} 
\begin{align}\label{lala}
\begin{split}
\omega^2 = \omega_A^2 + v^2 k^2 - i v^2 \lambda \, \sigma \, \omega \,, \qquad \omega_A := \frac{v}{\lambda_{GL}}= q \Psi_0 \sqrt{2 a \lambda}
\end{split}
\end{align}
where we have assumed the conductivity $\sigma(\omega)$ to be a constant. Frequency dependent terms in the conductivity are obviously present but will not affect the dispersion relation at leading order in $\omega$.\\
Notice that for $\omega_A=0$ this equation describes the propagation of electromagnetic waves in a conductor and implies the well-known skin-effect arising from the imaginary term $\propto \lambda$. Usually, this equation is solved by assuming a complex-valued wave-vector and a real-valued frequency. Here, we take the opposite approach and consider the wave-vector real and the frequency complex.
At zero wave-vector, the solutions of Eq.\eqref{lala} are given by
\begin{align}\label{MGDR}
\begin{split}
\omega = -i  \frac{v^2 \lambda \, \sigma}{2} \pm \sqrt{\omega_A^2 - \left( \frac{v^2 \lambda \, \sigma}{2}\right)^2} \,.
\end{split}
\end{align}
Because of the square root structure, there is clearly a competition between the dissipative effects and the mass term $\omega_A$ arising because of Anderson-Higgs mechanism which can result in a overdamped mode or an underdamped one. More precisely, the excitations of the gauge field show a real mass gap only when:
\begin{equation}
    \omega_A>\frac{v^2  \lambda \, \sigma}{2}\,.
\end{equation}
On the contrary, in the limit of strong dissipation, the frequencies are purely imaginary.

Let us consider the two limiting cases: (I) the near-critical region $T\approx T_c$ and (II) the low temperature region $T\approx 0$.
When $T \rightarrow T_c$, $\omega_{A}\rightarrow0$ because of Eq.\eqref{DISGA} (or equivalently $\lambda_{GL} \rightarrow \infty$), then Eq.\eqref{MGDR} gives two simple solutions
\begin{align}\label{}
\begin{split}
\text{$T \rightarrow T_c$:} \qquad \omega = -i v^2  \lambda \, \sigma \,, \qquad \omega = 0 \,,
\end{split}
\end{align}
where the decay time of the overdamped mode is $\tau=1/(v^2 \lambda \, \sigma)$. On the other hand, at $T\rightarrow0$, we do expect all dissipative effects, and in particular the conductivity $\sigma$, to vanish. The same equation gives rise to a pair of solutions which read
\begin{align}\label{}
\begin{split}
\text{$T \rightarrow 0$:} \qquad \omega = \pm \,\omega_{A} - i \frac{v^2 \lambda\,\sigma}{2}  = \pm \, \omega_p - i \frac{v^2 \lambda\,\sigma}{2} \,.
\end{split}
\end{align}
where we used \eqref{PLAFOR22} in the last equality.
In this opposite case, the excitations have a real gap with a small attenuation constant. Assuming that the conductivity vanishes at zero temperature, one would simply get $\omega=\pm \, \omega_p$ at exactly $T=0$. The crossover between the overdamped (high $T$) and underdamped (low $T$) regimes can be approximately found by equating the two terms, $\omega_A=\frac{v^2  \lambda \, \sigma}{2}$,
\begin{align}\label{}
\begin{split}
\lambda_{GL}(T) \, \sigma(T) = \frac{2}{v\,\lambda} \,.
\end{split}
\end{align}
Using common values for these quantities in weakly coupled superconductors, one obtains that the crossover temperature $T^*$ is approximately given by $T^*/T_c \sim 0.5$ \cite{Grigorishin2021}.

In a similar way, using the Rayleigh dissipation function formalism \cite{https://doi.org/10.1112/plms/s1-4.1.357}, the effects of the conductivity on the dispersion relation of the Higgs mode, Eq.\eqref{HGMO}, can be incorporated in the attenuation constant $\gamma$ given by \cite{Grigorishin2021}:
\begin{align}\label{}
\begin{split}
\gamma \sim \sigma \, \xi_{GL}^2  \, \Psi_0^2 \,.
\end{split}
\end{align}
where $\xi_{GL}$ is given in \eqref{CLNEWDEF}.
In this approximation, the dispersion relation of the mode is modified into
\begin{align}\label{HGMO2}
\begin{split}
 \omega^2 = \omega_H^2 + v^2 k^2  - i 2 \gamma \omega \,, \qquad \omega_H^2 := \frac{2 |b| v^2}{a} \,
\end{split}
\end{align}
which is of the same form of that for the gauge field fluctuations in Eq.\eqref{lala}.

Similarly, we consider the zero momentum solution of Eq.\eqref{HGMO2} which reads
\begin{align}\label{HGMO3}
\begin{split}
 \omega = -i \gamma \pm \sqrt{\omega_H^2-\gamma^2}  \,. 
\end{split}
\end{align}
In the near-critical regime, where $T \rightarrow T_c$,  we have that
\begin{align}\label{}
\begin{split}
\text{$T \rightarrow T_c$:} \qquad  \omega_H \sim (T-T_c)^{1/2} \,, \qquad \gamma\sim\xi_{GL}^2 \Psi_0^2 \sim const. 
\end{split}
\end{align}
Thus, around the critical point we have $\omega_H \ll \gamma$ which implies the appearance of two overdamped modes of the type
\begin{align}\label{hehe00}
\begin{split}
 \omega = -i\, \frac{\omega_H^2}{2 \gamma} \,, \qquad \omega = -2 i \gamma \,, 
\end{split}
\end{align}
where the relaxation time of the longest-living excitation is given by
\begin{align}\label{hehe}
\begin{split}
\tau_1 = \frac{2 \gamma}{\omega_H^2} \sim \frac{1}{|T-T_c|}\,.
\end{split}
\end{align}
Note that the strong effects of damping near $T_c$ render the observation of the Higgs mode problematic since the latter is strongly overdamped.
In the opposite limit of small temperature, we do expect the conductivity to vanish and we therefore expect the effects of dissipation to be negligible compared to the $\omega_H$ term. In particular, there, we do expect a pair of weakly attenuated modes with a real gap $\omega_H$
\begin{align}\label{T0HGEQ}
\begin{split}
\text{$T \rightarrow 0$:} \qquad \omega = \pm \,\omega_H -i \gamma \,.
\end{split}
\end{align}

Before concluding, let us present some remarks about the introduction of dissipative effects and the coupling to other conserved quantities. A standard way to promote the Ginzburg-Landau framework out of equilibrium and include dissipative effects is the so-called time-dependent complex Ginzburg-Landau theory \cite{RevModPhys.74.99}. Let us sketch the idea quickly by considering the dynamics of the complex order parameter $\Psi$ and the ungauged case (i.e., the superfluid). While the equilibrium solution is given by minimizing the free energy density introduced in Eq.\eqref{GL1}, the deviations from it are assumed to obey the simple time-dependent equation:
\begin{equation}
    \frac{\partial \Psi}{\partial t}=-\Gamma_0\,\frac{\delta \mathcal{F}}{\delta \Psi^*} \,,
\end{equation}
where $\Gamma_0$ is a phenomenological parameter which governs the relaxation of the order parameter. In general, the latter is taken to be a complex number. At the linearized level, and neglecting inhomogeneities, this equation also predicts the appearance of an overdamped amplitude (Higgs) mode near the critical point with dispersion:
\begin{equation}
    \omega=-i \,\mathrm{Re}\left[\Gamma_0\right]\,b+\dots \,.
\end{equation}
This result is qualitatively analogous to what was obtained before using the Rayleigh dissipation function. Indeed, also in this case, the imaginary gap of the amplitude mode vanishes at the critical temperature. Notice that in this language the mass of the Higgs mode is controlled by the imaginary part of the phenomenological parameter $\Gamma_0$, $\omega_H=\mathrm{Im}\left[\Gamma_0\right]\,b$, and also vanishes as expected at the critical point. 

More in general, in order to consider the near-critical dynamics of a superfluid, and in particular to obtain also propagating modes, one needs to couple the dynamics of the order parameter to the conserved charge density \cite{HOHENBERG20151}. For superfluids, this procedure will automatically end up in the so-called model F in the Hoenberg-Halperin classification \cite{RevModPhys.49.435} (see also \cite{Donos:2022qao} for a holographic derivation of this dynamics in the holographic superfluid model, \cite{Flory:2022uzp,Cao:2022mep} for a study of the nonlinear dynamics and \cite{Maeda:2009wv} for an analysis of the universality class of holographic superconductors). It would be interesting to extend the model F in order to account for a dynamical gauge field and the coupling between the different modes.

Notice that model F does not take into account the dynamics of energy and momentum fluctuations, which will be anyway irrelevant for our holographic model in the probe limit. One could also formally derive a hydrodynamic theory for the superconductor by matching together magneto-hydrodynamics with the spontaneous breaking of the U(1) symmetry. In this case, the most challenging question is how to incorporate in a precise way the presence of non-hydrodynamic modes therein, as it is the case for the fluctuations of the amplitude mode. Near the critical temperature, an approach similar to those used around the QCD critical point \cite{Grossi:2021gqi,Grossi:2020ezz,Stephanov:2017ghc} or those employed for pinned charge density waves \cite{Baggioli:2022pyb} might work.

We leave the construction of a complete and rigorous effective description of the superconducting critical dynamics in presence of dissipation as a task for the future. We will come back to this discussion in the outlook.

%%%%%%%%%%%%%%%%%%%%%%%%%%%%%%%%
%    
%%%%%%%%%%%%%%%%%%%%%%%%%%%%%%%%
\section{The holographic setup}
\label{secmod}
We consider the four dimensional Abelian-Higgs bulk action~\cite{Hartnoll:2008vx,Hartnoll:2008kx} 
\begin{equation}\label{GENMODELXX2222}
\begin{split}
S_{\text{bulk}} = \int \dd^4x \sqrt{-g} \left[ R + 6 - \frac{1}{4} F^2 -|D\Phi|^2 \, -M^2 |\Phi|^2  \right]  \,,
\end{split}
\end{equation}
in presence of a negative cosmological constant $ \Lambda=-3$. We have defined the bulk field strength $F:=dA$ and the covariant derivative $D_{\mu} := \grad_{\mu} -i Q A_{\mu}$, with $Q$ the charge of the complex bulk scalar field $\Phi$ and $M$ its mass.

For simplicity, we work in the probe limit in which the dynamics of the metric fluctuations is kept frozen. The corresponding equations of motion for the matter bulk fields are given by:
\begin{align}
&\nabla_\mu  F^{\mu\nu} - i Q ( \Phi^* D^\nu \Phi - \Phi D^\nu \Phi^*   )=0 \,, \label{eom1} \\
&\left(D^2 - M^2 \right) \Phi =0 \,, \quad \left(D^2 - M^2 \right) \Phi^{*} =0 \label{eom2}.
\end{align}
The background metric is chosen as:
\begin{equation}\label{METABG}
\dd s^2 =  \frac{1}{z^2}\left(-f(z)\, \dd t^2 +  \frac{\dd z^2}{f(z)}  +  \dd x^2 +  \dd y^2 \right) \,,
\end{equation}
with the emblackening factor which takes the Schwarzschild form:
\begin{equation}\label{BWLESD}
f(z) = 1 - \frac{z^3}{z_{h}^3}\,.
\end{equation}
The corresponding temperature and entropy density of the dual field theory are given by:
\begin{equation}
    T = \frac{3}{4 \pi z_{h}}\,,\quad s= \frac{4\pi}{z_h^{2}} \,.
\end{equation}
Finally, the ansatz for the bulk matter field is taken as:
\begin{equation}\label{PROBEFIELDSEQ}
A =  A_{t}(z) \, \dd t \,,  \qquad \Phi = \psi(z)\,.
\end{equation}
Note that, with $A_{z}=A_{x}=A_{y}=0$, the Maxwell equation of motion \eqref{eom1} implies that the phase of the scalar $\Phi$ is a constant~\cite{Hartnoll:2008vx}. Hence, for simplicity, we set the background phase to be zero and take $\Phi$ to be a real scalar in the background. 

Using the aforementioned notations, the bulk equations of motion can be written as 
\begin{equation}\label{EOMBG}
\begin{split}
A_{t}'' - \frac{2 Q^2 \psi^2}{z^2 \,f} A_{t} = 0 \,, \quad
\psi'' + \left( \frac{f'}{f} - \frac{2}{z} \right)\psi' + \frac{Q^2 A_t^2}{f^2}\psi - \frac{M^2}{z^2 f}\psi = 0 \,,
\end{split}
\end{equation}
and are solved numerically integrating them from the horizon $(z=z_{h})$ to the boundary $(z=0)$.
For the concrete numerical computations, we take $(z_{h}, Q, M^2)=(1, 1, -2)$. We assume standard quantization for the bulk scalar field and fix the conformal dimension of the dual operator to be $\Delta_\psi=2$. At the horizon, we impose the regularity conditions for both the gauge field, $A_{t}(z_{h}=1)=0$, and the scalar field.
Near the boundary, the matter fields behave as 
\begin{equation}\label{BGADSEX}
\begin{split}
A_{t} = \mu - \rho z + \mathcal{O} (z^2) \,, \quad \psi = \psi_1 z + \psi_2 z^2 + \mathcal{O} (z^3) \,.
\end{split}
\end{equation}
Using the holographic dictionary, $\mu$ can be interpreted as the chemical potential in the dual field theory and $\rho$ as the charge density.
Moreover, using standard quantization for the scalar field, $\psi_{1}$ represents the source for the dual scalar operator (the order parameter) and $\psi_2$ its the expectation value, i.e., the scalar condensate $\langle O_{2} \rangle$. 
In order to describe the spontaneous symmetry breaking of the dual $U(1)$ symmetry, we always set the source to be zero, $\psi_{1}=0$. We will describe the main physical properties of the broken phase in section \ref{phase}.
\subsection{Fluctuations and boundary conditions}
In order to study the dynamics of the low energy modes in the dual field theory, on top of the background solution Eq.\eqref{METABG}, we switch on the following bulk field fluctuations:
\begin{equation}\label{}
\begin{split}
    \delta A &=  \delta a_{t}(t, z, x) \, \dd t + \delta a_{x}(t, z, x) \, \dd x + \delta a_{y}(t, z, x) \, \dd y \,, \\ 
\delta \Psi &=  \delta \sigma(t, z, x) + i \, \delta \eta(t, z, x) \,,
\end{split}
\end{equation}
where the radial gauge $A_{r}=0$ is assumed. Importantly, we work in the probe limit in which the fluctuations of the metric are kept frozen. Moreover, we decompose all fluctuations in Fourier space using the notation:
\begin{equation}\label{}
\begin{split}
\xi(t, z, x) =   \bar{\xi}(z) \, e^{i (k x - \omega t)} \,.
\end{split}
\end{equation}
where for simplicity the wave-vector $\Vec{k}$ is aligned along the $x$ direction and $\xi$ is a collective label denoting a generic bulk field fluctuation.

The equations of motion for the fluctuations arising from Eqs.\eqref{eom1}-\eqref{eom2} decouple into two independent sectors:
\begin{equation}\label{FLUCSET}
\begin{split}
\text{Longitudinal sector:} &\quad \{\delta a_{t}(z), \delta a_{x}(z), \delta \sigma(z), \delta \eta(z) \} \,, \\
\text{Transverse sector:} &\quad \{\delta a_{y}(z) \}.
\end{split}
\end{equation}
Note that the complex scalar field fluctuation ($\delta \sigma, \delta \eta$) are \textit{only} coupled to the longitudinal vector components ($\delta a_{t}, \delta a_{x}$).
The equations in each sectors are as follows.
In the longitudinal sector, we have
\begin{eqnarray}\label{eq:coupledEOMsBroken}
0 &=&  f \delta \eta''
 +\left(f'-\frac{2f}{z}\right) \delta\eta'+\left(\frac{Q^2 A_{t}^2}{f}-\frac{M^2}{z^2}
 +\frac{{ \omega}^2}{f}-k^2 \right) \delta\eta 
 -\frac{2 Q \, i  \omega A_{t} }{f}\delta\sigma
 -\frac{i Q \, \omega \psi}{f}\delta a_t - i Q \, k \psi \, \delta a_x  \, , \nonumber \\
 &&\\
0 &=&  f \delta\sigma''
 +\left(f'-\frac{2f}{z}\right) \delta\sigma'+\left(\frac{Q^2 A_{t}^2}{f}-\frac{M^2}{z^2}
 +\frac{{ \omega}^2}{f}-k^2 \right) \delta\sigma 
 +\frac{2  Q^2 \, A_{t} \psi }{f} \delta a_t+\frac{2 Q \, i  \omega A_{t} }{f}\delta\eta \, , \\
0 &=& f {\delta a_t}''-\left(k^2 + 2 Q^2 \frac{\psi^2}{z^2}\right){\delta a_t}
-\omega k \, \delta a_x - \frac{2 Q i  \omega\psi}{z^2}\delta \eta-4 Q^2 \frac{A_{t}\, \psi}{z^2}\sigma  \, , \\
0 &=& f {\delta a_x}'' +f'{\delta a_x}'+\left(\frac{{ \omega}^2}{f}-2 Q^2 \frac{\psi^2}{z^2} \right){\delta a_x}
+\frac{ \omega  k}{f} \delta a_t +\frac{2 Q i  k\psi}{z^2} \delta \eta\,,
\end{eqnarray}
together with the constraint equation
\begin{equation}\label{eq:constraintBroken}
\frac{ \omega}{f}{\delta a_t}'+ k {\delta a_x}'= \frac{2 Q i}{z^2} \left(\psi'\,\delta \eta-\psi\,\delta\eta'\right)\, .
\end{equation}
In the transverse sector, the dynamics of the fluctuations is controlled by
\begin{eqnarray}\label{eq:transVecEOMBroken}
0&=& f\,{\delta a_y}'' +f'{\delta a_y}'+\left(\frac{{\omega}^2}{f}
-k^2 -2 Q^2 \frac{\psi^2}{z^2}\right)\,\delta a_y \, .
\end{eqnarray}

After defining the equations of motion, we need to specify the boundary conditions for the fluctuations and in particular for those of the bulk gauge field.
Following Ref.\cite{Domenech:2010nf} and our more recent work, Ref.\cite{Ahn:2022azl}, we promote the external gauge field in the boundary field theory to be a dynamical field. This is fundamental to describe a superconducting phase rather than a superfluid one.

Let us start by considering the bulk Maxwell action in (3+1) dimension as
\begin{equation}\label{GENMODELXX}
\begin{split}
S_{\text{bulk}} = -\frac{1}{4 e^2}\int \dd^4x \sqrt{-g}  F^2    \,,
\end{split}
\end{equation}
where $F=\dd A$ is the field strength for the $U(1)$ gauge field $A$ and the EM bulk coupling $e$ is re-introduced for clarity.
We then introduce the following boundary terms
\begin{equation}\label{BDRYAC}
\begin{split}
S_{\text{boundary}} = \int \dd^3x \, \left[ -\frac{1}{4 \lambda} F_{\mu\nu}^2 \,+\, J_{\text{ext}}^{\mu} \, A_{\mu}  \right] \,,
\end{split}
\end{equation}
where $\lambda$ parameterizes the strength of Coulomb interactions at the boundary (not to be confused with the bulk coupling $e$ in Eq.\eqref{GENMODELXX}) and the last term is just a Legendre transform in terms of an external current $J^\mu_{\text{ext}}$. 

The variation of the total action, $S_{\text{tot}}:= S_{\text{bulk}}+S_{\text{boundary}}$, reads
\begin{equation}\label{SUMBDY333}
\begin{split}
\delta_{A_\mu} S_{\text{tot}} &\,=\, \int \dd^3x \, \left[ \Pi^{\mu} \,-\, \frac{1}{\lambda} \partial_{\nu} F^{\mu\nu} + J_{\text{ext}}^{\mu}  \right]  \delta A_{\mu}   \,,
\end{split}
\end{equation}
where the conjugate momenta of the gauge field, $\Pi^{\mu}$, is given by\footnote{Note that $\Pi^{\mu}$ is the radially conserved bulk current obtained from the Maxwell equation: $0 \,=\, \partial_z \left( \sqrt{-g} \, F^{z \mu} \right)  \,=\,  \partial_z \,\Pi^{\mu}$.}
\begin{equation} \label{CONJ2}
\begin{split}
\Pi^{\mu} \,=\, \frac{\delta S_{\text{bulk}}}{\delta A_{\mu}} \,=\, \frac{\sqrt{-g}}{e^2}\, F^{z \mu} \big|_{z\rightarrow0} \,.
\end{split}
\end{equation}
Eq.\eqref{SUMBDY333} is equivalent to the boundary Maxwell equations
\begin{equation} \label{BMEQS}
\begin{split}
\partial_{\nu} F^{\mu\nu}  = \lambda \left(\Pi^{\mu} + J_{\text{ext}}^{\mu}\right)\,,
\end{split}
\end{equation}
which implies that the gauge field, $A_{\mu}$, is now dynamical in the boundary field theory description. Following this prescription, the external sources can be determined as 
\begin{equation}\label{BCHO}
\begin{split}
\delta J_{\text{ext}}^{\,x\,(L)} = -\frac{\omega}{\lambda}Z_{A_{x}}^{(L)} - \frac{1}{e^2}\frac{\omega}{\omega^2-k^2} Z_{A_{x}}^{(S)} \,, \qquad \delta J_{\text{ext}}^{\,y\,(L)} = - \frac{\omega^2-k^2}{\lambda}Z_{A_{y}}^{(L)} - \, \frac{1}{e^2}Z_{A_{y}}^{(S)} \,,
\end{split}
\end{equation}
where $Z_{A_{x}} := k \delta a_t + \omega \delta a_x$, $Z_{A_{y}} : = \delta a_y$. (L) and (S) respectively stand for leading and subleading terms. Additionally, the conservation equation $\nabla_\mu J^{\mu}_{\text{ext}}=0$ holds and it implies that the time component, $\delta J_{\text{ext}}^{\,t\,(L)}$, is fixed by the others appearing in Eq.\eqref{BCHO}.

Near the AdS boundary ($z\rightarrow 0$), the fluctuations behave as 
\begin{align}\label{FLADSEXP}
\begin{split}
\delta a_{\mu} &= \delta a_{\mu}^{(L)} + \delta a_{\mu}^{(S)} z + \dots \,,   \\
\delta \sigma &= \delta \sigma^{(L)} z + \delta \sigma^{(S)} z^2 + \dots \,,   \\
\delta \eta &= \delta \eta^{(L)} z + \delta \eta^{(S)} z^2 + \dots \,,  
\end{split}
\end{align}
and the gauge-invariant combinations $Z_{A_{\mu}}^{(L) \text{\,or\,} (S)}$ are constructed accordingly.

We will derive the dispersion relations of the low-energy modes using the determinant method \cite{Kaminski:2009dh}. For this purpose, we define the source matrix for the longitudinal/transverse sector as
\begin{align}\label{APPENSMATA}
\begin{split}
\mathcal{S_{\text{long}}} = \left(
\begin{array}{ccc} 
   \delta J_{\text{ext}}^{\,x\,(L)\,(I)} \quad & \delta J_{\text{ext}}^{\,x\,(L)\,(II)}  \quad & \delta J_{\text{ext}}^{\,x\,(L)\,(III)} 
\\ \delta \eta^{(L)(I)} & \delta \eta^{(L)(II)} & \delta \eta^{(L)(III)}
\\ \delta \sigma^{(L)(I)} & \delta \sigma^{(L)(II)} & \delta \sigma^{(L)(III)}
\end{array}
\right) \,, \qquad \mathcal{S_{\text{trans}}} = \delta J_{\text{ext}}^{\,y\,(L)} \,,
\end{split}
\end{align}
where the indices $I,II,III$ denote the $n$-th independent solution. 
The dispersion relation of the modes are then obtained by imposing the determinant of the source matrix to vanish:
\begin{equation}\label{DETMATSOUMAT}
    \det \mathcal{S_{\text{long}}}\left(\omega,k\right)=0\,,\qquad \mathcal{S_{\text{trans}}}\left(\omega,k\right)=0\,.
\end{equation}
In what follows, we set $e=1$ and keep $\lambda$ as a free parameter to control the ratio between the strength of Maxwell interactions in the bulk and those at the boundary.

%%%%%%%%%%%%%%%%%%%%%%%%%%%%%%%%
\section{The equilibrium superconducting state}
\label{phase}

\begin{figure}[]
\centering
     {\includegraphics[width=8.5cm]{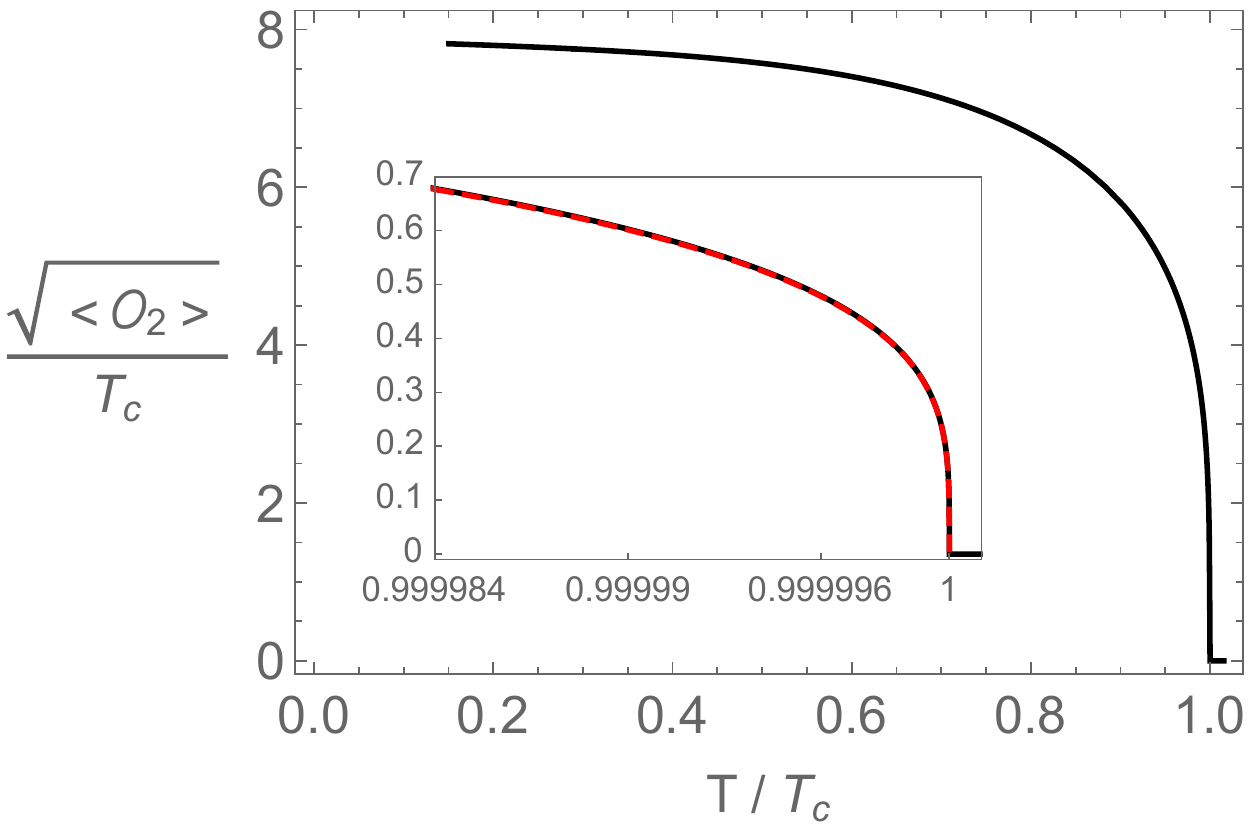} \label{}}
 \caption{Order parameter $\langle O_{2} \rangle$ vs. reduced temperature $T/T_c$. The critical temperature is $T_c/\mu=0.0587$. The inset shows the near-critical mean field behavior: numerical result (solid black), fitting result \eqref{NEARTCCON} (dashed red).}\label{STSKF}
\end{figure}

By numerically solving equations \eqref{EOMBG} with the boundary conditions defined in the previous section, one observes the appearance of a bulk solution with a non-trivial profile for the bulk complex scalar field above a certain critical value of the chemical potential. This is the broken phase in which the $U(1)$ symmetry is spontaneously broken. For our choice of parameters, we find  $\mu_c \, z_{h} \approx 4.062$ which corresponds to a critical temperature $T_{c}/\mu=0.0587$ consistent with the results in Ref.\cite{Amado:2009ts}.
We plot the profile of the scalar condensate as a function of the reduced temperature in  Fig. \ref{STSKF}.
As expected, close to the critical point we observed the typical mean-field behavior
\begin{equation}\label{NEARTCCON}
\begin{split}
\langle O_{2} \rangle \,\sim\, \, \sqrt{1-T/T_c}  \,.
\end{split}
\end{equation}

\begin{figure}[h]
\centering
     \includegraphics[width=7.0cm]{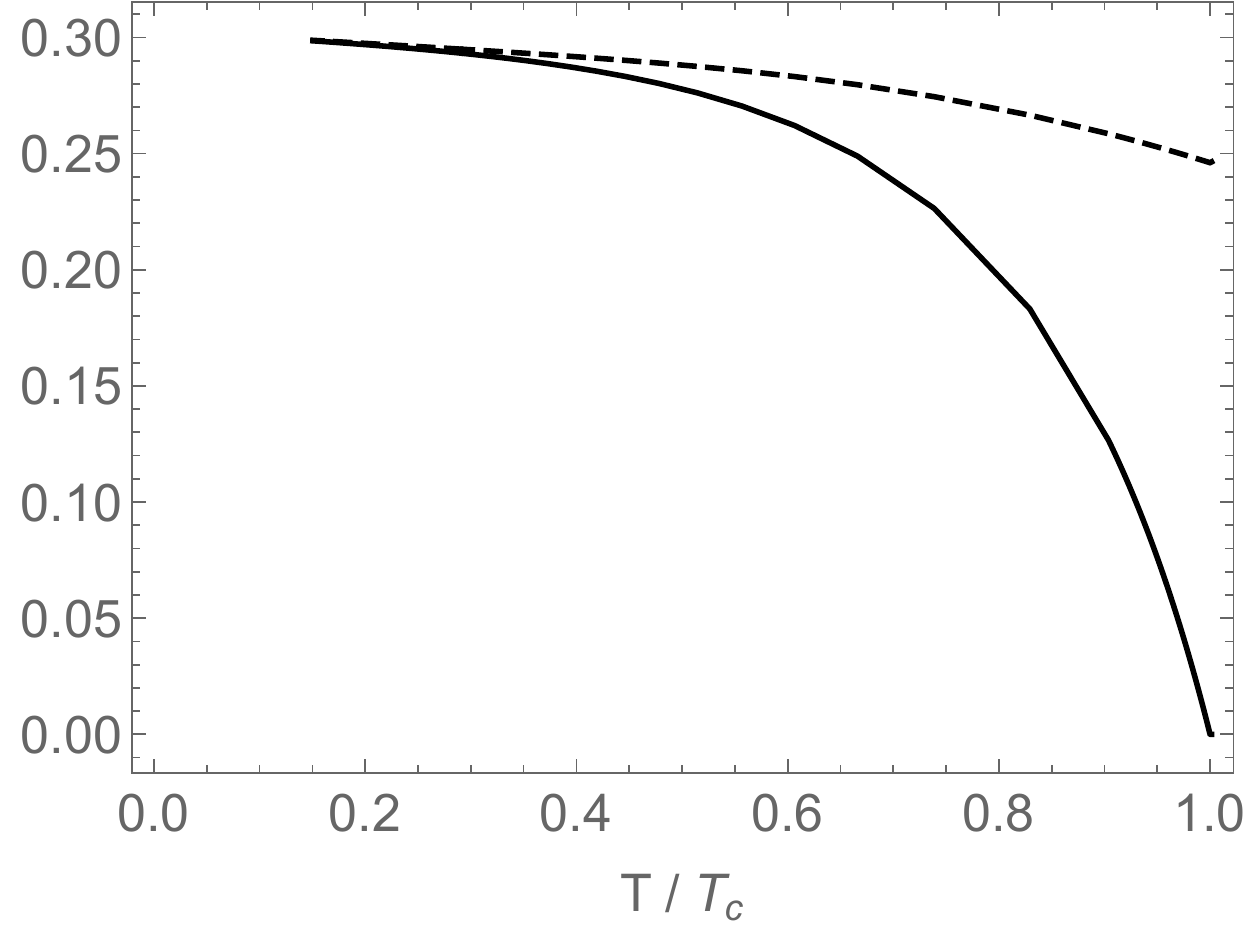} \label{SIGM1} \quad \includegraphics[width=7.2cm]{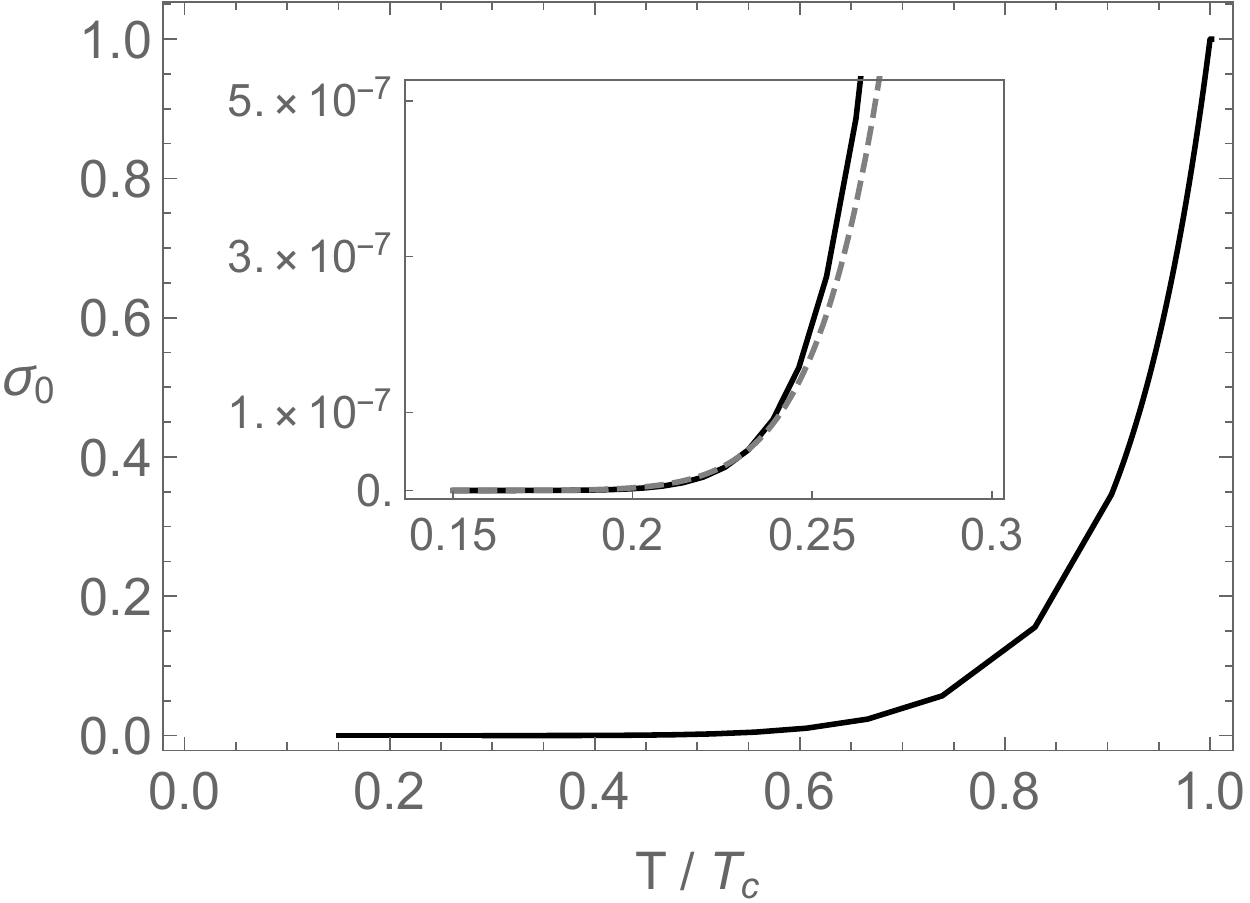}
 \caption{\textbf{Left:} Total density $\rho/\mu^2$ (dashed) and superfluid density $\rho_s/\mu^2$ (solid). The total density $\rho$ is evaluated from Eq.\eqref{BGADSEX} while the superfluid density is from the optical conductivity data. \textbf{Right:} $\sigma_0$ as a function of the reduced temperature. The inset shows the low temperature behavior: numerical result (solid black), fitting result \eqref{LOTTCONE} (dashed gray).}\label{CONFIG}
\end{figure}

Before moving to the dynamics of the fluctuations at finite frequency and wave-vector, we can study the electric response of the system in the broken phase. The electric conductivity \footnote{Previous studies of the conductivity in presence of Coulomb interactions in holography can be found in \cite{Mauri:2018pzq,Romero-Bermudez:2018etn}.} can be defined holographically using
\begin{equation}\label{defd}
\begin{split}
\sigma (\omega) =  \frac{1}{i \omega}\frac{\delta a_x^{(S)}}{\delta a_x^{(L)}} \,,
\end{split}
\end{equation}
where $\delta a_x^{(L)}$ is the leading coefficient of the fluctuation $\delta a_x$, while $\delta a_x^{(S)}$ the subleading coefficient. In absence of coupling to momentum (i.e., in the probe limit), the optical conductivity takes the simple form 
\begin{equation}\label{OPTI}
\begin{split}
    \sigma(\omega)=\sigma_0+\left(\frac{i}{\omega}+\delta(\omega)\right)\frac{\rho_s}{\mu} \,,
\end{split}
\end{equation}
where $\rho_s$ is the superfluid density.
The superfluid density approaches the total density at low temperature, as shown in the left panel of Fig. \ref{CONFIG}.

From the formula above, we can also extract the parameter $\sigma_0$.
In the small temperature regime, $T/T_c \ll 1$, it was shown~\cite{Hartnoll:2008vx,Hartnoll:2008kx,Herzog:2009xv,Hartnoll:2009sz} that $\sigma_0 := \lim\limits_{\omega \to 0} \text{Re}[\sigma(\omega)]$ is associated with the superconducting energy gap $\Delta$ via
\begin{equation}\label{LOTTCONE}
\begin{split}
\sigma_0 \,\sim\, e^{-\Delta/T} \,,\qquad \Delta:= \sqrt{\langle O_{2} \rangle}/{2}  \,,
\end{split}
\end{equation}
i.e., the low temperature behavior of conductivity 
$\sigma_0$ is exponentially suppressed by the condensate $\langle O_{2} \rangle$. 
 We show this behavior in the right panel of Fig. \ref{CONFIG}, proving that the formula above works very well. Notice that, as well known, the energy gap extracted is given by $2\Delta \sim 8 T_c$ and much larger than the BCS prediction $2\Delta \sim 3.5 T_c$.

%%%%%%%%%%%%%%%%%%%%%%%%%%%%%%%%
\section{Transverse collective modes}
\label{sectran}
In this section, we study the dispersion relation of the transverse low-energy collective modes. Unless otherwise mentioned, we set $\lambda/T=0.1$.

\subsection{Massive electromagnetic waves}

In order to understand the dynamics in the transverse sector, we utilise the following equation: 
\begin{align}\label{TRANSFIT1}
\begin{split}
\omega^2 = \tilde{\omega}_A^2 + \tilde{v}^2 k^2 - i \, \tilde{\sigma} \, \omega \,,
\end{split}
\end{align}
which is exactly of the same form as the one derived in the dissipative Ginzburg-Landau framework in the previous section, Eq.\eqref{lala}, i.e.,
\begin{align}\label{TRANDUAL}
\begin{split}
\tilde{\omega}_A \,\leftrightarrow\, \omega_A \,,\qquad \tilde{v} \,\leftrightarrow\, v \,, \qquad \tilde{\sigma} \,\leftrightarrow\, v^2 \, \lambda \, \sigma \,.
\end{split}
\end{align}
Using these notations, $\tilde v$ parameterizes the velocity of propagation of EM waves, $\tilde \sigma$ the dissipative effects coming from the conductivity and $\tilde{\omega}_A$ the emergent mass arising because of the Anderson-Higgs mechanism.

\begin{figure}[ht!]
\centering
   \includegraphics[width=7.2cm]{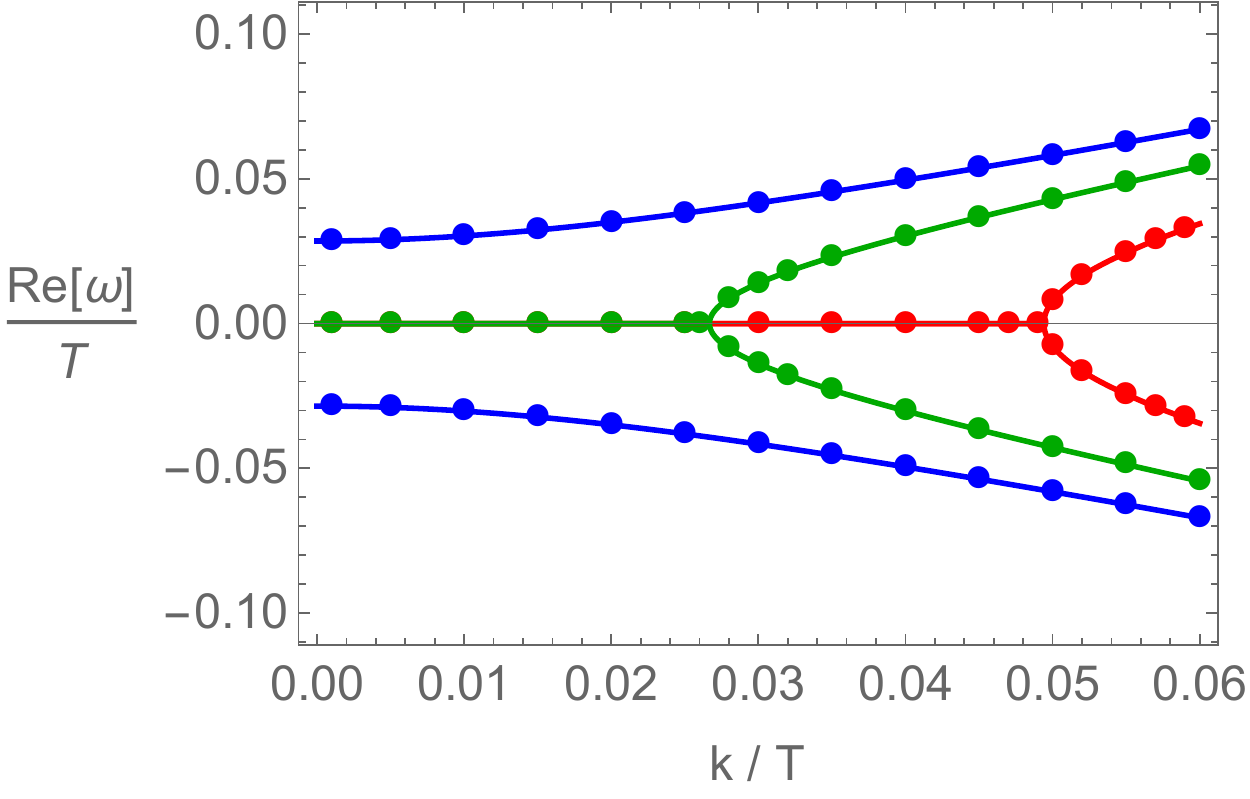} \quad
     \includegraphics[width=7.2cm]{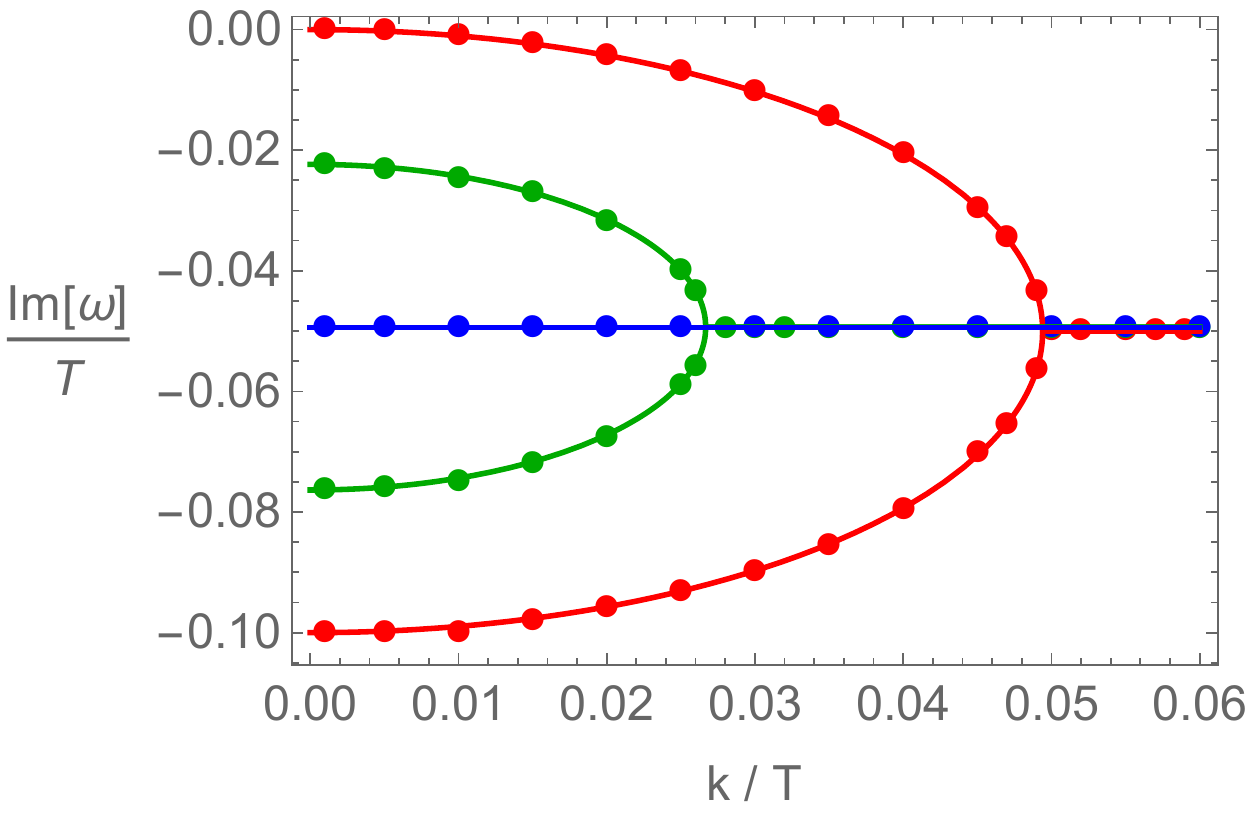} 
 \caption{The dispersion relation of the lowest collective modes in the transverse sector for different values of the reduced temperature $T/T_c = (1, 0.999, 0.998)$ (red, green, blue). Symbols represent the numerical values and solid lines are fits using Eq.\eqref{TRANSFIT1}.}\label{TRANFIG}
\end{figure}

\paragraph{Transverse excitations in the normal phase.}
In the normal phase, $T \geq T_c$, the equation \eqref{TRANSFIT1} can be formally derived using magnetohydrodynamics \cite{Hernandez:2017mch} and has been verified holographically in \cite{Baggioli:2019sio,Ahn:2022azl}. In particular, because of the probe limit, in the normal phase we do expect 
\begin{align}\label{vtilde2}
\begin{split}
\text{$T \geq T_c$:} \qquad \tilde{\omega}_A=0 \,,  \qquad \tilde {v}^2 = 1 - \lambda \, \chi_{BB} \,, \qquad \tilde \sigma=\sigma_0 \lambda \,,
\end{split}
\end{align}
together with $\chi_{BB}=-3/(4\pi T)$, as proved explicitly in \cite{Ahn:2022azl} (see also appendix \ref{appenaaa} for the derivation of $\tilde{v}$).
Furthermore, let us recall that above $T_c$, the bulk field $A_t$ (associated with the chemical potential) is absent in the transverse sector \eqref{eq:transVecEOMBroken}, which implies that the transverse dispersion relation of the normal phase is independent of the value of $\mu$. As a consequence, the dispersion data shown in red color in Fig. \ref{TRANFIG} are representative for all the temperatures $T \geq T_c$ (or $\mu \leq \mu_c$).

\paragraph{Transverse excitations in the superconducting phase.}
The dispersion relation of the lowest collective modes in the transverse sector is shown in Fig. \ref{TRANFIG} for different values of temperature in the superconducting phase. At the critical temperature (red data), we observe the standard behavior for EM waves in a conductor, in which the effects of screening induce a gap in the wave-vector \cite{Baggioli:2019jcm}. The dynamics of EM waves displays a crossover between an overdamped diffusive behavior for long wave-lengths to a propagating behavior at short wave-lengths. The crossover between the two regimes is controlled by the conductivity of the system and the value of the electromagnetic coupling $\lambda$. We refer to \cite{Ahn:2022azl} for a complete study of this behavior.

By decreasing the temperature and moving deeper into the superconducting phase (green and blue data), we observe that the critical wave-vector becomes smaller. At a critical value of the temperature, the gap of the dispersion relation changes its nature and becomes a real energy gap, while the imaginary part of the dispersion becomes approximately constant.\footnote{The dynamics of the real part of the dispersion relation is reminiscent of what found in \cite{Baggioli:2018nnp,Baggioli:2018vfc} with the difference that therein no hydrodynamic mode survives.}

In Fig. \ref{TRANFIG}, we also display the fitting curves (solid lines) using \eqref{TRANSFIT1}, which are in good agreement with the numerical values (symbols).
Note that in general there are three fitting parameters ($\tilde{\omega}_A\,, \tilde {v} \,,\tilde \sigma$), while the numerical quasi-normal mode data has only two independent degrees of freedom at a given wave-vector. Therefore, for practical purposes, we fix $\tilde {v}^2 = 1 - \lambda \, \chi_{BB}$ even in the superconducting phase and we only fit for the two parameters $\tilde{\omega}_A\,, \tilde \sigma$. We then verify a posteriori the validity of this assumption. In the following subsections, we discuss their temperature and EM coupling dependence in detail.

Before continuing, we remind the reader that, in the case of holographic superfluids, the spectrum does not display any transverse hydrodynamic mode (see \cite{Amado:2009ts} for details).

\subsection{Zero wave-vector excitations}
We are ready to investigate the dispersion relation of the transverse EM waves in the superconducting phase. For simplicity, we start with the homogeneous case, $k=0$. The solutions of Eq.\eqref{TRANSFIT1} at $k=0$ read
\begin{equation}\label{eqk0}
    \omega = -\frac{i}{2} \tilde{\sigma} \pm \frac{1}{2}\sqrt{4 \, \tilde{\omega}_A^2-\tilde \sigma^2} \,
\end{equation}
and will be analyzed in detail below. Let us remind that in the normal phase we have $\tilde \sigma $ finite and $\tilde{\omega}_A=0$.
\begin{figure}[]
\centering
     {\includegraphics[width=8.2cm]{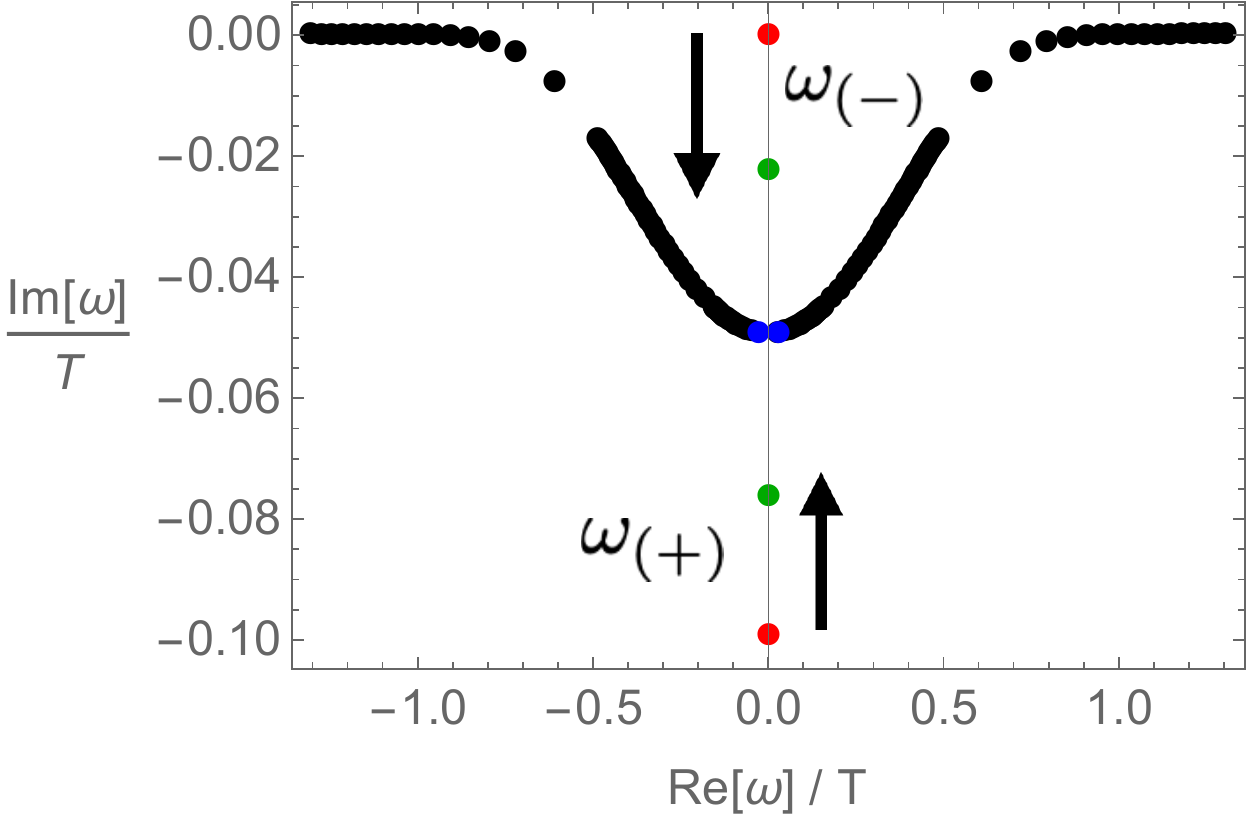} \label{}}
 \caption{The dynamics of the lowest collective modes in the transverse sector at $k=0$ by decreasing the reduced temperature from $T/T_c=1$ to $0.3$ (in the direction of the arrows). The (red, green, blue) data are the same as in Fig. \ref{TRANFIG}.}\label{TRANFIG2}
\end{figure}

Depending on the value of $\tilde{\omega}_A$ and $\tilde{\sigma}$, the dispersion in Eq.\eqref{eqk0} can give purely imaginary or complex modes. More precisely, we have three distinct cases. Whenever the dissipative effects are dominating, $4 \, \tilde{\omega}_A^2 < \tilde{\sigma}^2$, the modes are purely imaginary, with dispersion relation
\begin{equation}\label{TISMM0}
 \omega_{(\pm)} \,=\, -\frac{i}{2} \left( \tilde{\sigma} \pm \sqrt{\tilde{\sigma}^2 - 4 \, \tilde{\omega}_A^2} \right)  \,.
\end{equation}
In the small $\tilde{\omega}_A$ limit, these imaginary poles are just given by
\begin{equation}\label{TISMM}
    \omega_{(+)} \,\approx\, - i \left( \tilde{\sigma} - \frac{\tilde{\omega}_A^2}{\tilde{\sigma}} \right) \,, \qquad \omega_{(-)} \,\approx\,   -i \frac{\tilde{\omega}_A^2}{\tilde{\sigma}}\,.
\end{equation}
At a critical value of the mass,  $4\,\tilde{\omega}_A^2 =\tilde{\sigma}^2$, these two poles collide on the imaginary axes at $\omega_{\text{collision}}  \,=\, -\frac{i}{2} \tilde{\sigma}$. After the collision, they split into two complex poles and move away from the imaginary axes towards the real axes in a symmetric fashion. 

In the opposite limit, in which the mass dominates over the dissipative effects, $4 \, \tilde{\omega}_A^2 > \tilde{\sigma}^2$, we have the complex poles
\begin{align}\label{TISMM2}
\begin{split}
\omega \,=\, \omega_{(C)} \,:=\, \pm \, \omega_{(R)} \,-\, i \, \omega_{(I)}  \,=\, \pm \frac{1}{2}\sqrt{4 \, \tilde{\omega}_A^2-\tilde \sigma^2} \,-\, \frac{i}{2} \tilde{\sigma}  \,.
\end{split}
\end{align}
 As a general rule, dissipative effects become subdominant at low temperature. Therefore, the dynamics just described is what we do expect by decreasing the temperature from the critical point down to zero temperature. This is exactly what we observe in Fig. \ref{TRANFIG2}.  

\begin{figure}[]
\centering
     \includegraphics[width=6.8cm]{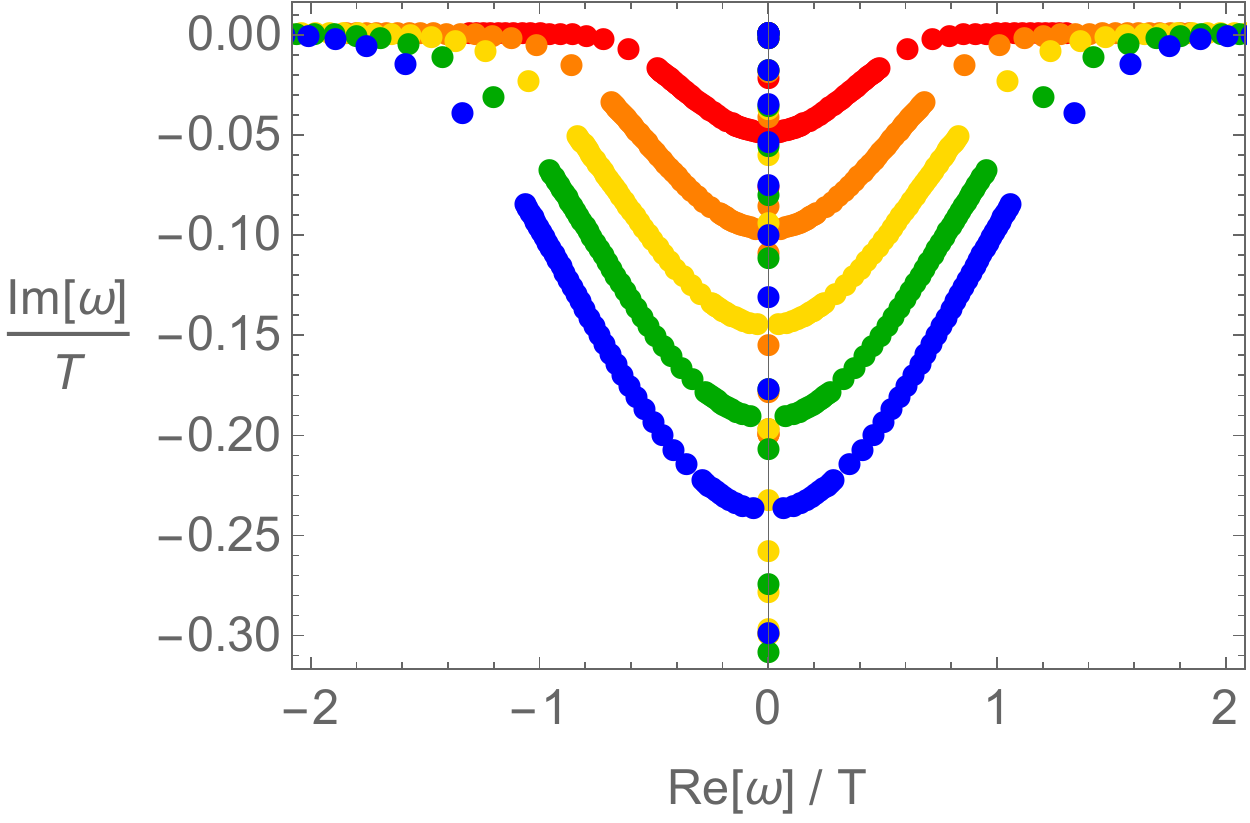} \quad 
     \includegraphics[width=7.6cm]{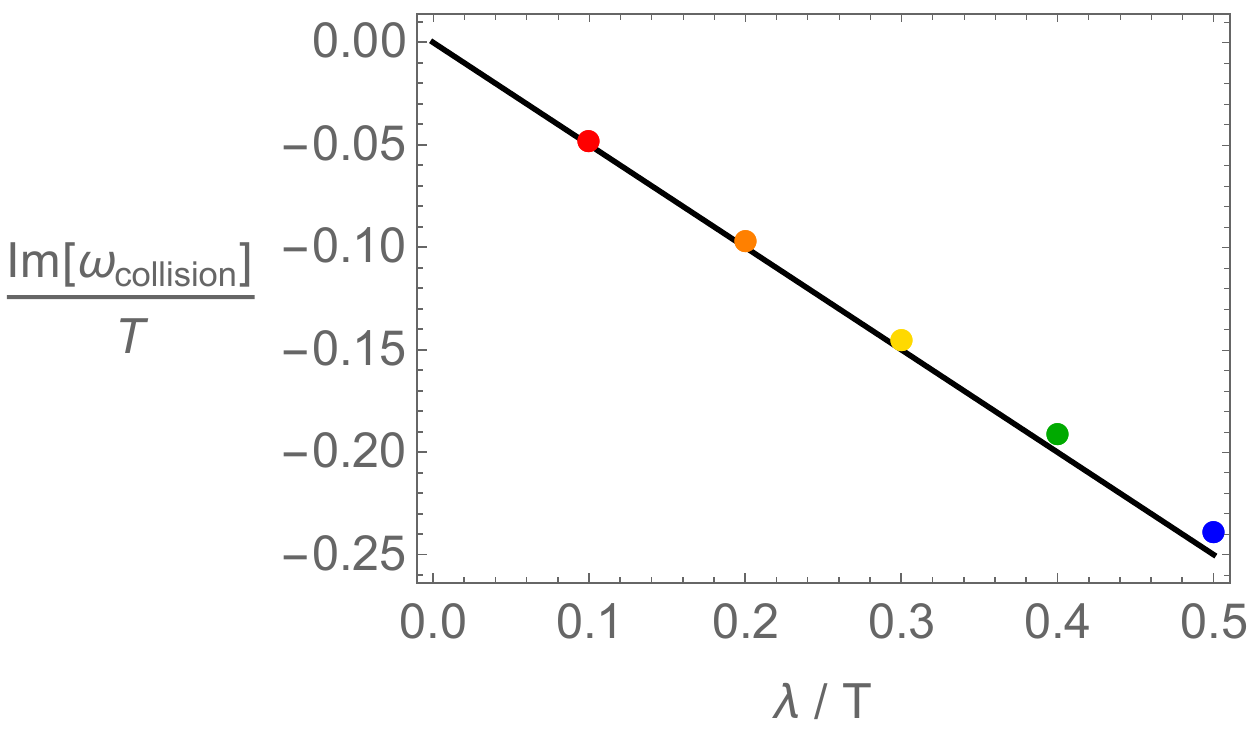} 
 \caption{\textbf{Left:} The dynamics of the lowest QNMs in the transverse spectrum at $k=0$ for $T/T_c \in [1,0.3]$ with $\lambda/T = 0.1-0.5$ (from red to blue data). The red data corresponds to Fig. \ref{TRANFIG2}. \textbf{Right:} The collision frequency, $\omega_{\text{collision}}$, as a function of the EM coupling $\lambda/T$. The black solid line is the phenomenological finding in Eq.\eqref{COLLLAM}.}\label{TRANFIG5}
\end{figure}

As an interesting observation, we find that, at leading order in the EM coupling $\lambda$, the collision between the two modes occurs at the specific value
\begin{align}\label{COLLLAM}
\begin{split}
\frac{\text{Im} \, [\omega]}{T} \bigg|_{\text{collision}} = -\frac{1}{2}\frac{\lambda}{T} \,,
\end{split}
\end{align}
which is confirmed numerically in Fig. \ref{TRANFIG5}. As shown explicitly in the right panel, this expression represents only an approximation in the regime of small EM coupling and it fails above $\lambda/T \approx 0.4$.

\begin{figure}[]
\centering
     \includegraphics[width=7.2cm]{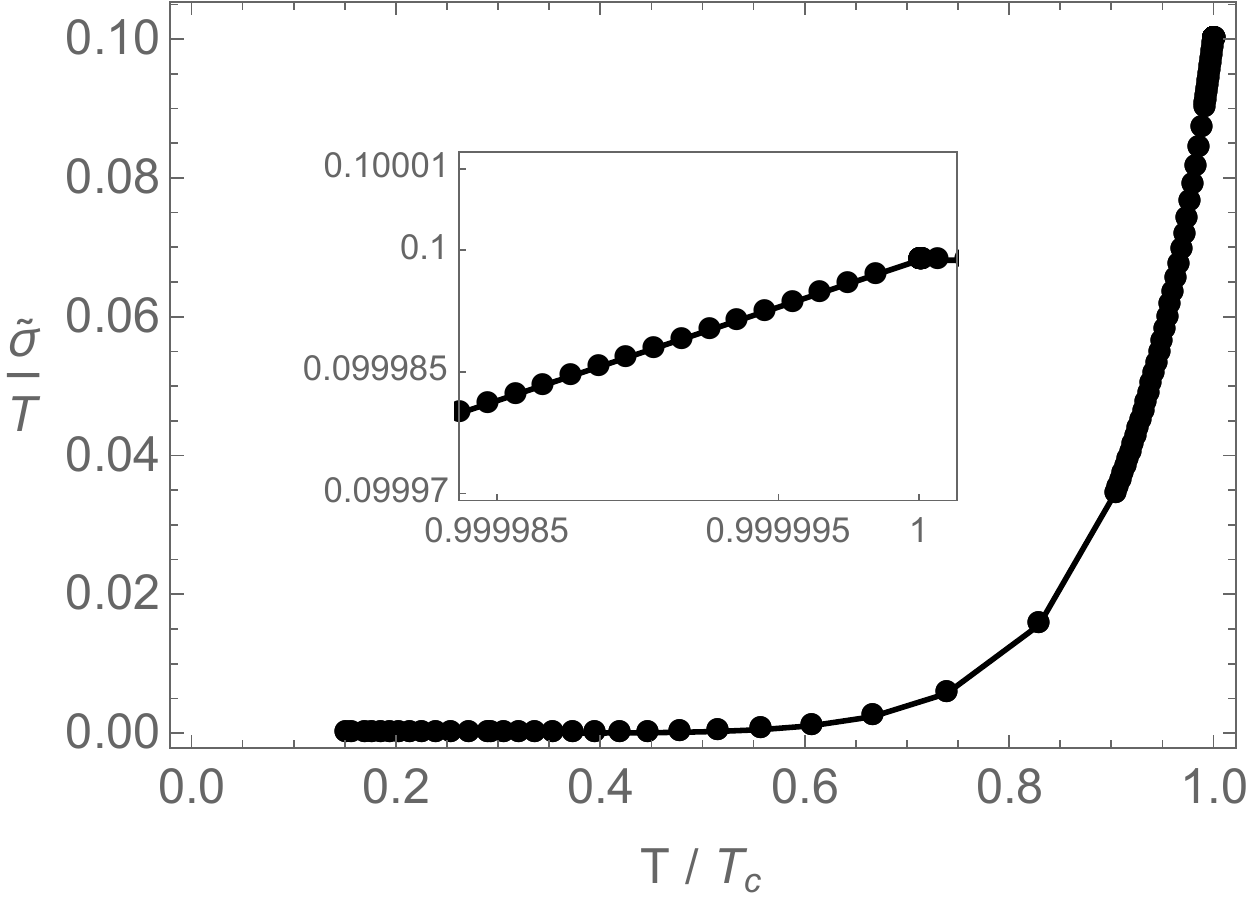} 
     \quad
      \includegraphics[width=7.2cm]{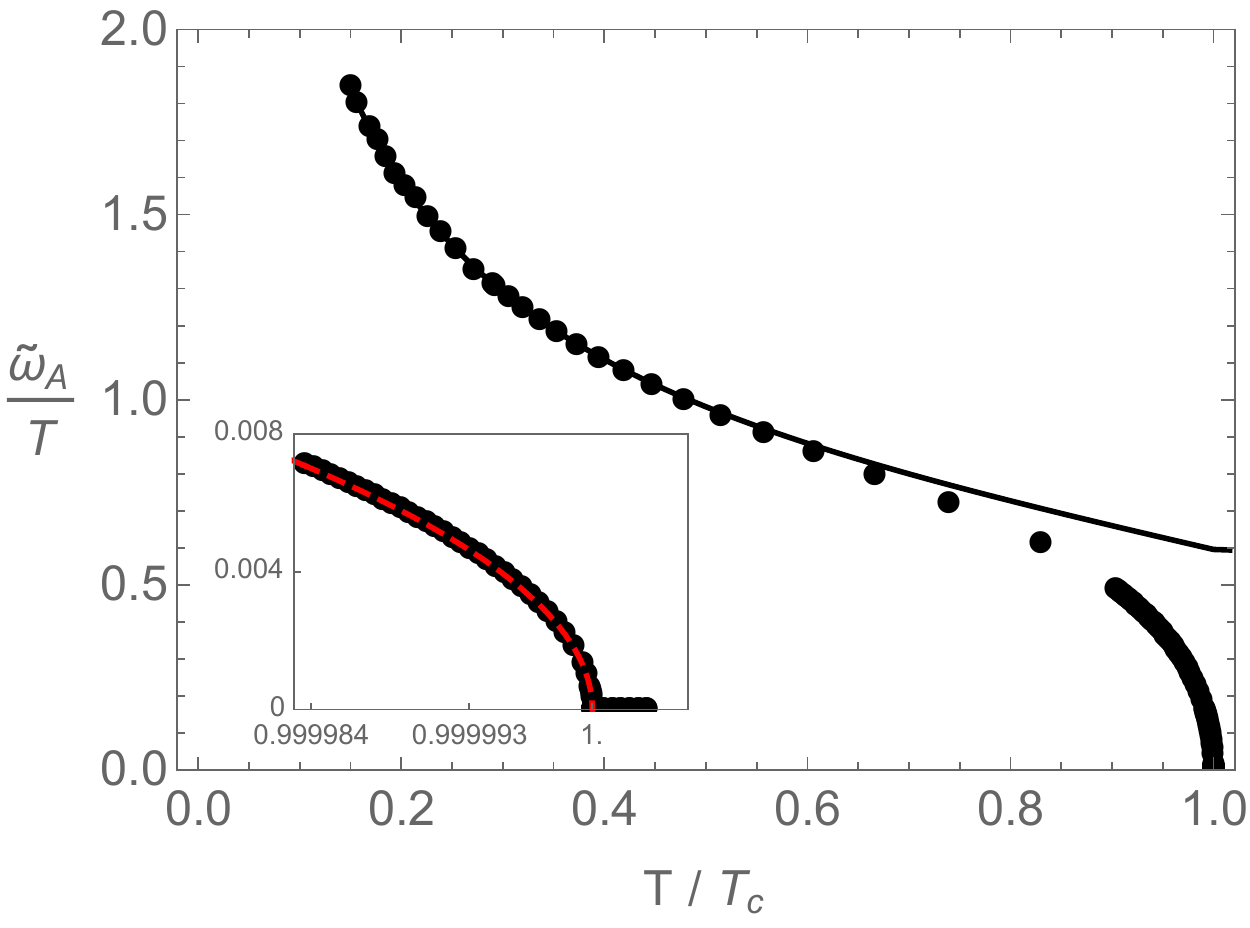} 
 \caption{The temperature dependence of the phenomenological paramters $\tilde \sigma$ and $\tilde{\omega}_A$ in Eq.\eqref{eqk0}. Dots are evaluated from the numerical fits. Solid lines represent the analytical expression in Eq. \eqref{TRANSFIT2} (left panel) and Eq.\eqref{TRANSFIT3} (right panel). The insets show the data near the critical point, $T=T_c$. The dashed red line in the right panel is the expression in Eq.\eqref{TRANSFIT4}.}\label{TraFIG12}
\end{figure}

By fitting the data at $k=0$, we can extract the temperature dependence of the phenomenological parameters $(\tilde{\sigma}/T, \, \tilde{\omega}_A/T)$. Their behavior is shown in Fig. \ref{TraFIG12}.
Interestingly, we find that the dissipative parameter $\tilde \sigma$ takes the same form as in the normal phase and does not receive corrections in the superconducting state. 
In particular, for all the values of the temperature, within the probe limit approximation, we find that
\begin{align}\label{TRANSFIT2}
\begin{split}
\tilde{\sigma} \,=\, \sigma_0 \, \lambda \,,
\end{split}
\end{align}
where:
\begin{equation}
    \sigma_0 := \lim\limits_{\omega \to 0} \text{Re}[\sigma(\omega)]\,,
\end{equation}
and $\sigma(\omega)$ is the conductivity defined in Eq.\eqref{defd} and shown in the right panel of Fig. \ref{CONFIG}.

The dynamics of the other fitting parameter, $\tilde{\omega}_A$, is more complex. In the regime of small temperature, $T \ll T_c$, we find that this parameter is well fitted by the plasma frequency value 
\begin{align}\label{TRANSFIT3}
\begin{split}
 \omega_p \,:=\,  \sqrt{\lambda\, \frac{\rho^2}{\epsilon+p}} \,,
\end{split}
\end{align}
where $\epsilon$ is the energy density, and $p$ the thermodynamic pressure which can be evaluated using the Smarr relation $\epsilon+p = sT+\mu \rho$. The low temperature behavior of $\tilde{\omega}_A$ is shown in the right panel of Fig. \ref{TraFIG12} using a solid line.
On the contrary, near the critical point, the value of $\tilde{\omega}_A$ strongly deviates from the plasma frequency value, Eq.\eqref{TRANSFIT3}, and vanishes at the critical point with a square root behavior,
\begin{align}\label{TRANSFIT4}
\begin{split}
\tilde{\omega}_A \,=\, \alpha \, \sqrt{1-T/T_c} \,,
\end{split}
\end{align}
where $\alpha$ is a $\lambda$-dependent constant.

 In particular, the mass of the EM waves vanishes at the critical point since $\lambda_{GL}\rightarrow \infty$ in Eq.\eqref{GAGM}. At the same time, it is expected that in the limit of small temperature, the mass of the gauge field fluctuations approaches the plasma frequency value, see Eq.\eqref{PLAFOR22}. In other words, our holographic results are perfectly compatible with the GL picture reviewed in Section \ref{secGL}. In principle, using perturbative methods, one could extract analytically the value of the parameter $\alpha$ which determines the near-critical behavior of the mass $\tilde{\omega}_A$, as done in \cite{Donos:2022qao}. We leave this analysis for the future.

\begin{figure}[]
\centering
    \includegraphics[width=7.0cm]{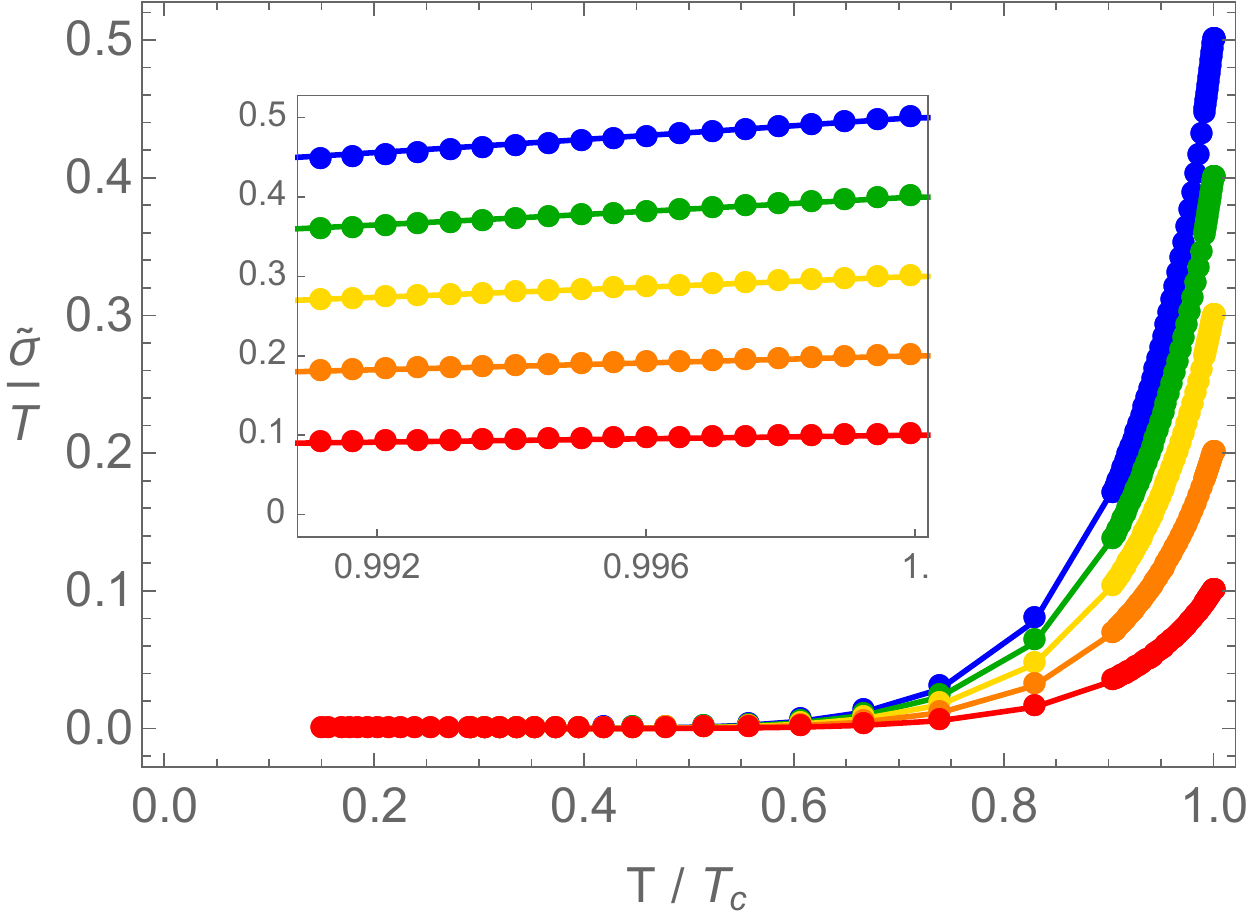} \quad
      \includegraphics[width=7.0cm]{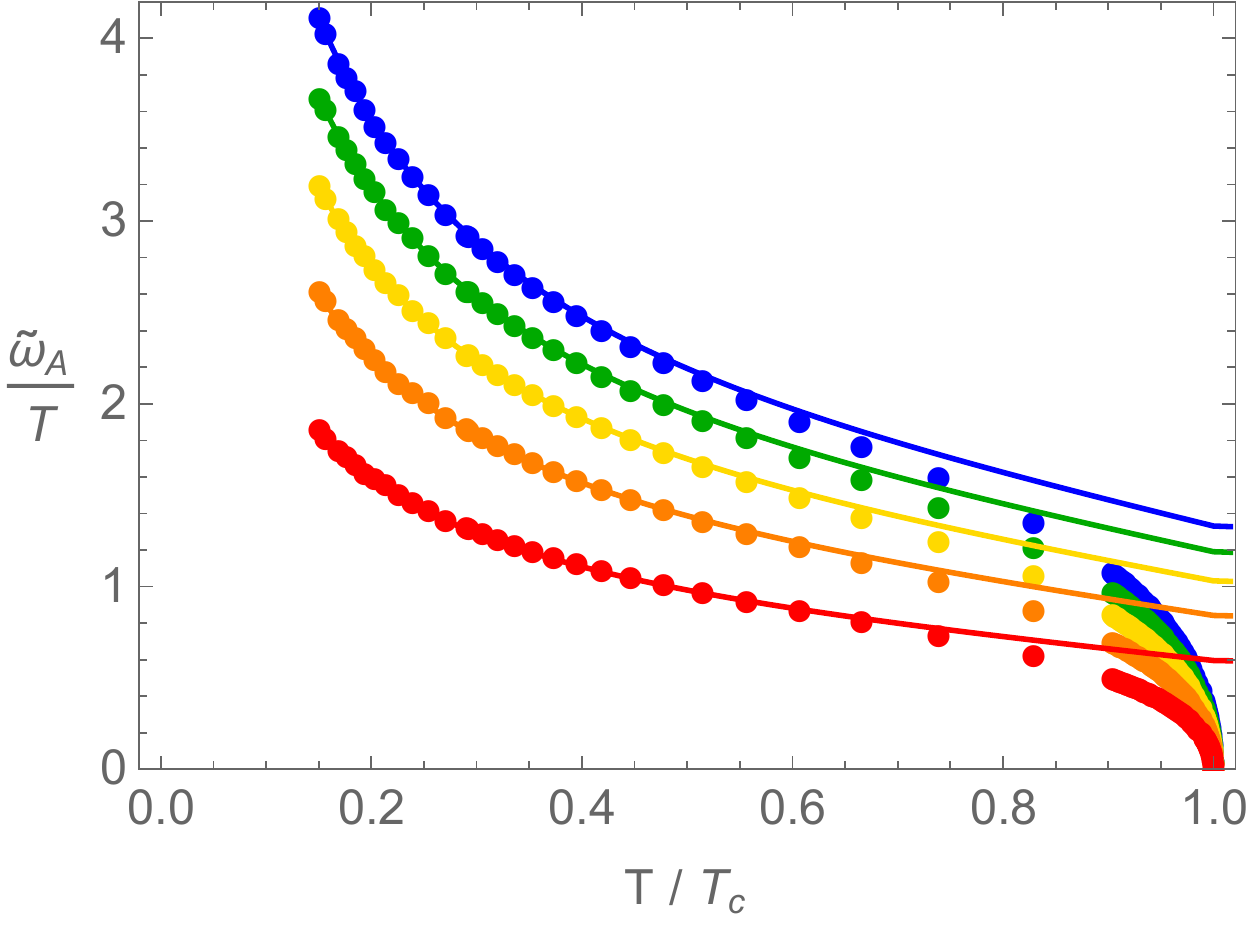} 
     \includegraphics[width=7.2cm]{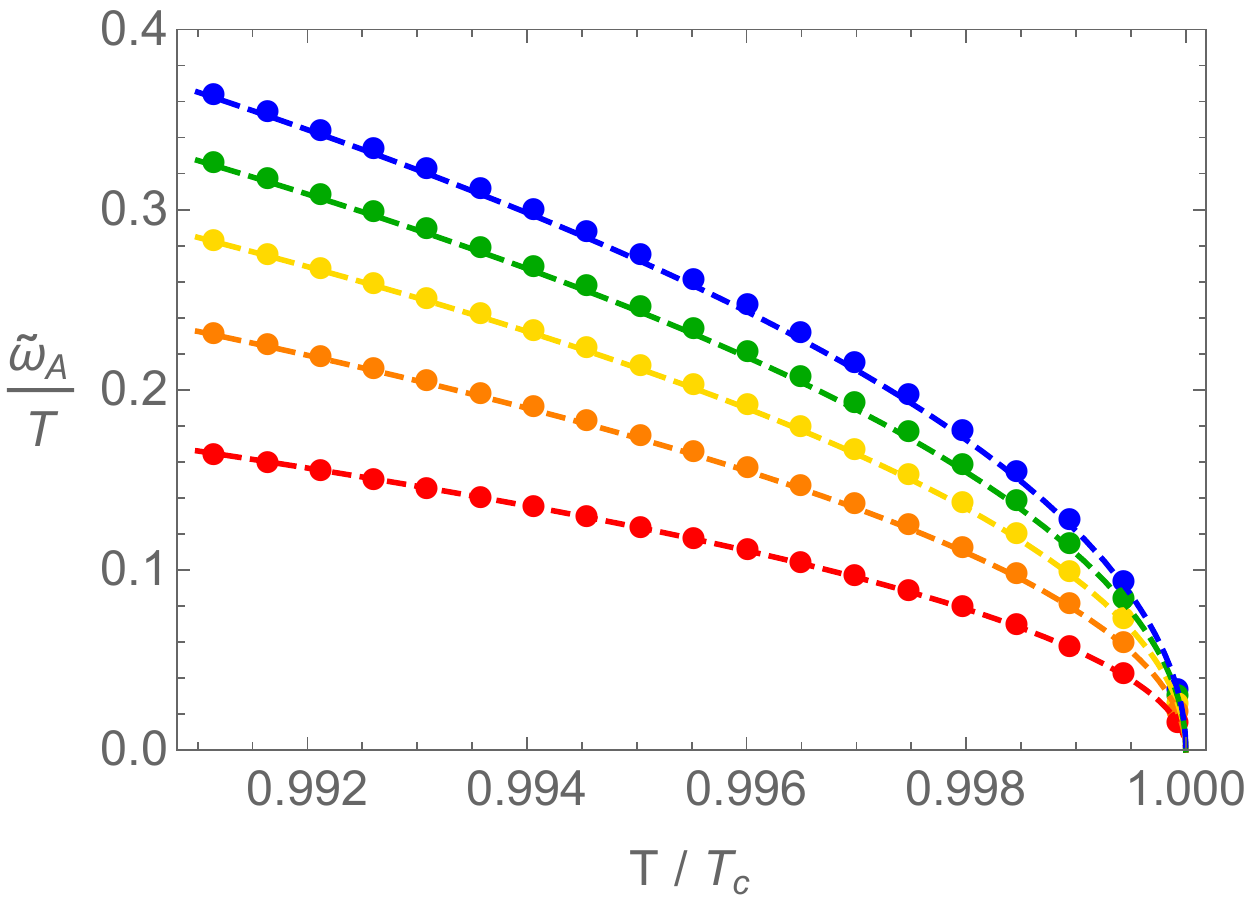} \quad
     \includegraphics[width=6.7cm]{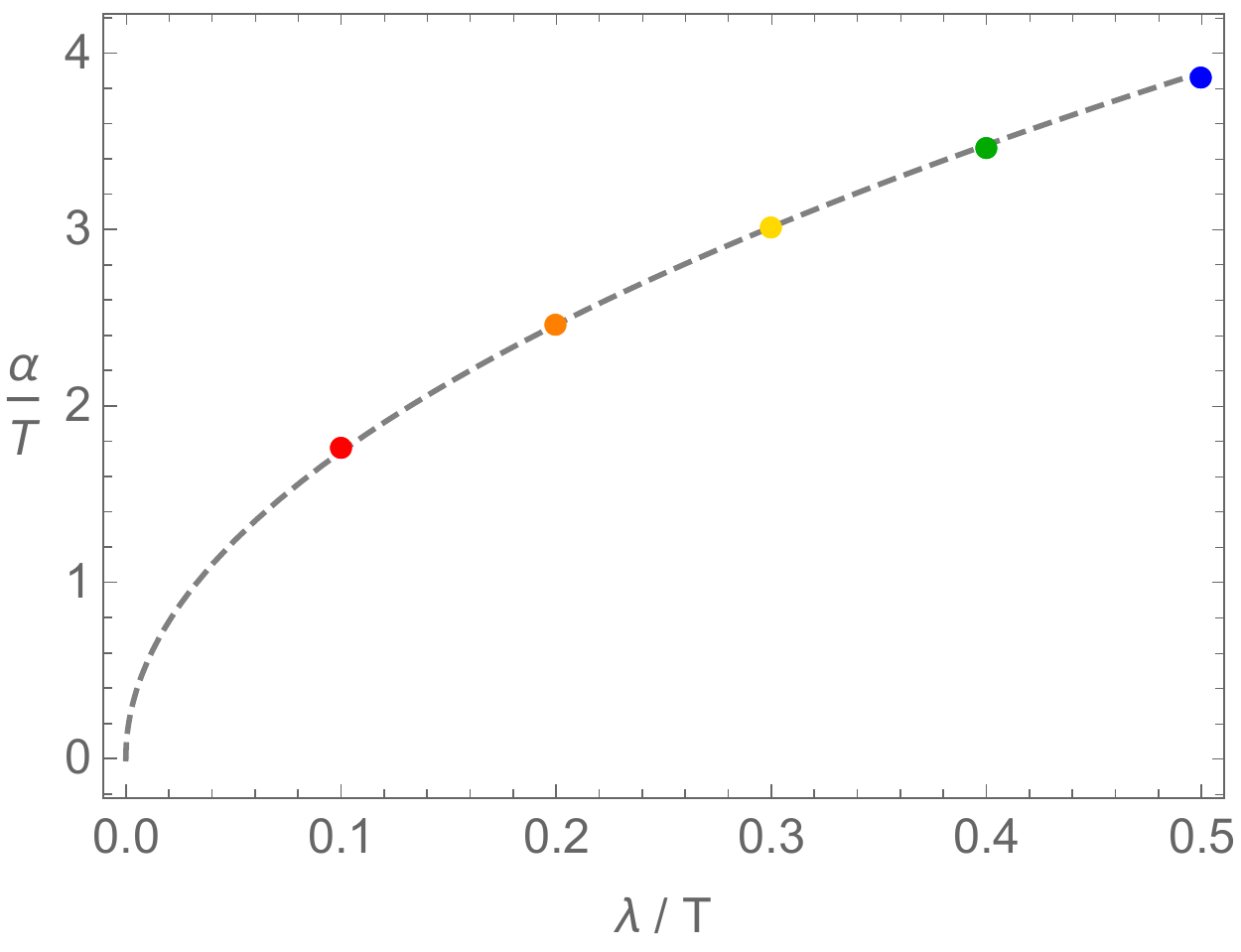} 
 \caption{The phenomenological parameters appearing in the dispersion relation of the lowest QNMs in the transverse sector at $k=0$. Different colors from red to blue correspond to $\lambda/T = 0.1-0.5$. \textbf{Top left: }the conductivity $\tilde{\sigma}$ and the expression in Eq.\eqref{TRANSFIT2} (solid lines). \textbf{Top right: }the phenomenological mass $\tilde{\omega}_A$ together with the plasma frequency value in Eq.\eqref{TRANSFIT3} (solid lines). \textbf{Bottom left: }the behavior of the mass close to the critical point, $T\sim T_c$ and the fitting formula in Eq.\eqref{TRANSFIT4} (dashed lines). \textbf{Bottom right: } the phenomenological parameter $\alpha$ as a function of the EM coupling. The dashed line is the fitting formula $\alpha/T= 5.5 \sqrt{\lambda/T}$. }\label{TRANFIG6}
\end{figure}
\subsection{EM coupling dependence}
To find the EM coupling dependence, we have performed the same analysis for different values of $\lambda$. The results are shown in Fig. \ref{TRANFIG6}. First, we observe that for all the values of the electromagnetic coupling and temperature, the parameter $\tilde \sigma$ obeys the expression in Eq.\eqref{TRANSFIT2}. Second, we find that independently of the value of the EM coupling, the mass of the gauge field fluctuations approaches the plasma frequency value at low temperatures. Interestingly, we observe that the mass $\tilde{\omega}_A$ reaches the plasma frequency value at a larger temperature for smaller values of the EM coupling (see top right panel in Fig.\ref{TRANFIG6}). Finally, near the critical temperature, the mass always vanishes following the mean-field behavior in Eq.\eqref{TRANSFIT4}. The constant of proportionality $\alpha$ depends on the electromagnetic coupling and, at least for this choice of parameters, it is well approximated by the fitting expression $\alpha/T= 5.5 \sqrt{\lambda/T}$. As already mentioned, the value of this constant should be related to the GL parameters which can be computed directly from the holographic picture, as done in \cite{Donos:2022qao} for the superfluid case.

\paragraph{Comparison with the perturbative analytical results near the critical point.}
Recently, Ref.\cite{Natsuume:2022kic} studied the holographic Meissner effect in the holographic superconductor model using perturbative analytical techniques valid in the near-critical regime (see \cite{Herzog:2010vz,Donos:2022qao} for similar analyses in the case of holographic superfluids).
In particular, a closed-form for the London penetration length, $\lambda_{\text{holo}}$, was obtained. Using that expression, we can  extract an analytical formula for the mass of the gauge field $\tilde{\omega}_A$ which is given by
\begin{align}\label{MAKOTOFOR}
\begin{split}
\tilde{\omega}_A \,=\, \frac{1}{\lambda_{\text{holo}}} \,=\,  \sqrt{\frac{2\lambda}{1+\lambda} \, I} \,, \qquad I := \int_{0}^{1} \dd z \, \left(\frac{\psi(z)}{z}\right)^2 \,.
\end{split}
\end{align}
In the expression above, $\psi(z)$ is the bulk complex scalar field (see Eq.\eqref{PROBEFIELDSEQ}). The limits of integration are the location of the boundary $z=0$ and that of the horizon $z=1$.

\begin{figure}[]
\centering
     \includegraphics[width=7.0cm]{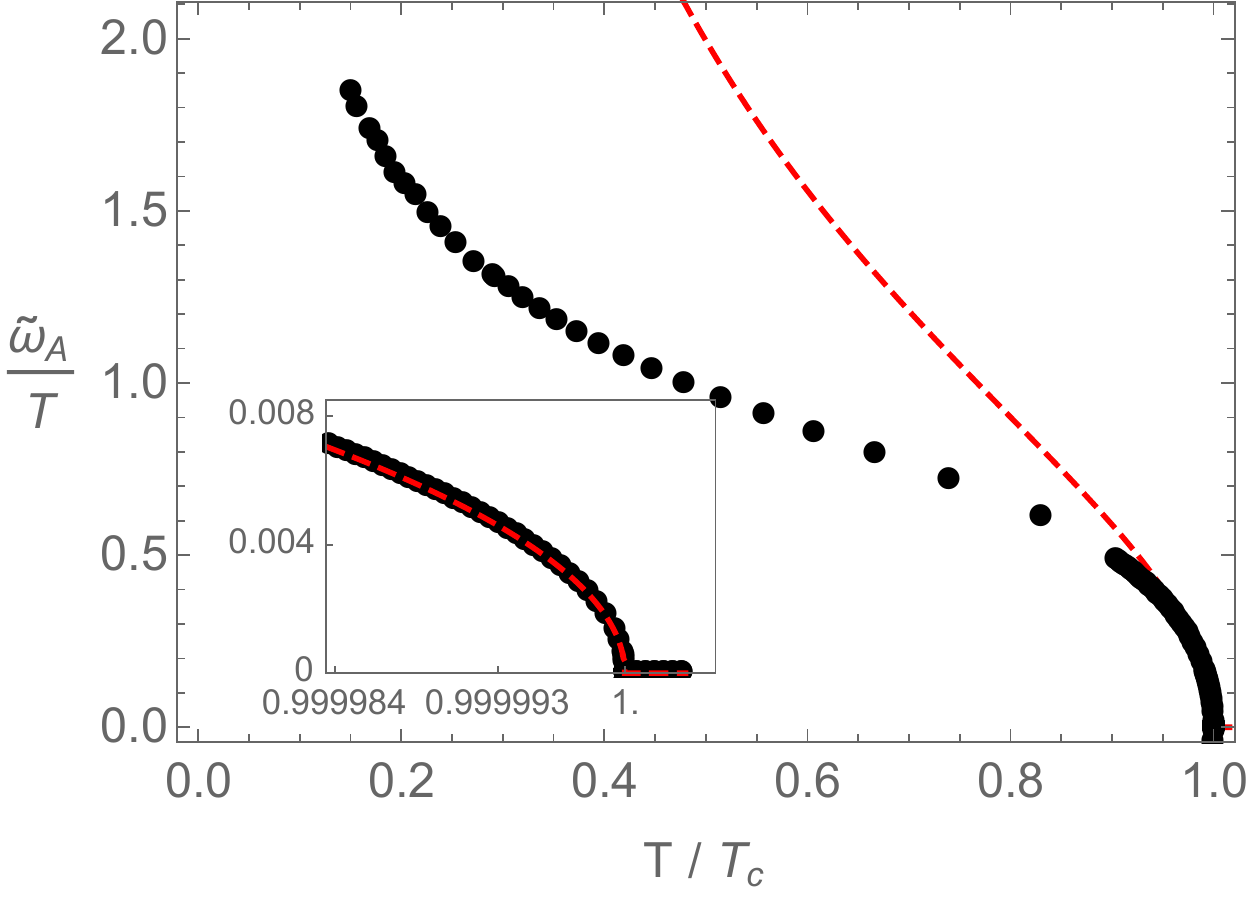}
     \quad 
     \includegraphics[width=7.2cm]{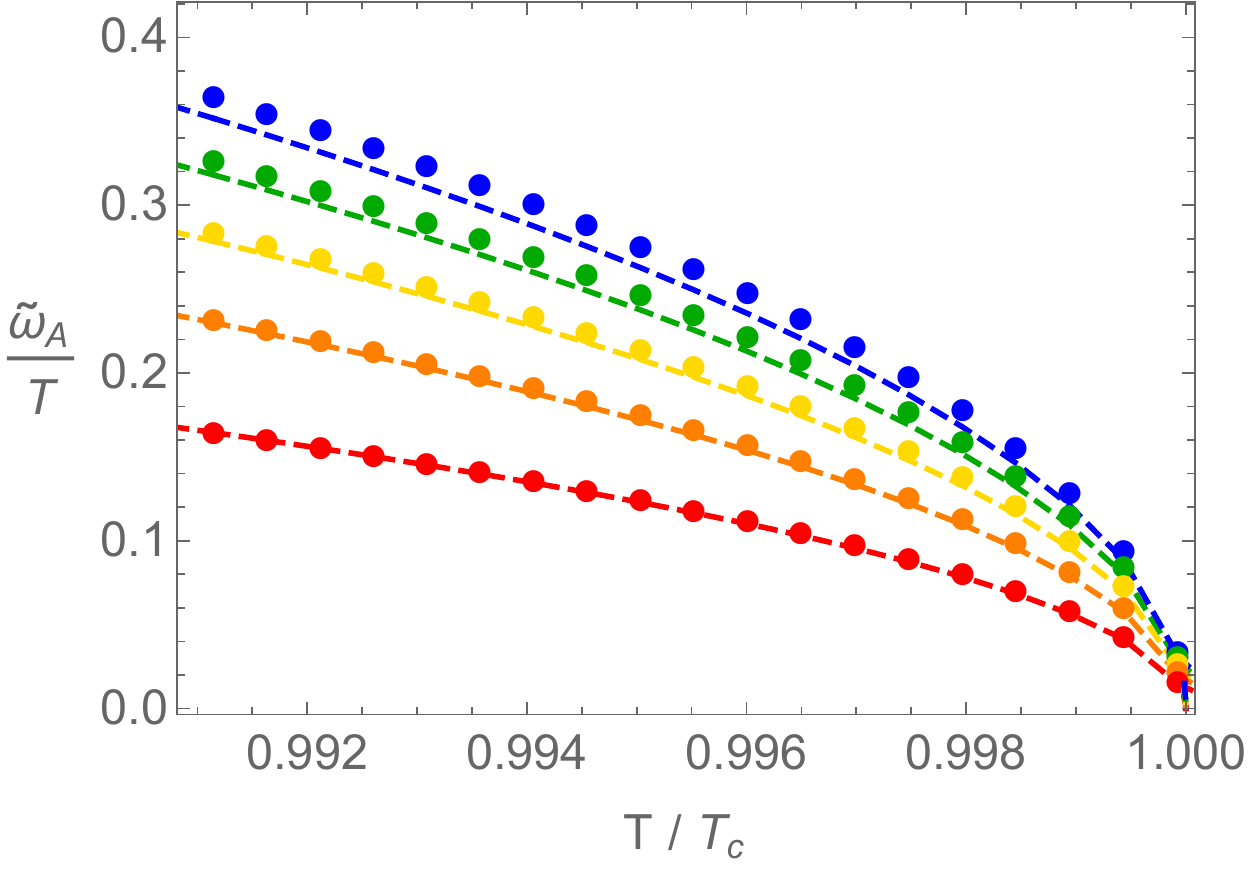} 
 \caption{The gauge field mass $\tilde{\omega}_A$. Dots are numerically obtained by fitting the dispersion relation. Dashed lines represent the analytical expression in Eq.\eqref{MAKOTOFOR}.  \textbf{Left:} $\lambda/T = 0.1$. \textbf{Right:} $\lambda/T = 0.1-0.5$ (red-blue).}\label{MAKOTOFIG}
\end{figure}

The comparison between our numerical data and the expression \eqref{MAKOTOFOR} derived in \cite{Natsuume:2022kic} is presented in  Fig. \ref{MAKOTOFIG}. The agreement near the critical point, $T/T_c\approx 1$, is excellent. Interestingly, we notice that the validity of Eq.\eqref{MAKOTOFOR} extends to lower temperatures when the EM coupling is small. On the contrary, for large values of the EM coupling $\lambda$, the analytical formula approximates well the numerical data only very close to the critical point.
%

%%%%%%%%%%%%%%%%%%%%%%%%%%%%%%%%
\section{Longitudinal collective modes and the Anderson-Higgs mechanism}
\label{seclong}

We now move to the discussion of the longitudinal sector. Once again, unless otherwise mentioned, we set the value of the EM coupling to $\lambda/T=0.1$.

\subsection{Collective excitations}
\begin{figure}[]
\centering
     \includegraphics[width=12cm]{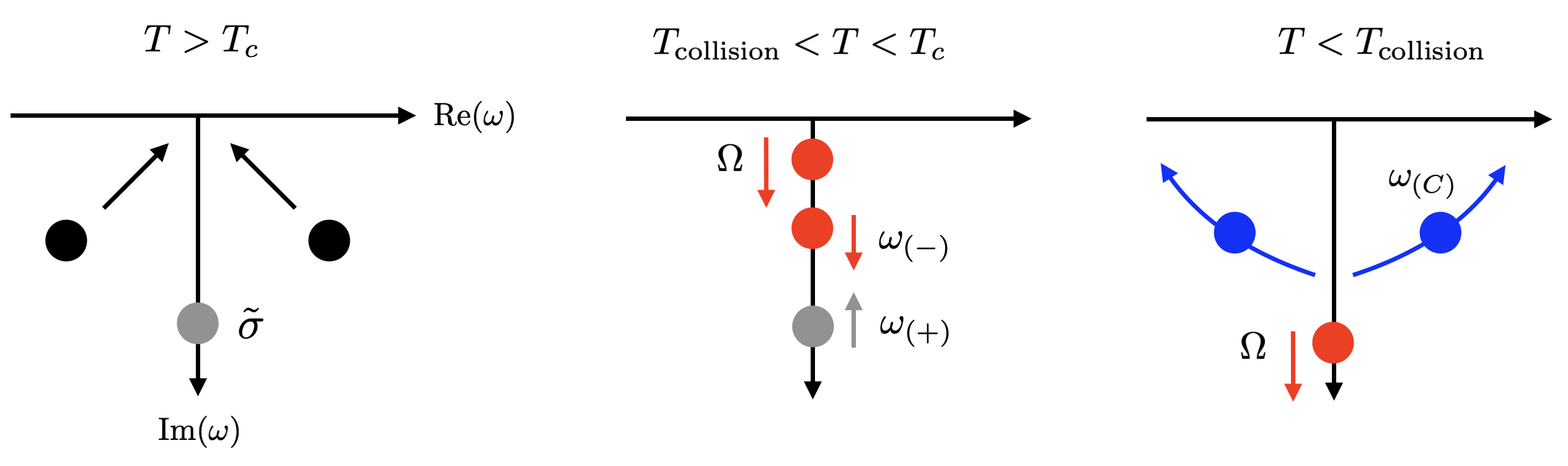}

     \vspace{0.65cm}
     
      \includegraphics[width=7.3cm]{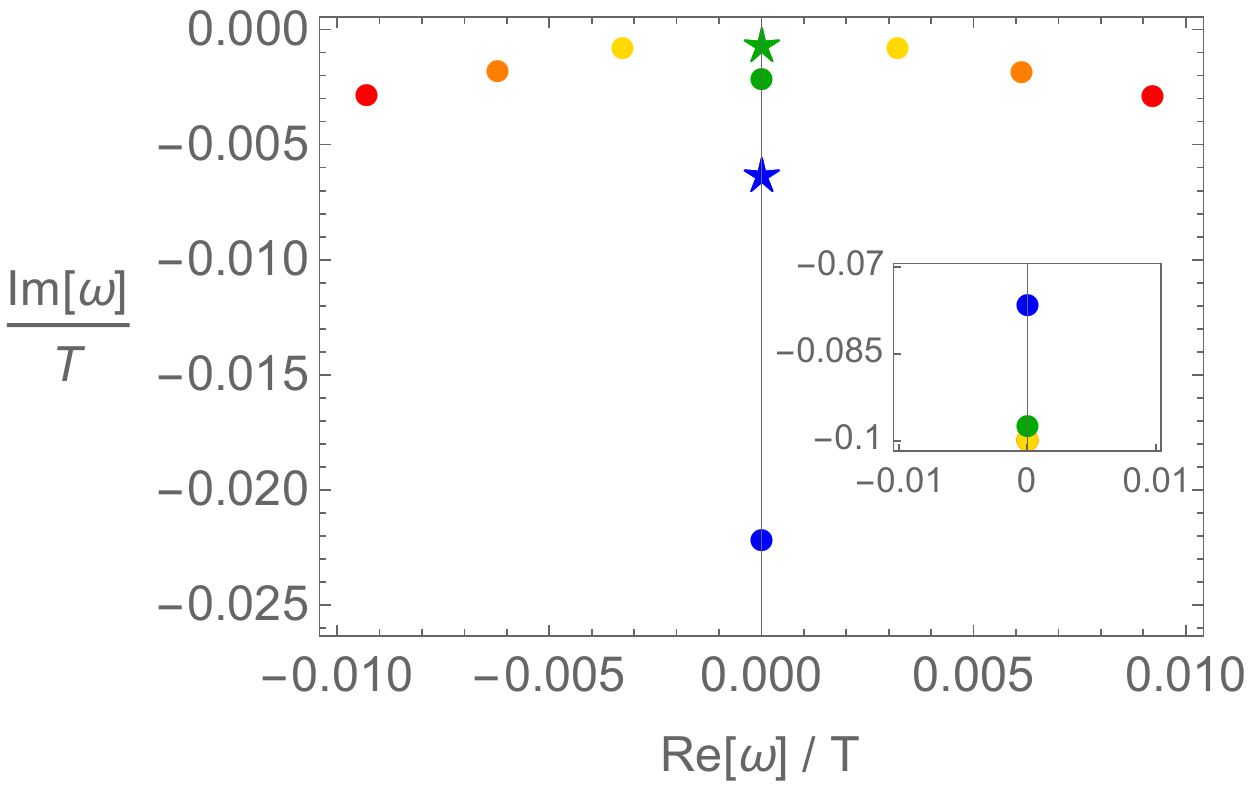} 
     \includegraphics[width=7.0cm]{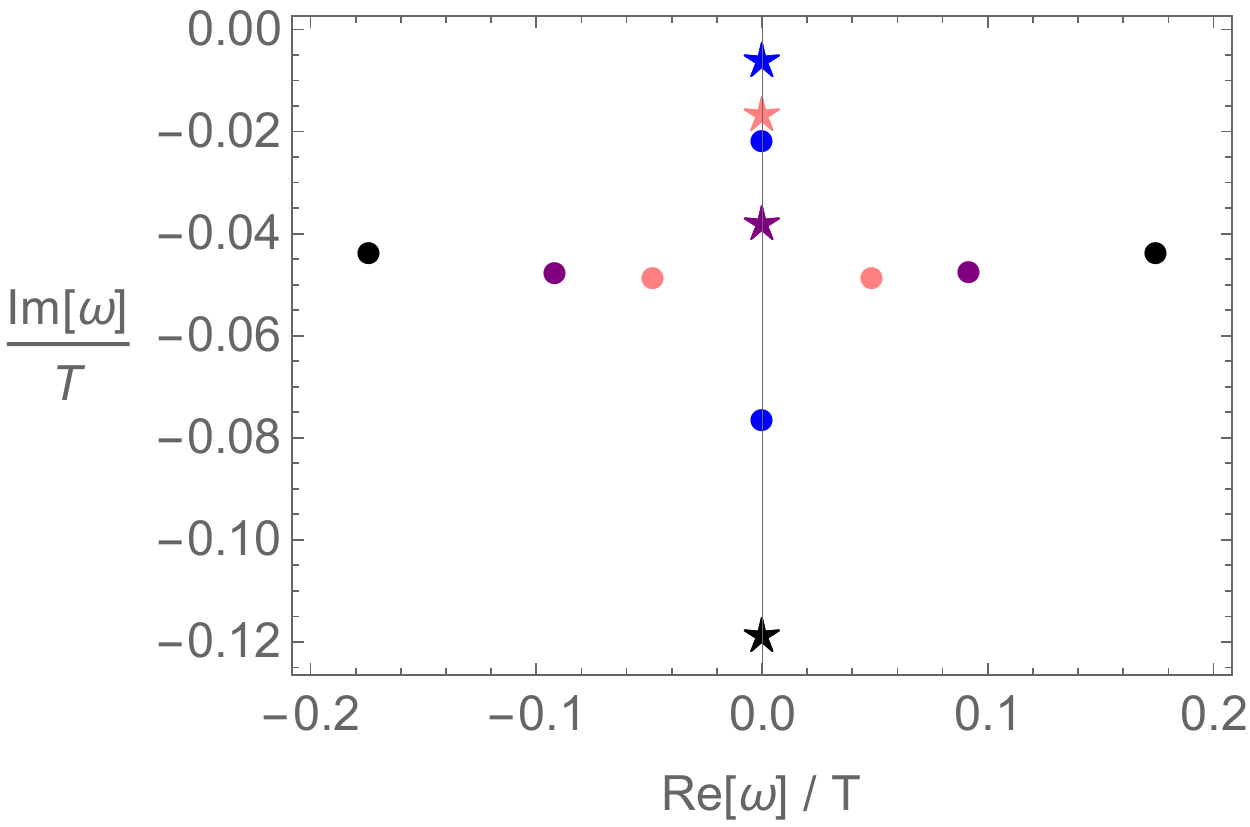}
     \caption{\textbf{Top panel: }Schematic plot of the poles in the longitudinal channel at zero momentum. In the normal phase, the black symbols represent the scalar fluctuations ($\delta \sigma, \delta \eta$), while the gray symbol is the damped charge diffusion from $(\delta a_x, \delta a_t)$. Below $T_c$, the scalar sector couples to the gauge sector and three damped poles appear ($\Omega, \omega_{(+)}, \omega_{(-)}$).  As the temperature is lowered, the $\omega_{(+)}$ pole collides with $\omega_{(-)}$ and generates a coupled of complex modes (blue symbols). The other pole $\Omega$ becomes more and more overdamped. \textbf{Bottom panel: } the numerical data. \textbf{Bottom left:} the near critical region, $T/T_c = 1.001 - 0.999$ (red-blue). The $\omega_{(\pm)}$ poles are represented with circles while the $\Omega$ one with stars.  The inset shows the behavior of $\omega_{(+)}$ pole.  \textbf{Bottom right:} the collision regime, $T/T_c = 0.999 - 0.989$ (blue, pink, purple, black). The $\omega_{(\pm)}$ poles (circles) collide on the real axes, while the $\Omega$ pole (star) moves down along the imaginary axes.}\label{LONGPLOT0}
\end{figure}

For simplicity, let us start with the homogeneous case, $k=0$. Given that the dynamics is complicated, we find instructive to first present a schematic description which refers to the top panel of Fig. \ref{LONGPLOT0}.

In the normal phase, above the critical temperature, the fluctuations of the scalar order parameter at zero wave-vector decouple from those of the gauge field. The modes associated with the scalar fluctuations, sometimes referred as critical modes, have both a real and imaginary gap which vanish at the critical temperature. This dynamics is exactly equivalent to that presented in the probe holographic superfluid in \cite{Amado:2009ts} (see also \cite{Donos:2022qao}) and can be easily derived using the time-dependent Ginzburg-Landau theory.

In addition to the scalar critical modes, there is a non-hydrodynamic mode that corresponds to damped charge diffusion. Here, charge fluctuations are damped (rather than diffusing) 
because of the effects of dynamical electromagnetism (see \cite{Hernandez:2017mch,Ahn:2022azl}), i.e., $\omega = -i \tilde{\sigma} = -i \sigma_0 \, \lambda$. At the critical temperature, $T=T_c$, the two critical modes approach the origin. However, differently from the case of the superfluid, the mode corresponding to the fluctuations of charge does not go to the origin at the critical point as it remains overdamped. As a consequence, just below the critical temperature, no massless propagating degree of freedom appears (cfr. second sound in superfluids), but rather one observes three different modes with a purely imaginary frequency which we denote as $\Omega,\, \omega_{(+)},\, \omega_{(-)}$. Decreasing further the temperature, two of these three modes, $\omega_{(+)}$ and $\omega_{(-)}$, collide on the imaginary axes and create a pair of complex modes which move towards the real axes and become underdamped. We will refer to those (complex) modes as $\omega_{(C)}$. The other third mode, $\Omega$, remains on the imaginary axes, and its (negative) imaginary part increases by decreasing temperature. 

\begin{figure}[]
\centering
     \includegraphics[width=4.8cm]{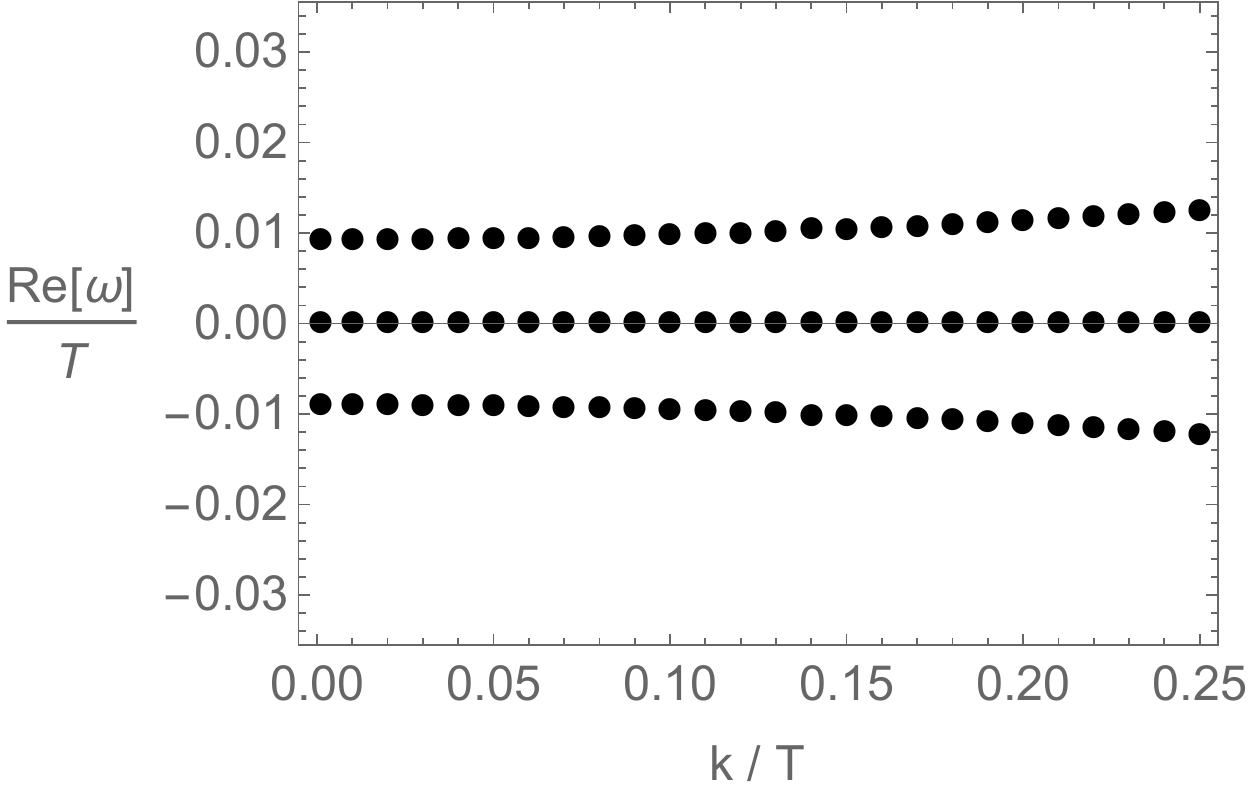} 
     \includegraphics[width=4.8cm]{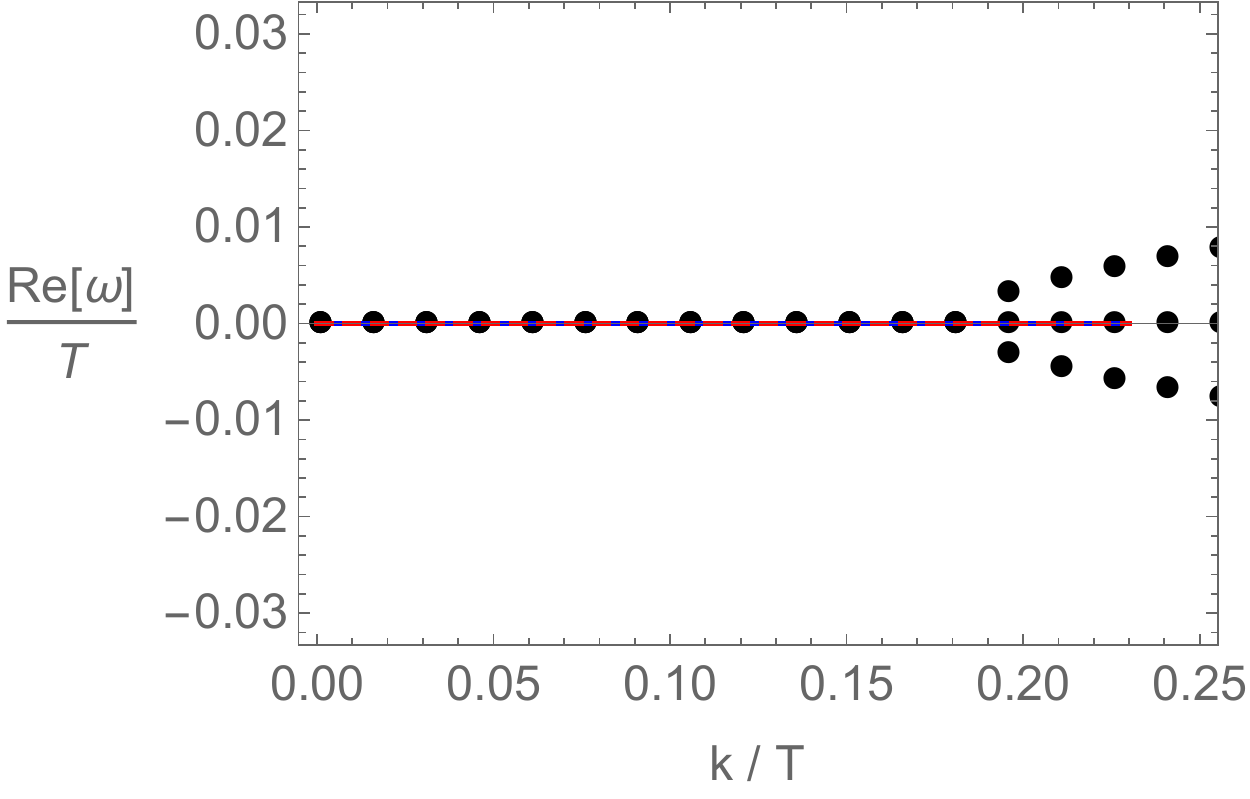} 
     \includegraphics[width=4.8cm]{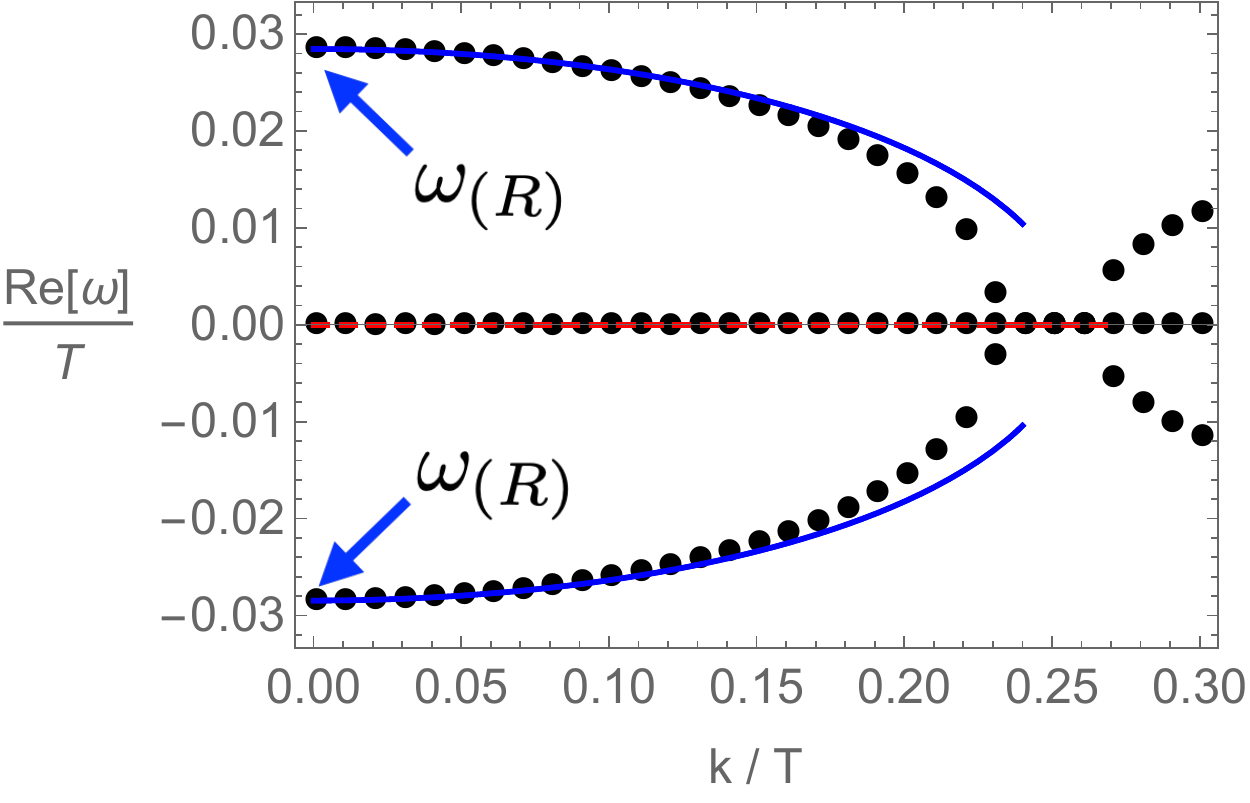} 

     \vspace{0.2cm}
     
     \includegraphics[width=4.8cm]{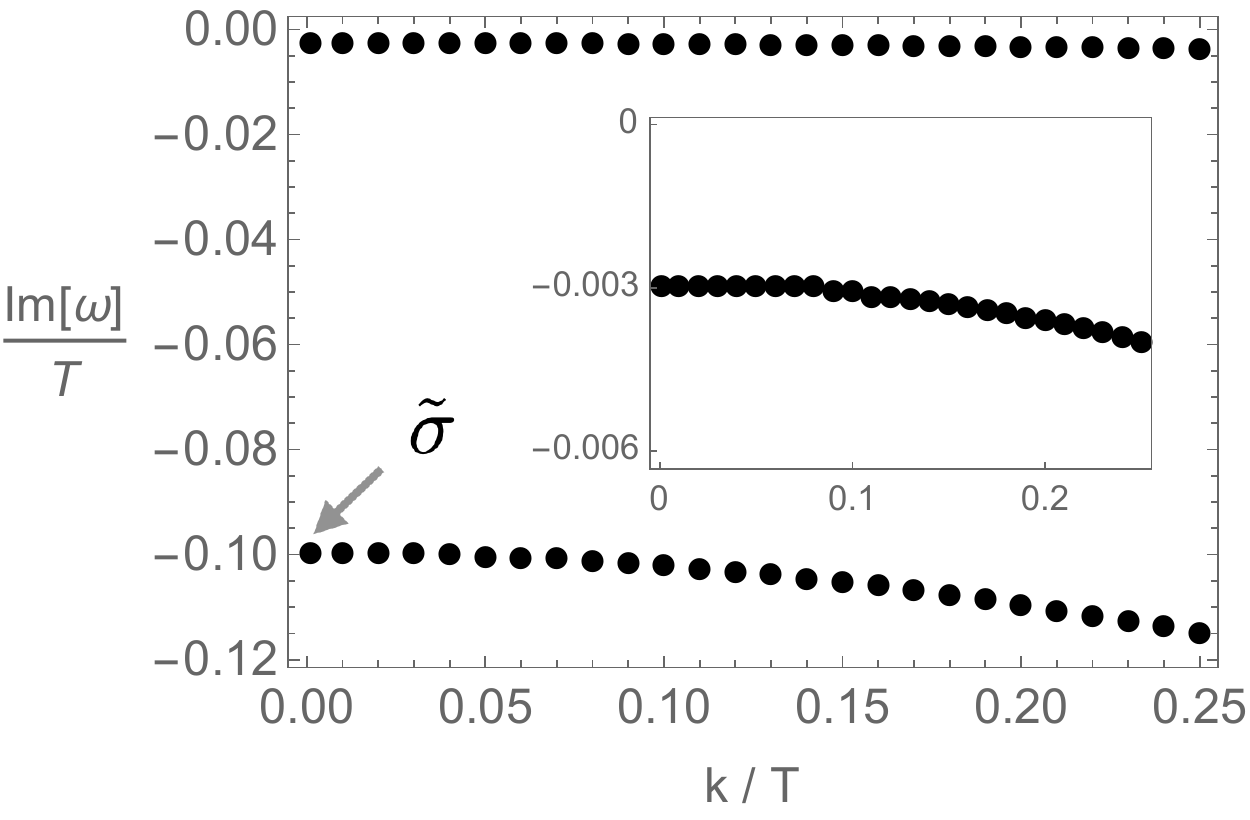} 
     \includegraphics[width=4.8cm]{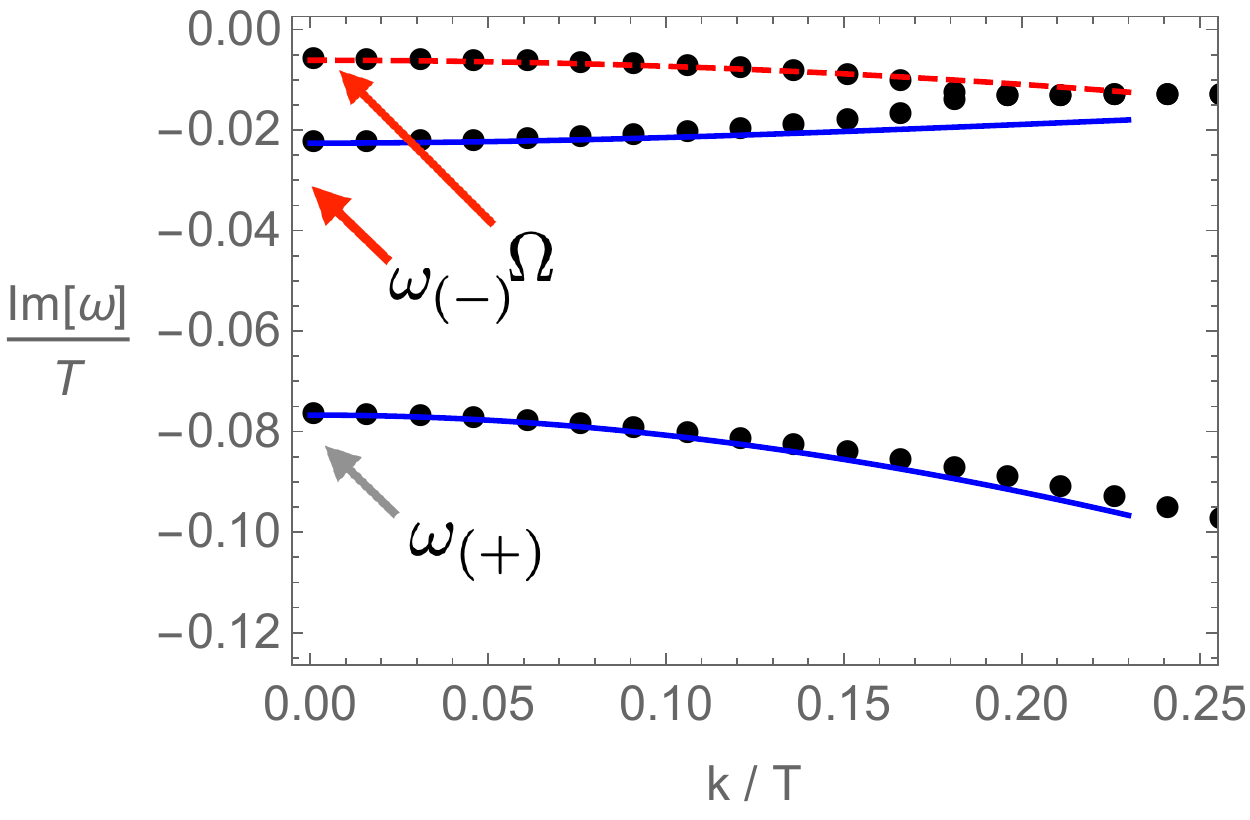} 
     \includegraphics[width=4.8cm]{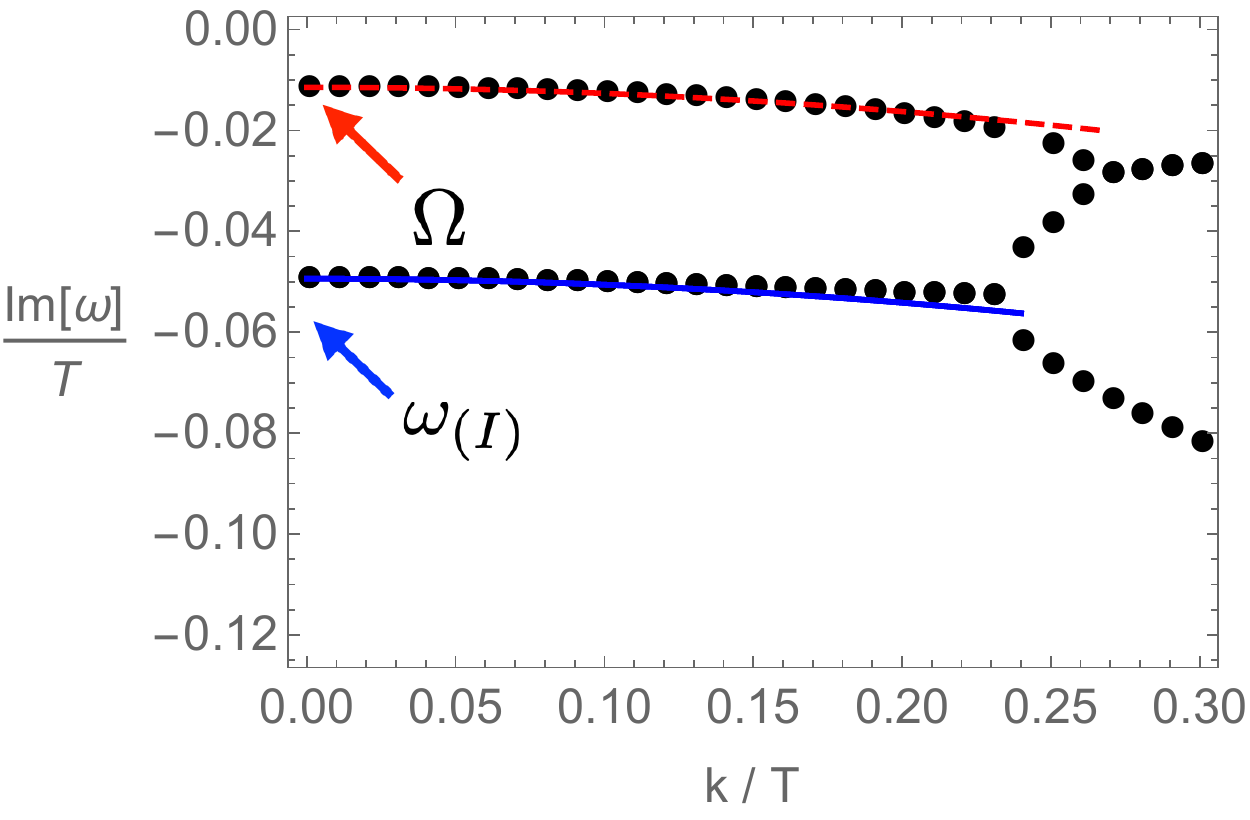} 

      \vspace{0.5cm}

      \includegraphics[width=4.8cm]{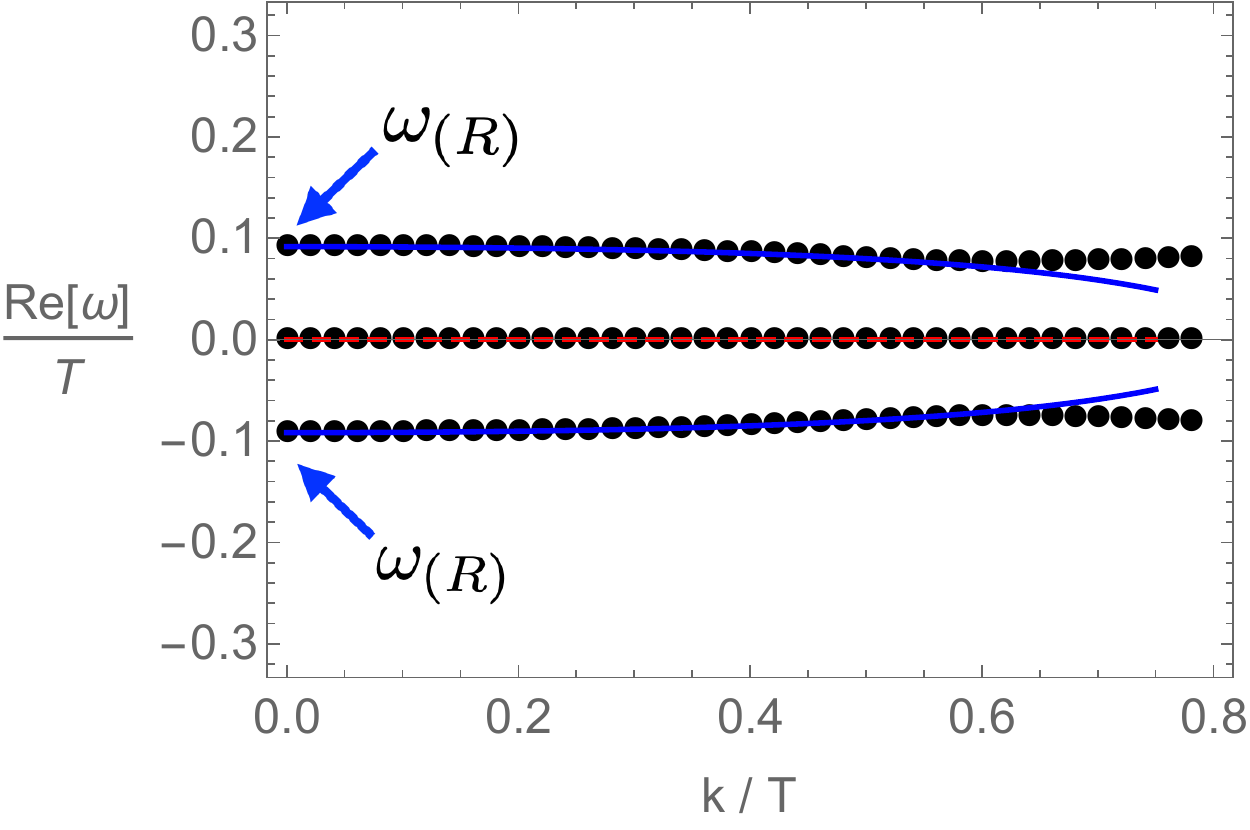} 
     \includegraphics[width=4.8cm]{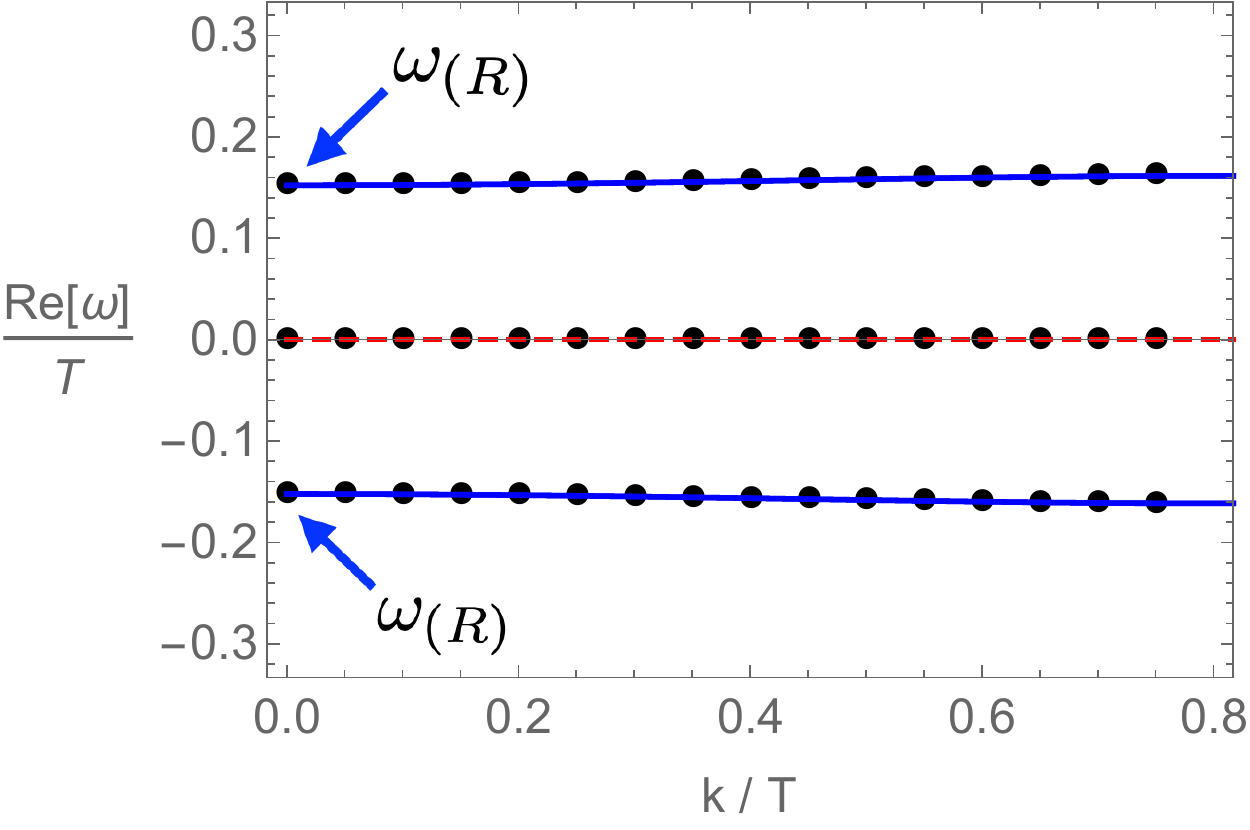} 
     \includegraphics[width=4.8cm]{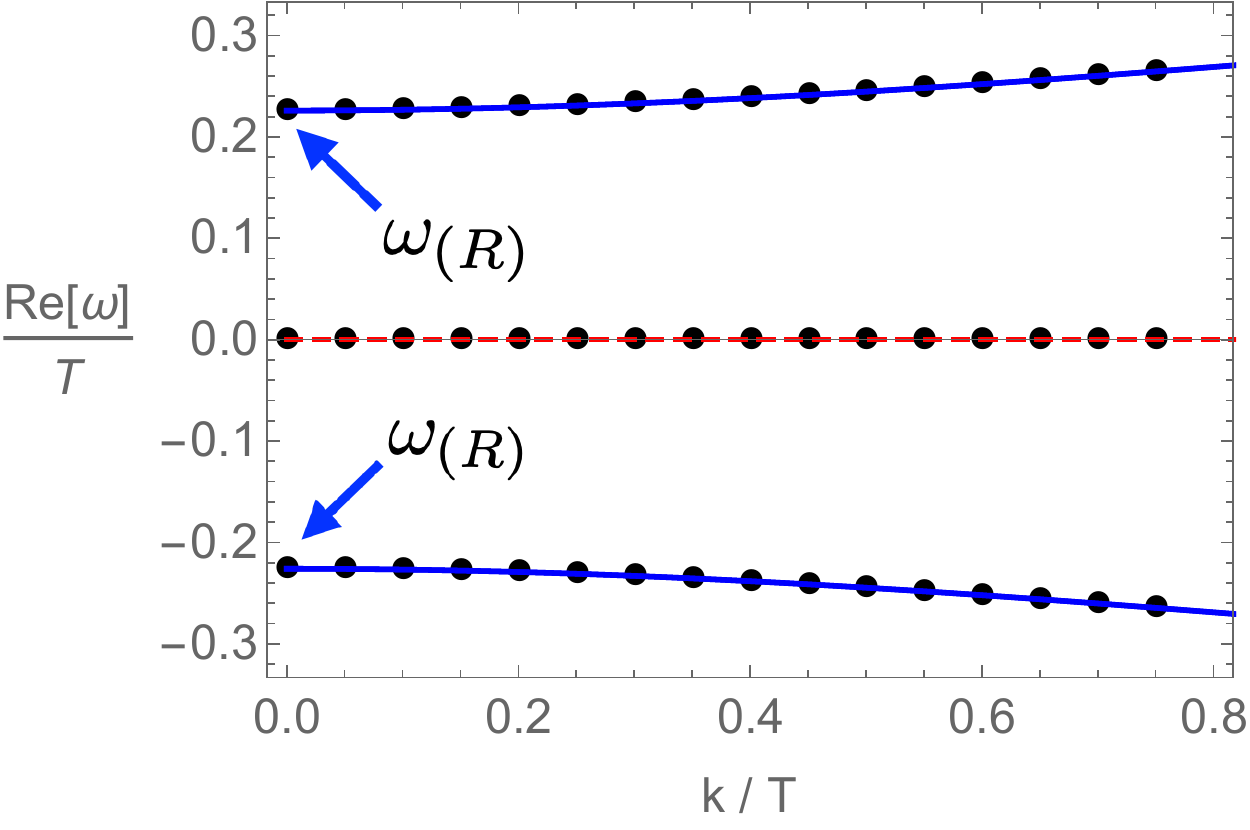} 

      \vspace{0.2cm}
     
     \includegraphics[width=4.8cm]{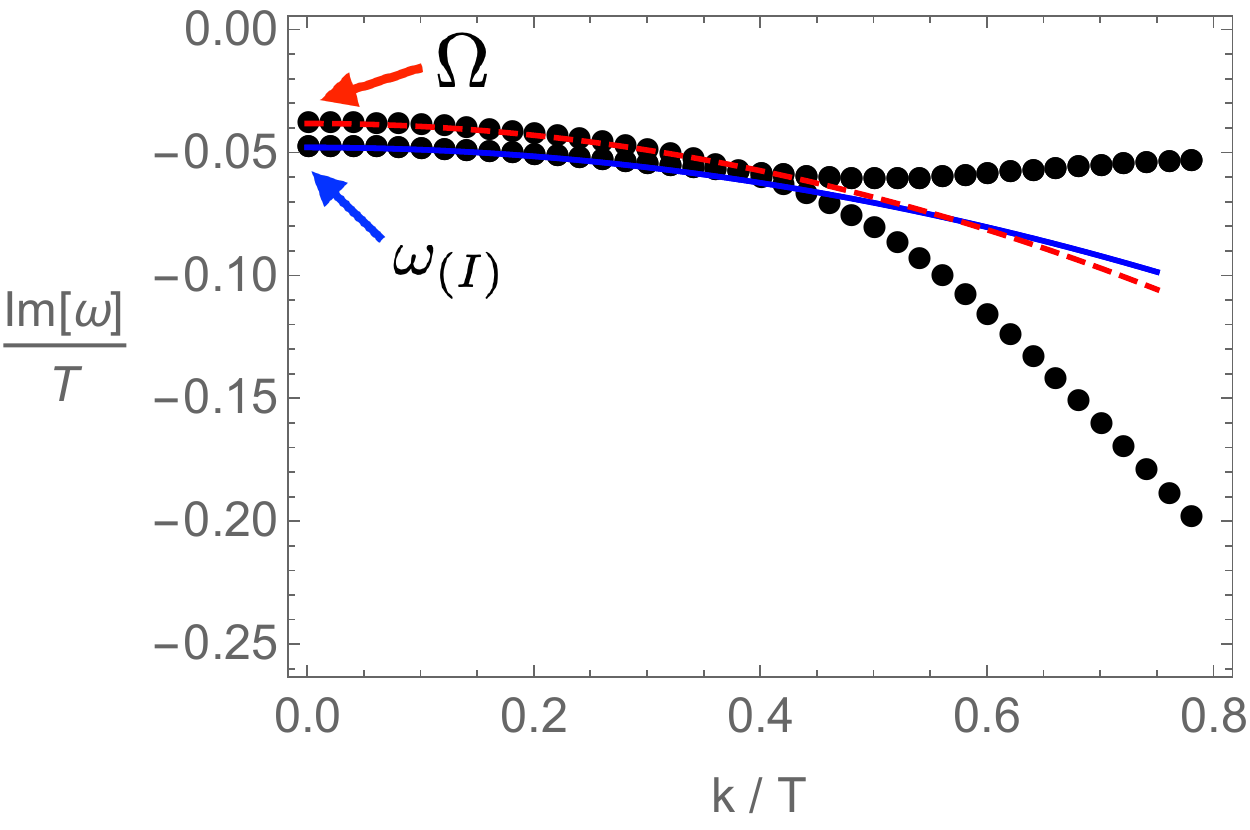} 
     \includegraphics[width=4.8cm]{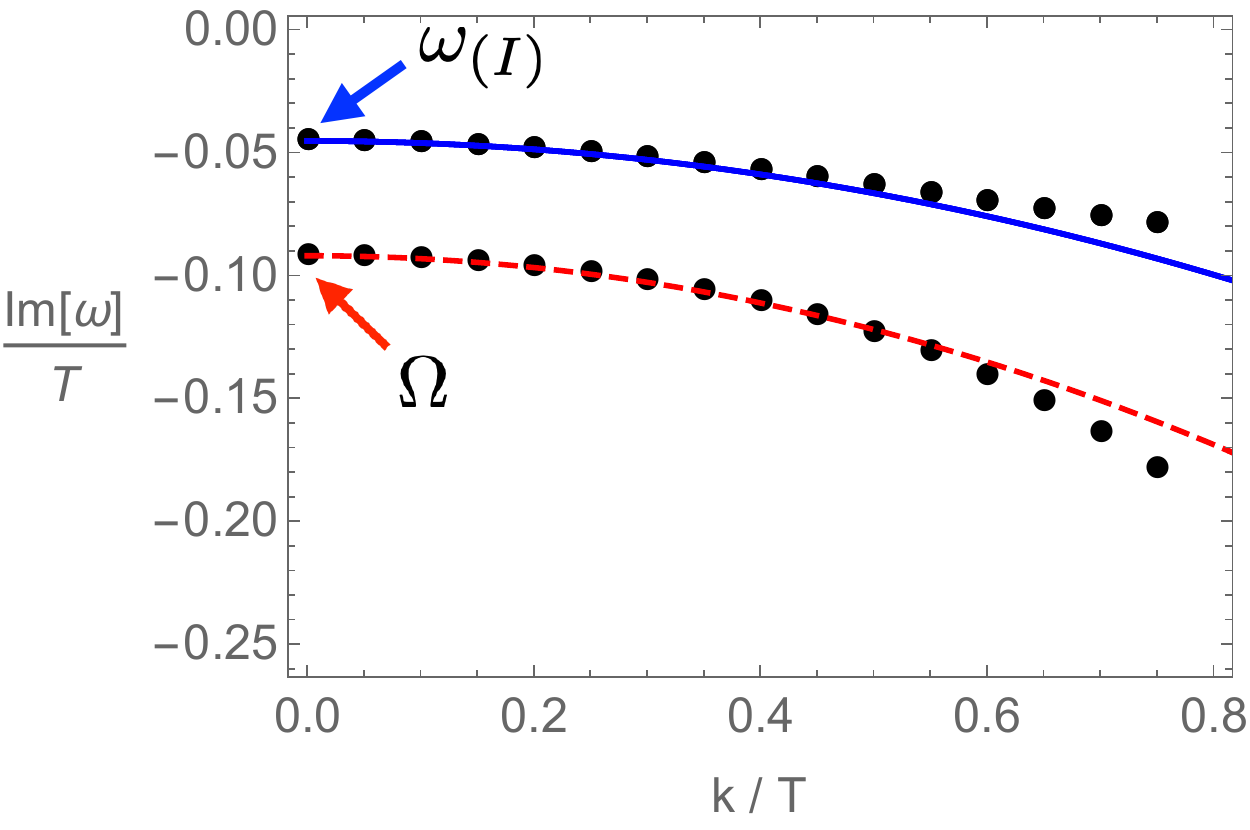} 
     \includegraphics[width=4.8cm]{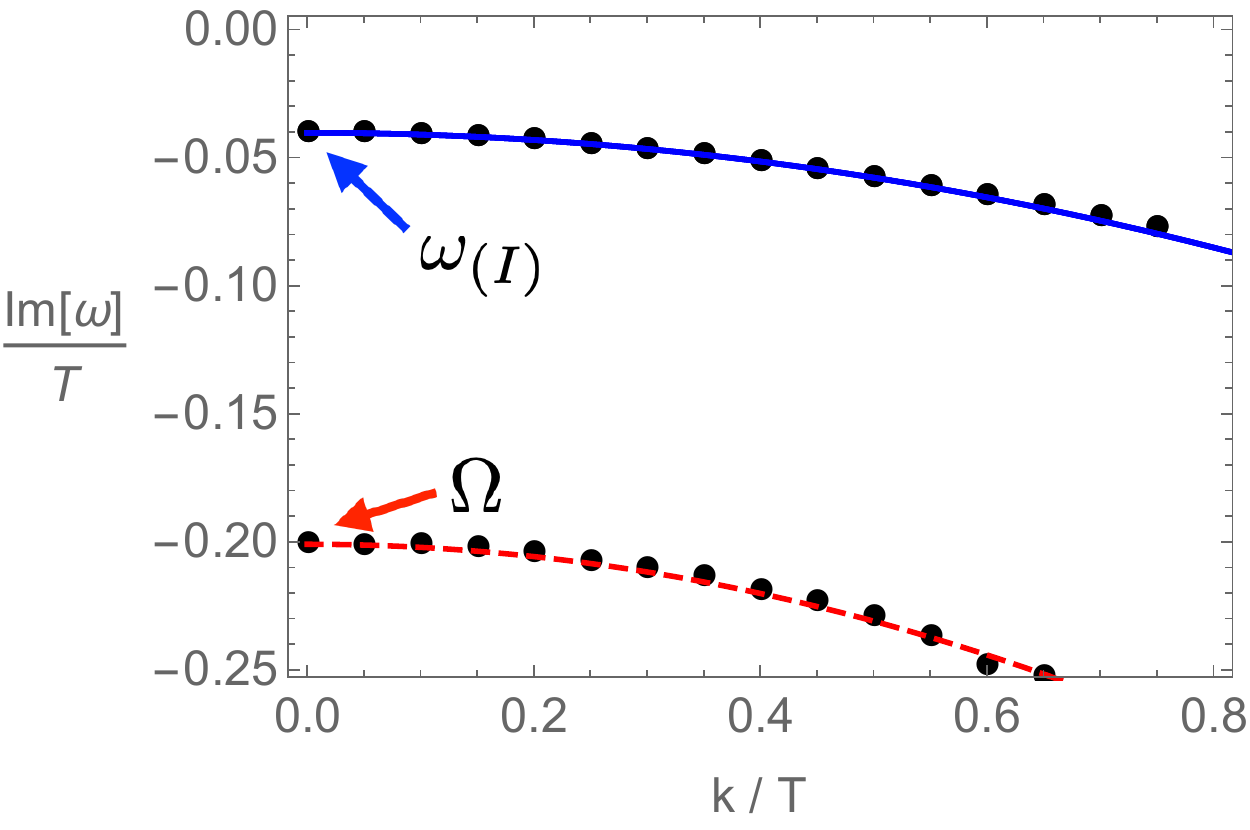} 
     
 \caption{Dispersion relation of the low-energy modes in the longitudinal spectrum. From top left to bottom right, the temperature decreases, $T/T_c = 1.00141, \, 0.999438, \, 0.998946,\, 0.996497, \, 0.991633, \, 0.982048$. Each panel is associated to the corresponding imaginary part located below it.}\label{LONGPLOT1}
\end{figure}
Notice that, at zero wave-vector, the dynamics of the pair of modes $\omega_{(\pm)}$ is the same as in the transverse sector (e.g. Fig. \ref{TRANFIG2}). This simply reflects the fact that, when the momentum is \textit{zero}, the equations of motion for the longitudinal fluctuations, Eq.\eqref{eq:coupledEOMsBroken}, can be decomposed into two decoupled sectors: i) ($\delta \sigma, \delta \eta, \delta a_t$); ii) $\delta a_x$. Then, the equation for $\delta a_x$ is exactly the same as the one in the transverse sector, Eq.\eqref{eq:transVecEOMBroken}.

We now move to the case of finite wave-vector, $k \neq 0$. The dynamics is more complicated as all the fluctuations are now coupled. Phenomenologically, at least in the limit of small wave-vector, $k/T\ll 1$, we find that the lowest quasi-normal modes are well approximated by the following equations with six phenomenological parameters ($\tilde{\sigma}, \tilde{\omega}_A, \mathcal{V}, \Gamma, \Omega, D_{\Omega}$):
\begin{align}\label{LONGFOR1}
\begin{split}
\omega  \left(\omega + i \, \tilde{\sigma} + i \, \Gamma \, k^2 \right) = \mathcal{V}^2 \, k^2 + \tilde{\omega}_A^2 \,, \qquad \omega + i \, \Omega + i \, D_{\Omega} \, k^2 = 0 \,.
\end{split}
\end{align}
Solving the two equations above gives the  dispersion of the modes in the limit of small wave-vector,
\begin{align}
&\omega=\pm \frac{1}{2}\sqrt{4 \, \tilde{\omega}_A^2-\tilde \sigma^2}-\frac{i}{2} \tilde{\sigma} \,+\, \left( \pm \frac{2 \mathcal{V}^2 - \Gamma \tilde{\sigma}}{2 \sqrt{4 \, \tilde{\omega}_A^2 - \tilde{\sigma}^2}} - \frac{i}{2} \Gamma \right) k^2    \,,\label{LONGFOR2a} \\
&\omega = - i \, \Omega - i \, D_{\Omega} \, k^2 \,.\label{LONGFOR2b}
\end{align}
As we will see shortly,  Eq.\eqref{LONGFOR2a} is related to the second sound in the superfluid case and reduces to $\omega = \omega_{(\pm)}$ or $\omega_{(C)}$ at $k=0$. The  Eq.\eqref{LONGFOR2b} is related to the ``Higgs'' mode.

Let us stress that the first equation of \eqref{LONGFOR1} is not derived from a formal effective description (e.g., hydrodynamics) but just an educated guess from two limiting cases. i) for $k=0$, the transverse mode is equivalent to the longitudinal mode ii) for $\lambda=0$, the dispersion relation of the superfluid ($\tilde{\sigma}=\tilde{\omega}_A =0$) is recovered. 
The second equation of \eqref{LONGFOR1} is the same as the superfulid case because it is supposed to be the Higgs mode, which is $\lambda$ independent. We will come back to these points in the following paragraph.
%a phenomenological perspective and using the guidance of the GL framework presented in Section \ref{secGL}. 
More generally, we do expect the above two modes (Eqs.\eqref{LONGFOR1}) to couple. Nevertheless, as we will see, at least in the limit of small wave-vector, the decoupling results become a reasonable approximation. This indicates that the coupling between the two equations above generates corrections to the dispersion relations which are higher-order in $k$.

Before continuing with our analysis, let us pause and discuss first the superfluid limit in which the gauge field is not dynamical at the boundary. For superfluids (see e.g. \cite{Amado:2009ts}), one finds the \textit{second} sound waves and the damped charge diffusive mode or amplitude mode. This corresponds to assume that (I) our parameters ($\tilde{\sigma},\, \tilde{\omega}_A$) vanish in Eq.\eqref{LONGFOR1} and (II) that $\Gamma$ and $\mathcal{V}$ are exactly the attenuation constant and the speed of propagation of second sound in the superfluid. As we will see, this is indeed the case. In the superfluid limit, the $\omega_{(\pm)}$ modes combine into second sound and the $\Omega$ mode becomes the Higgs mode.

The dispersion relation of the lowest QNMs in the longitudinal spectrum is shown in Fig. \ref{LONGPLOT1} from high temperature (top left), in the normal phase, to the lowest temperature accessible (bottom right).\footnote{Let us remind that the probe limit approximation ceases to be trustable at low temperature.} In solid/dashed lines, we display the fitting formulas using Eqs.\eqref{LONGFOR2a}-\eqref{LONGFOR2b}.
In what follows, we discuss in detail the coefficients, ($\tilde{\sigma}, \tilde{\omega}_A, \mathcal{V}, \Gamma, \Omega, D_{\Omega}$) appearing in Eqs.\eqref{LONGFOR1}. Finally, we will discuss the similarities and differences with the GL picture presented in Section \ref{secGL}.

\subsection{The fate of second sound}\label{PLASEC}
\begin{figure}[]
\centering
     \includegraphics[width=7.0cm]{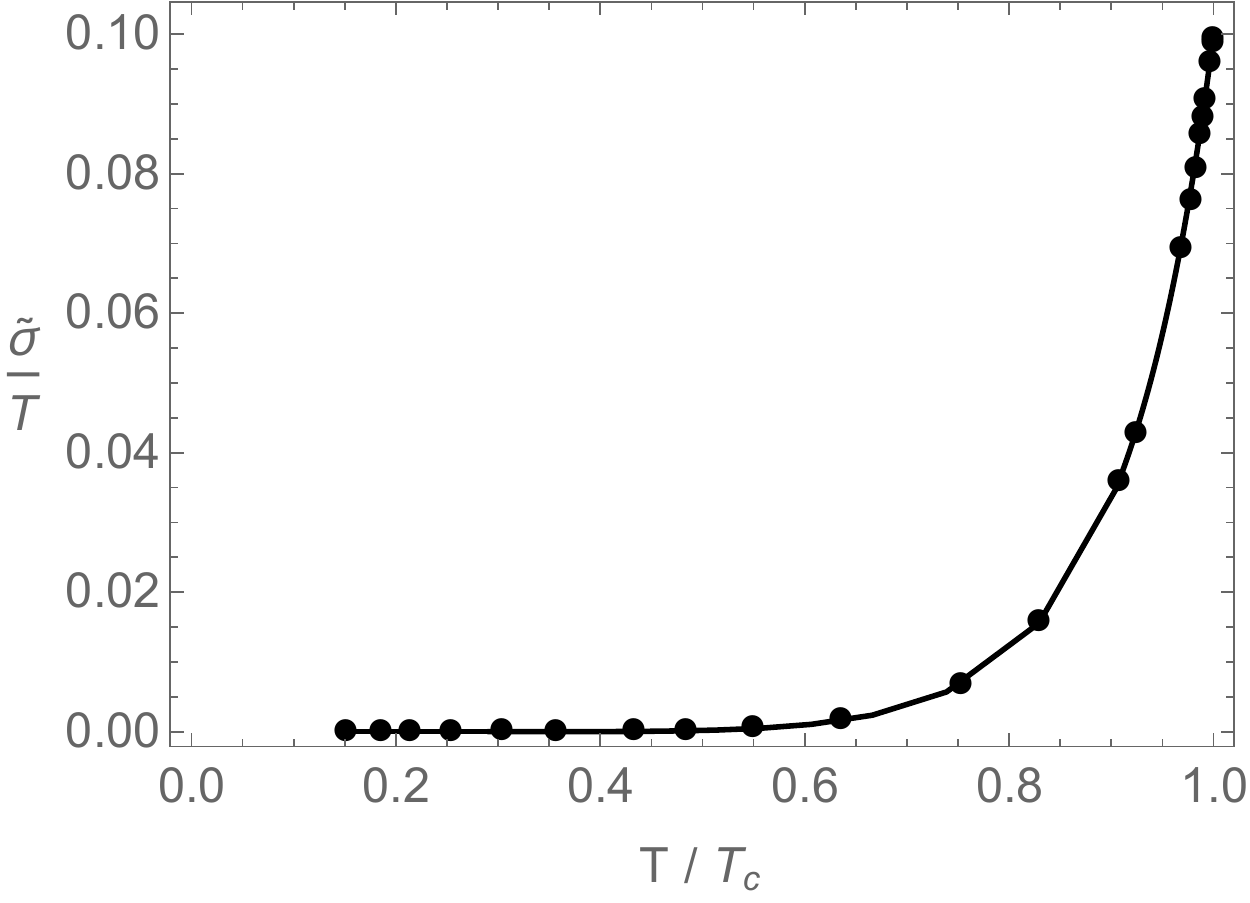}\quad 
     \includegraphics[width=7.0cm]{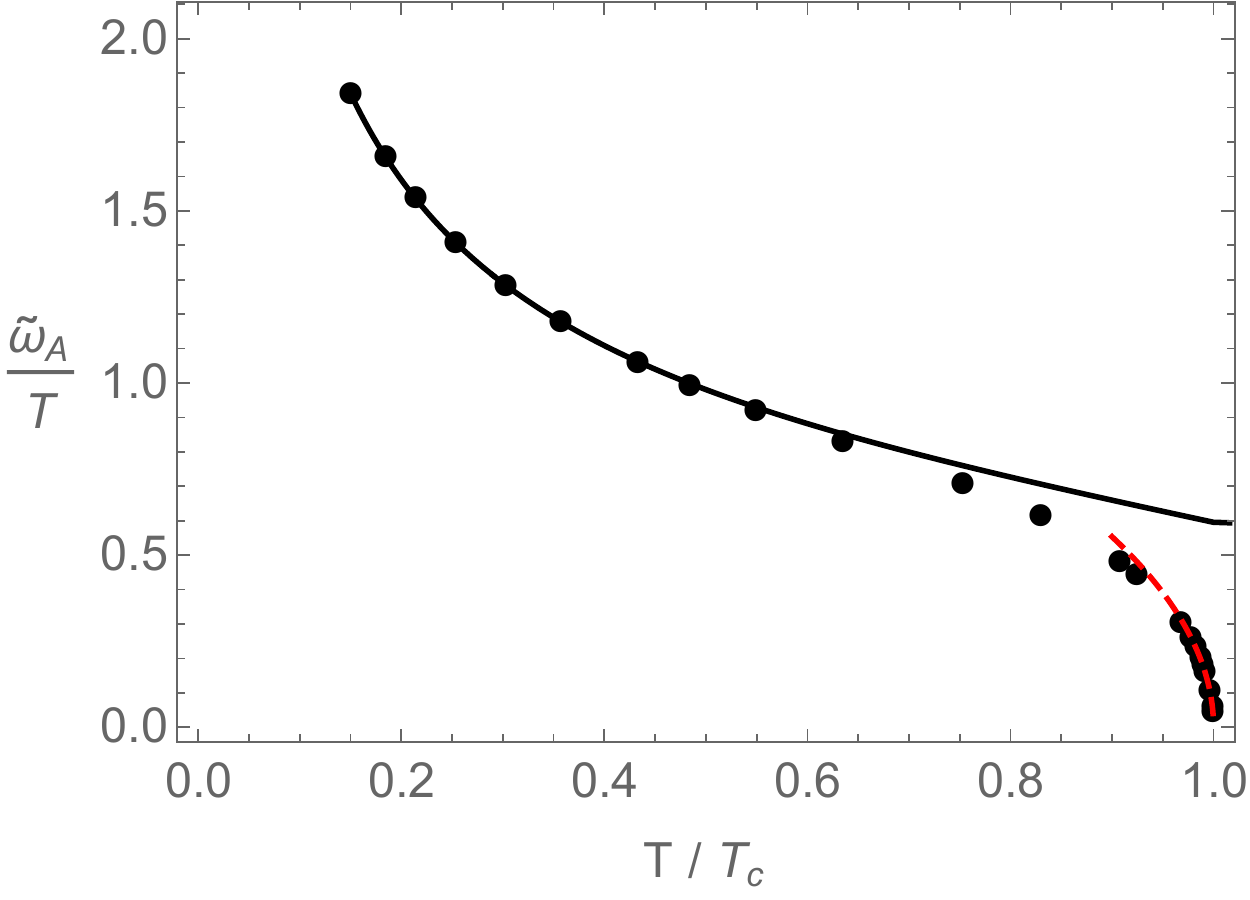}

     \vspace{0.2cm}
     
     \includegraphics[width=7.0cm]{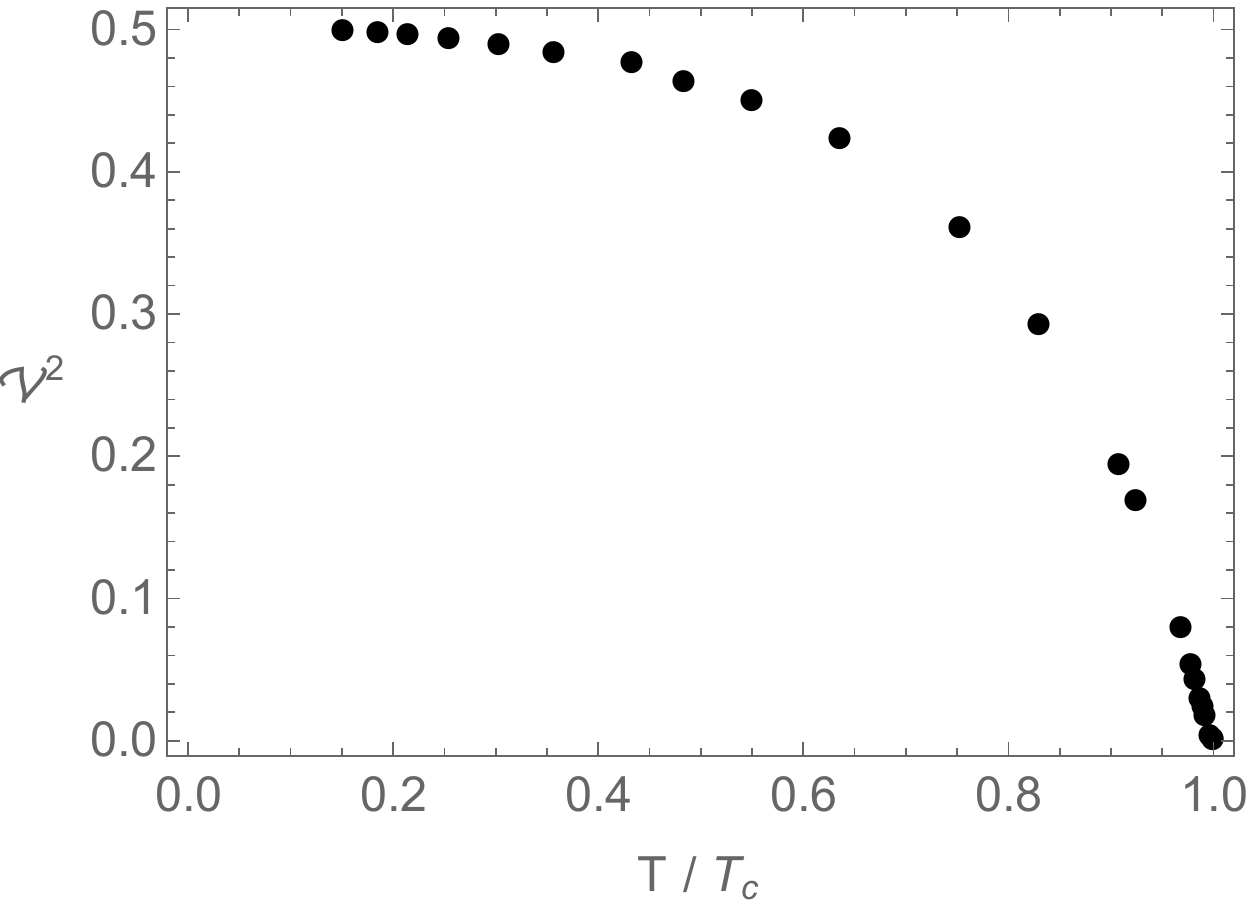} \quad
     \includegraphics[width=7.2cm]{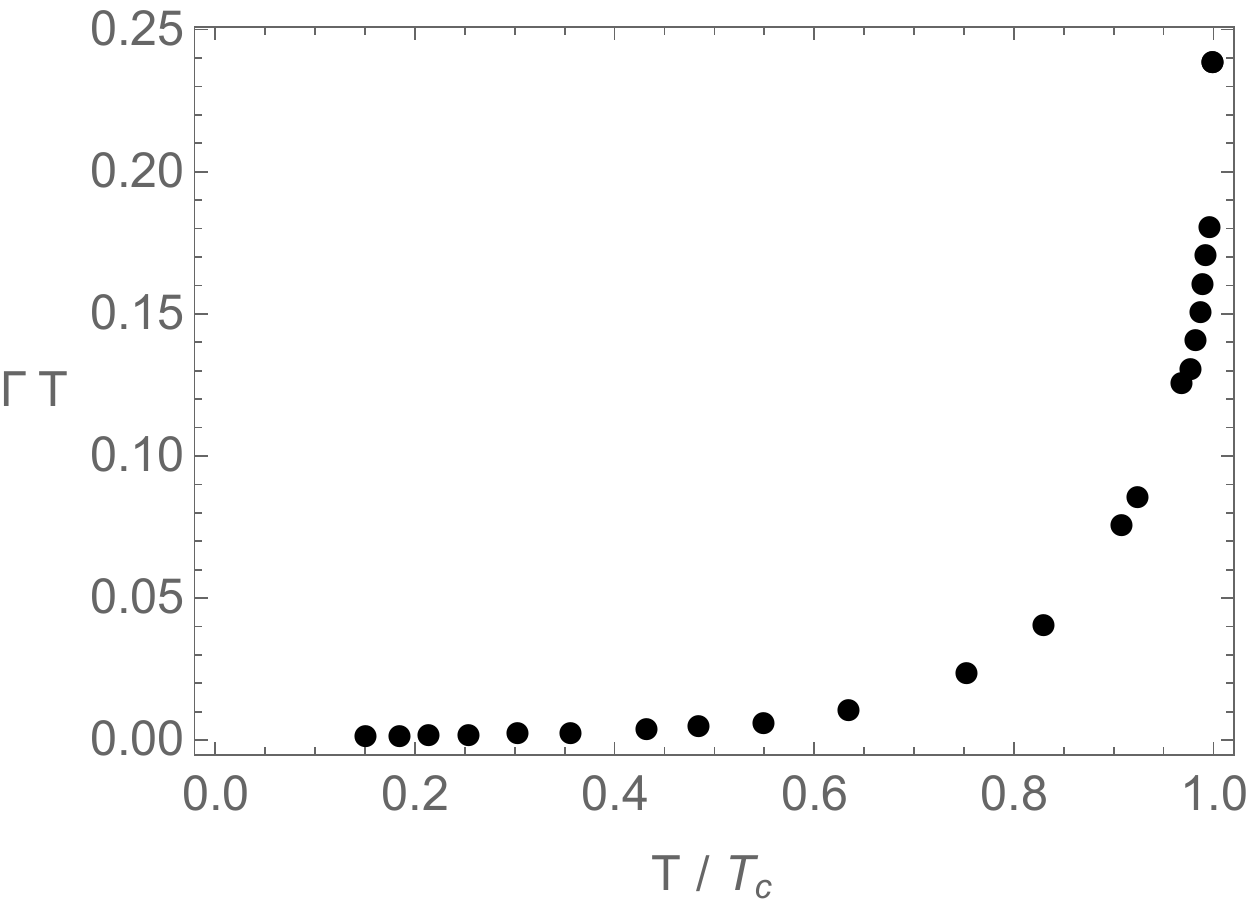} 
     
 \caption{Coefficients appearing in the dispersion relation, Eq.\eqref{LONGFOR2a}. Solid black lines are drawn using  Eq.\eqref{TRANSFIT2} for $\tilde{\sigma}$ and Eq.\eqref{TRANSFIT3} for $\tilde{\omega}_A$. The red dashed line near $T_c$ is Eq.\eqref{TRANSFIT4}. The speed of propagation $\mathcal{V}$ and the attenuation constant $\Gamma$ coincide exactly with those reported for second sound in the holographic superfluid \cite{Amado:2009ts} .}\label{PLOTFW1}
\end{figure}

Let us first analyze the dynamics of the second sound in our holographic superconductor. Its dispersion relation at low $k$ is given by Eq.\eqref{LONGFOR2a}. The temperature dependence of the coefficients ($\tilde{\sigma}, \tilde{\omega}_A, \mathcal{V}, \Gamma$) is presented in Fig. \ref{PLOTFW1}.
Interestingly, the coefficients ($\tilde{\sigma}, \tilde{\omega}_A)$ are exactly the same as the ones appearing in the dispersion relation for EM waves in the transverse sector (cfr. Eqs.\eqref{TRANSFIT2}-\eqref{TRANSFIT4}). 
At the same time, as perhaps expected, the speed of sound $\mathcal{V}$ and the attenuation constant $\Gamma$ coincide with those of second sound in the holographic superfluid model \cite{Amado:2009ts}. Notice how the speed of propagation approaches the conformal sound speed $\mathcal{V}^2=1/2$ at low temperature and vanishes at the critical point.

Also notice that as $T\rightarrow T_c$, both $\tilde{\omega}_A$ and $\mathcal{V}$ vanish. As a consequence, the dispersion relation therein becomes
\begin{align}\label{awa}
\begin{split}
\omega  = - i \tilde{\sigma} - i \,\Gamma\, k^2    \,,
\end{split}
\end{align}
which is the damped charge diffusion mode at $T\geq T_c$. In other words, the attenuation constant $\Gamma$, in the limit $T\rightarrow T_c$ , becomes the charge diffusion constant in the normal phase, $D_c T= 3/(4\pi)\sim0.238$. This is consistent with our data in Fig. \ref{PLOTFW1}.

Let us also notice that the dynamics of this mode is completely missing in the GL formalism presented in Section \ref{secGL} since we have not considered the coupling to the conserved charge density nor the corresponding charge fluctuations. In order to include this mode into the EFT framework, one should extend the GL theory and promote it as in model F in the Hoenberg-Halperin classification \cite{RevModPhys.49.435} for superfluids. In the context of holographic superfluids, the matching with model F has been proved explicitly in \cite{Donos:2022qao}. It would be interesting to repeat this analysis in the case of a superconductor with dynamical Coulomb interactions.

\begin{figure}[]
\centering
     \includegraphics[width=7.0cm]{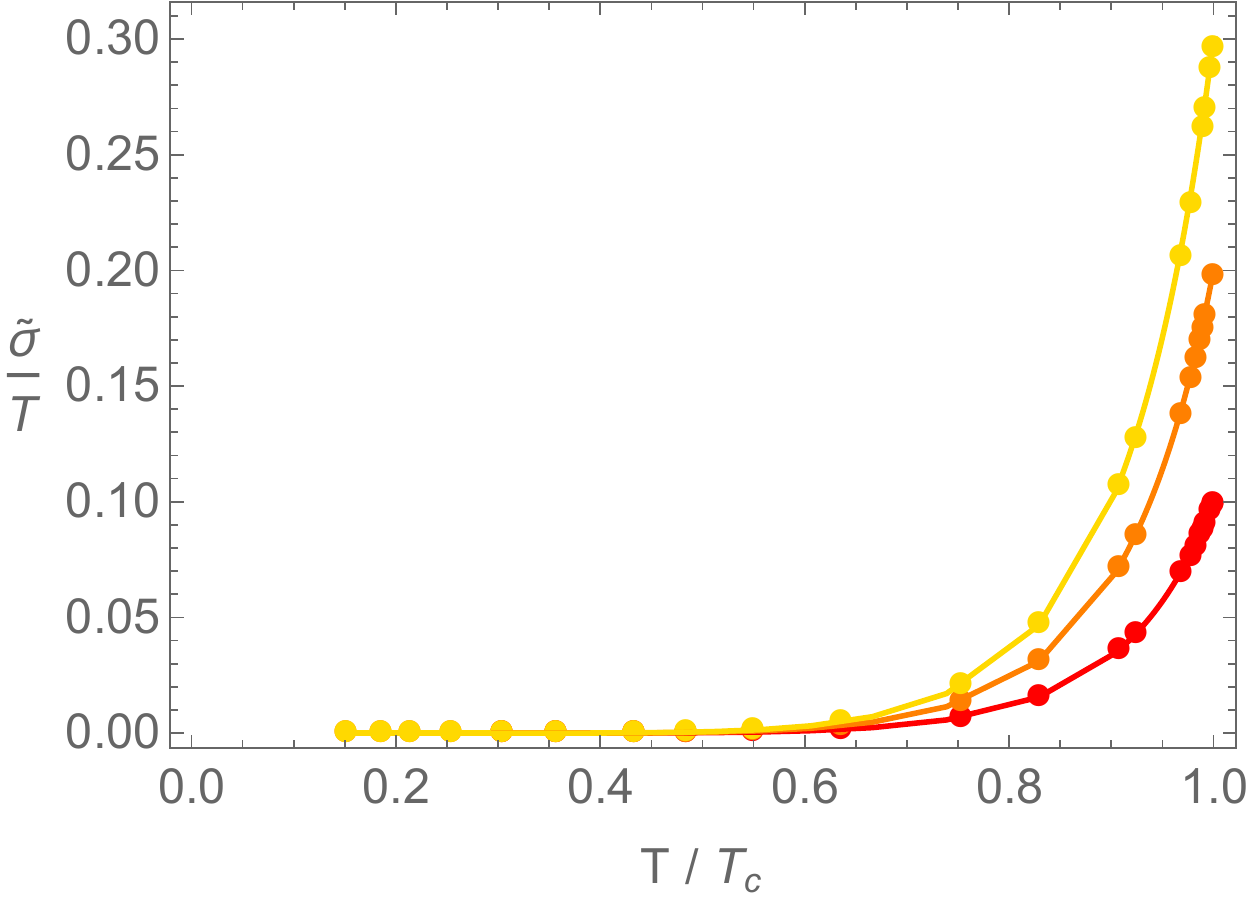} \quad
     \includegraphics[width=7.0cm]{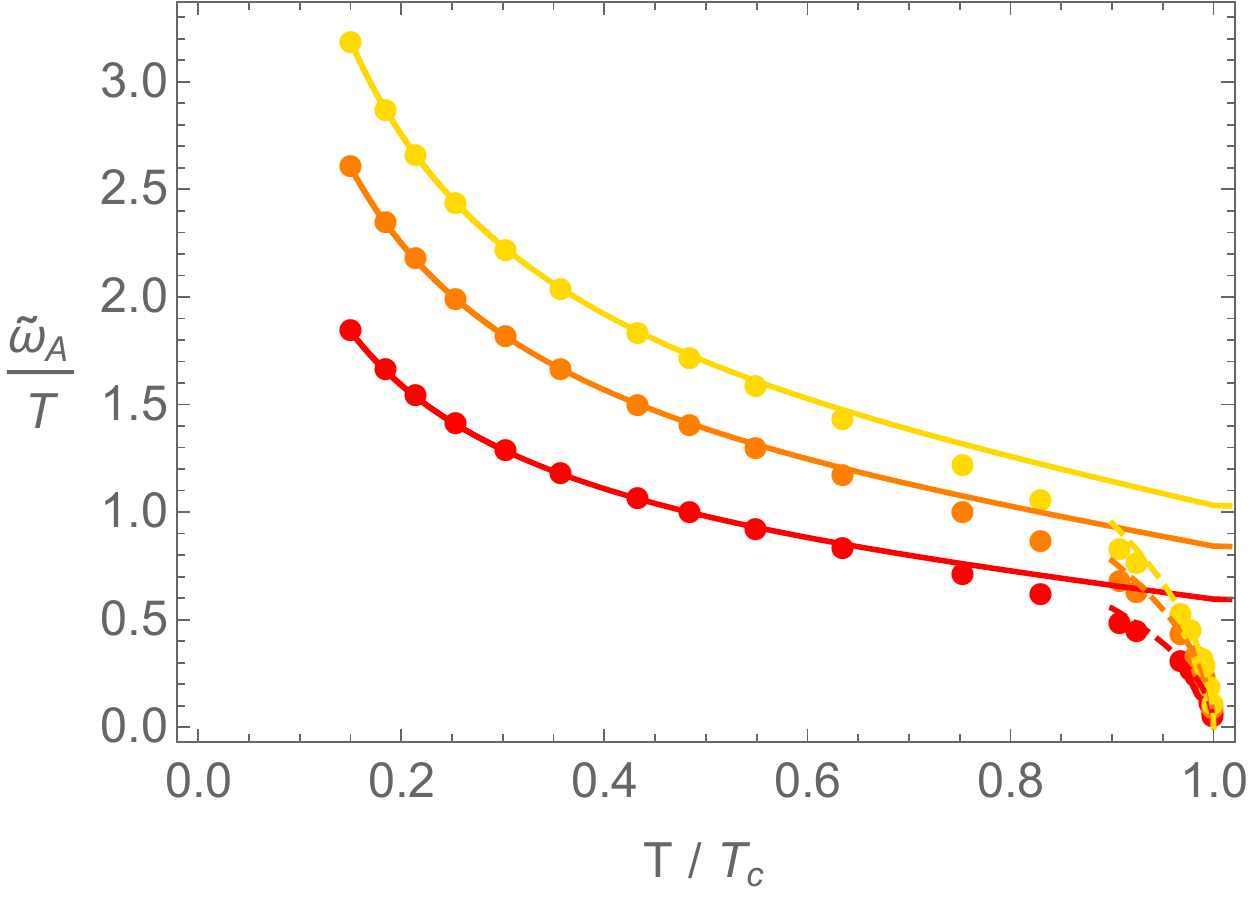} 

     \vspace{0.2cm}
     
     \includegraphics[width=7.0cm]{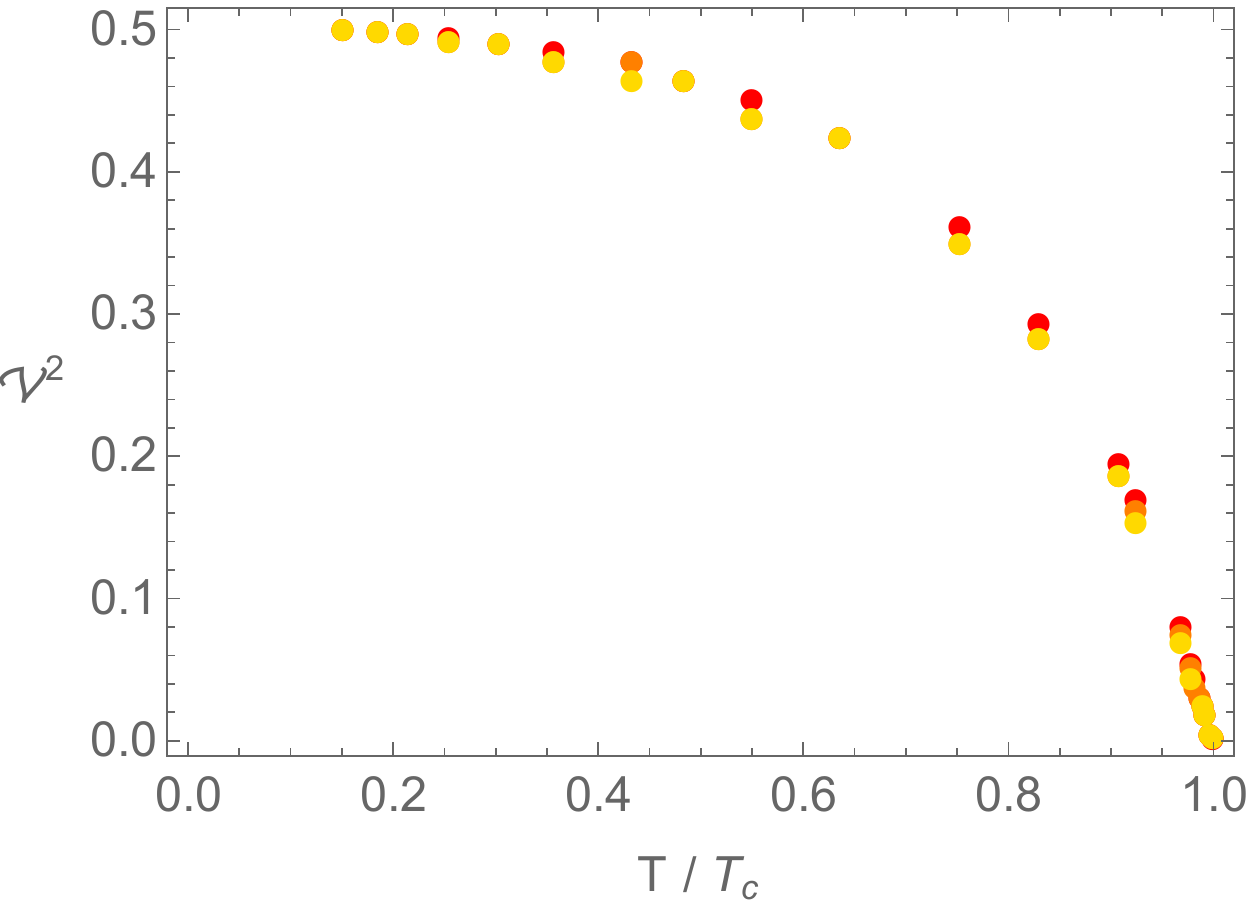} \quad
     \includegraphics[width=7.2cm]{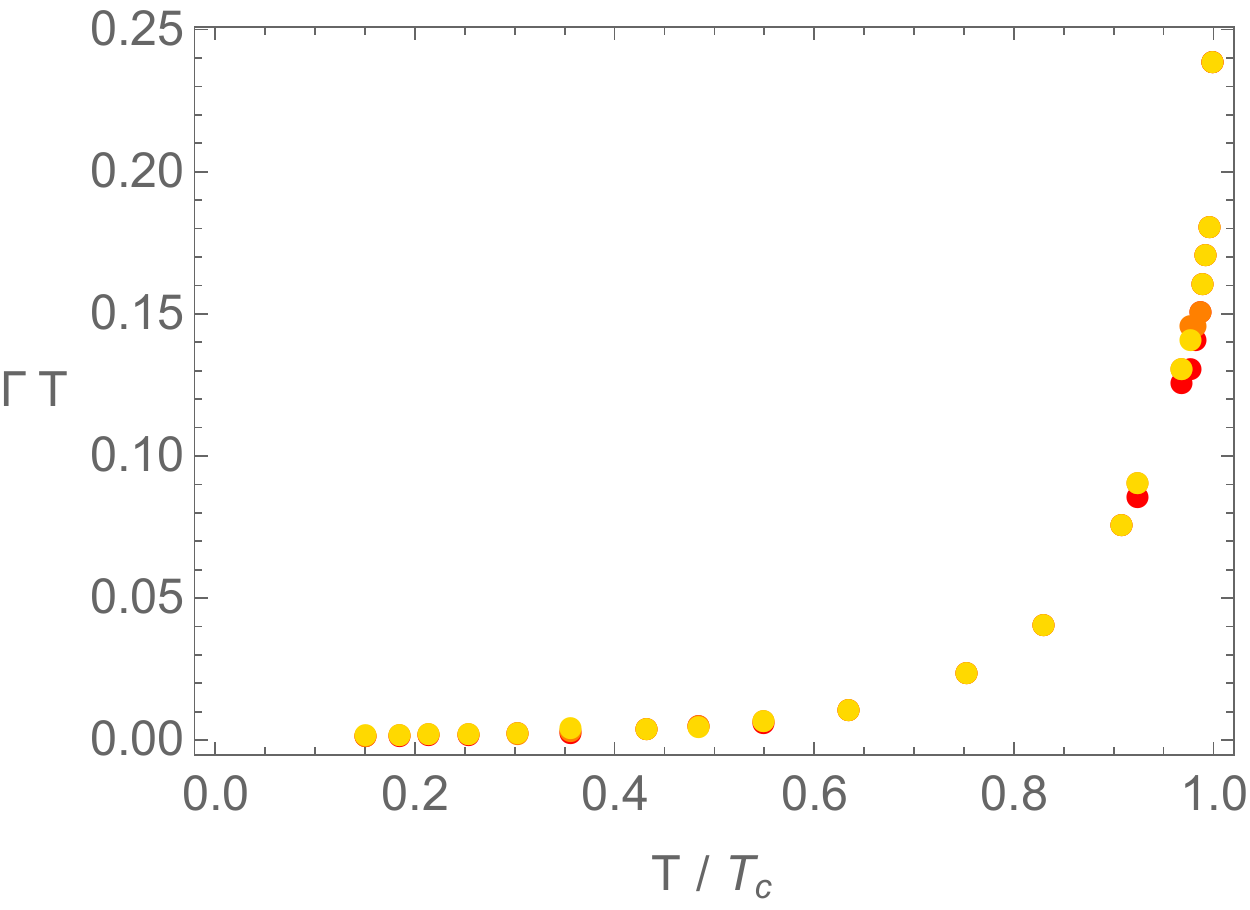} 
     
 \caption{Coefficients appearing in the dispersion relation \eqref{LONGFOR2a} as a function of the reduce temperature for $\lambda = 0.1, 0.2, 0.3$ (red, orange, yellow). Solid lines represent Eq.\eqref{TRANSFIT2} for $\tilde{\sigma}$ and Eq.\eqref{TRANSFIT3} for $\tilde{\omega}_A$. The dashed lines are Eq.\eqref{TRANSFIT4}.}\label{PLOTFW4}
\end{figure}

\begin{figure}[]
\centering
     \includegraphics[width=7.3cm]{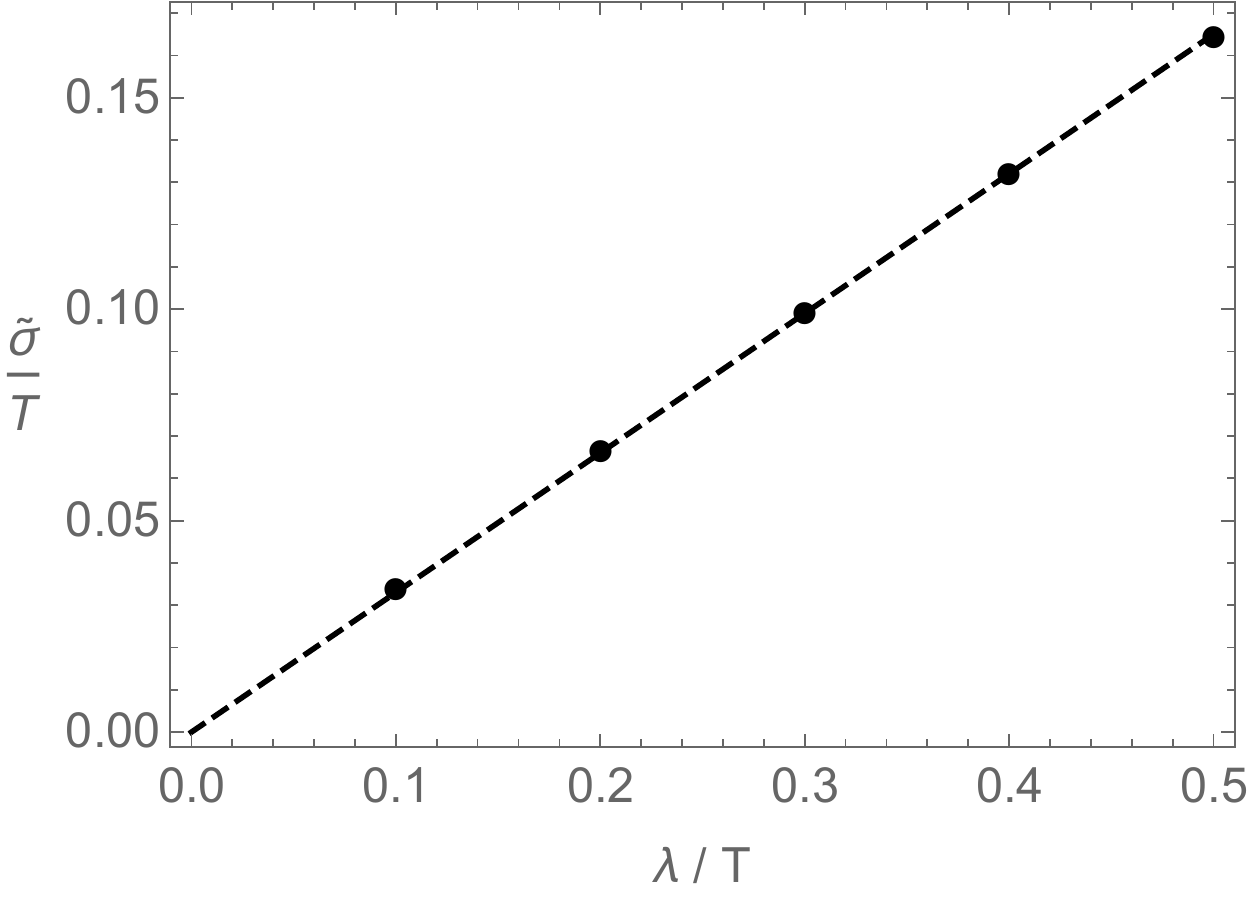} \quad
     \includegraphics[width=7.3cm]{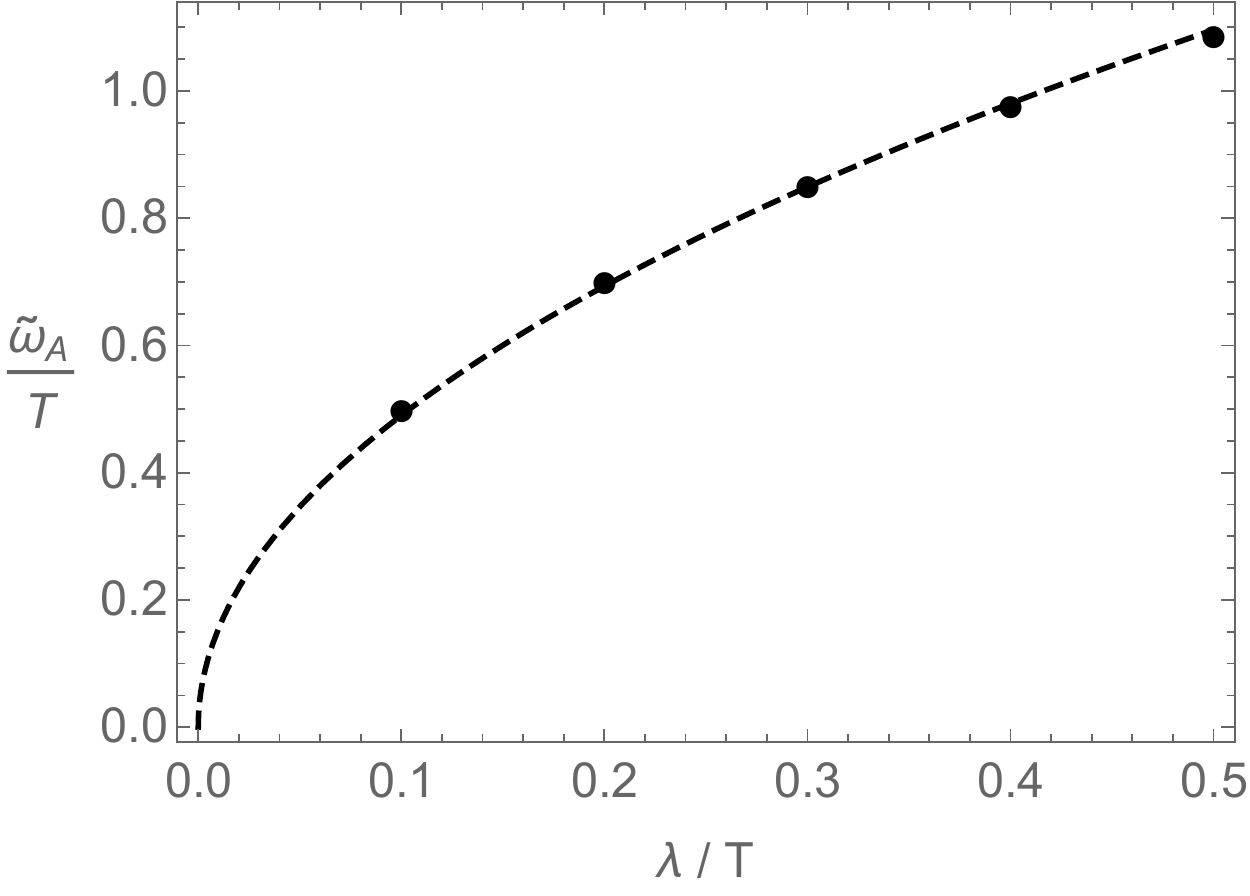} 
 \caption{The $\lambda$ dependence of the dissipative coefficient $\Tilde{\sigma}$ and the mass $\tilde{\omega}_A$ at $T/T_c=0.9$. \textbf{Left:} $\tilde{\sigma}$ vs. $\lambda$. The dashed line is the fitting formula $\tilde{\sigma}/T= 0.33 {\lambda/T}$. \textbf{Right:} $\tilde{\omega}_A$ vs. $\lambda$. The dashed line is the fitting formula $\tilde{\omega}_A/T= 1.55 \sqrt{\lambda/T}$.}\label{lambplt}
\end{figure}

Finally, we want to discuss the effects of the EM coupling on the coefficients appearing in the dispersion relations. The behavior of the various coefficients as a function of temperature for different values of the EM coupling is shown in Fig. \ref{PLOTFW4}.
Interestingly, we find that the velocity $\mathcal{V}$ and the attenuation constant $\Gamma$ are independent of the EM coupling $\lambda$. %
On the contrary, as expected, the dissipative coefficient $\Tilde{\sigma}$ and the mass $\tilde{\omega}_A$ depends on the EM coupling $\lambda$. Their dependence is shown in Fig. \ref{lambplt} for a value of the temperature close to the critical point. The first shows a linear behavior, while the mass shows a square root behavior with $\lambda$.


\subsection{The ``Higgs" mode and its mass}\label{HIGSEC}

\begin{figure}[]
\centering
     \includegraphics[width=7.3cm]{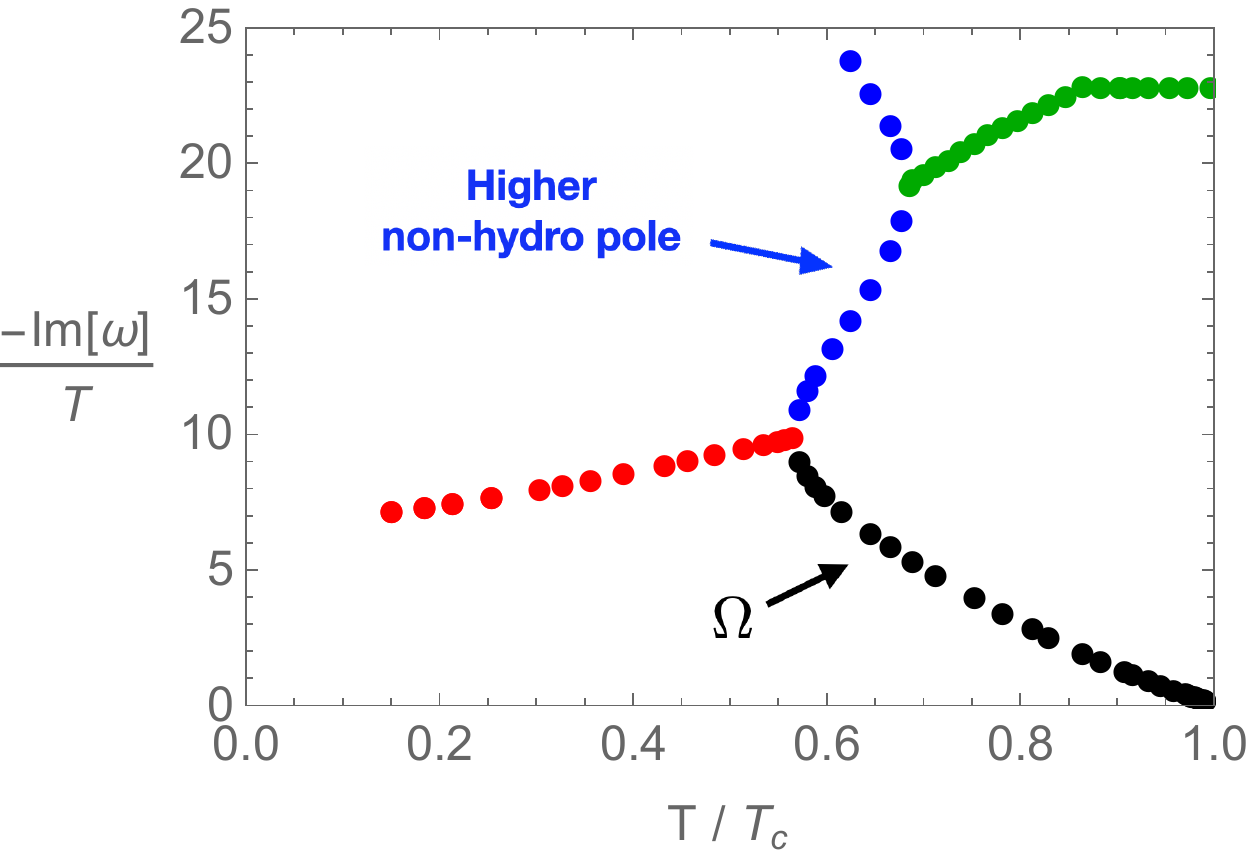} 
     \includegraphics[width=7.0cm]{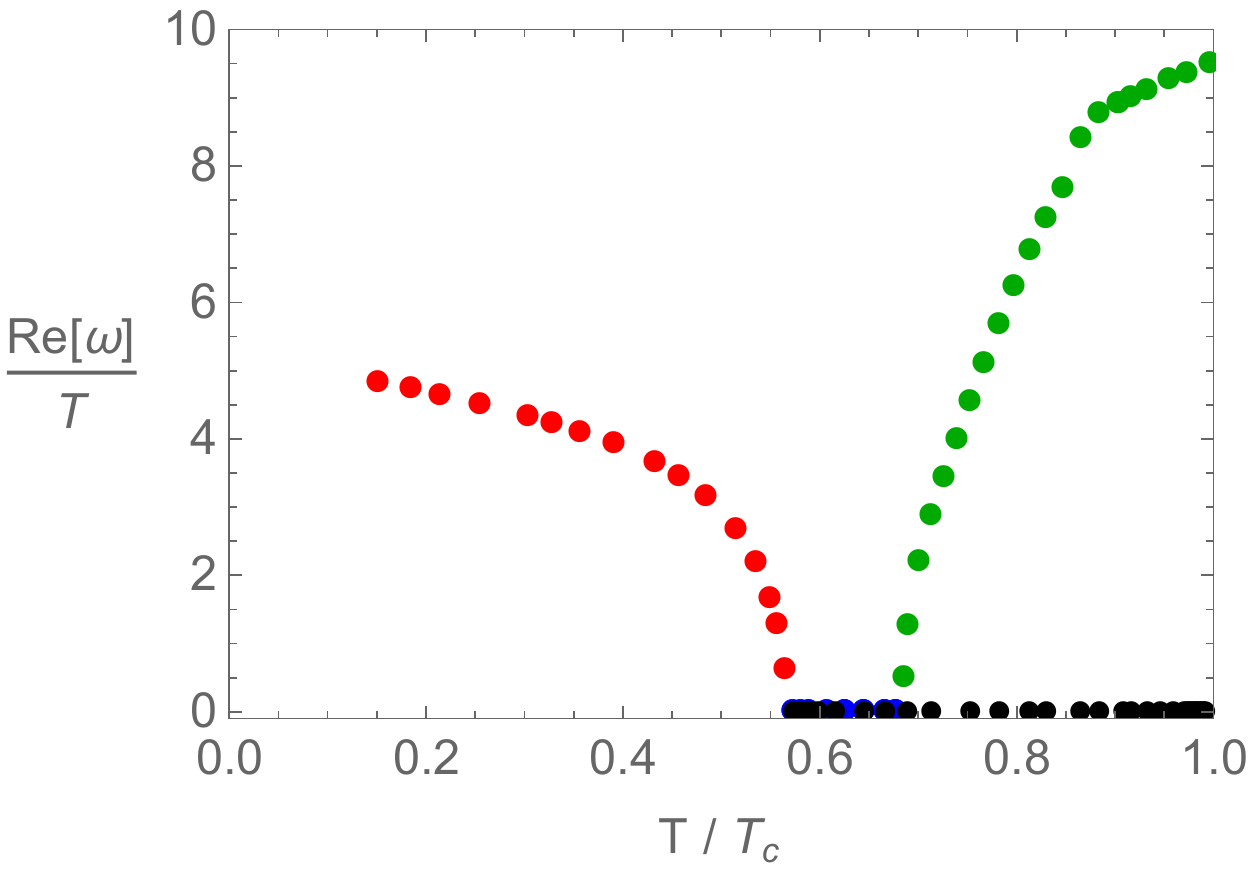} 
 \caption{The dynamics of the Higgs mode as a function of the reduced temperature. The black dots encode the fluctuations of the amplitude of the order parameter close to the critical point.}\label{PLOTFW2}
\end{figure}

\paragraph{Higgs mode at zero wave-vector.}
Next, we discuss the fate of the damped diffusive mode in Eq.\eqref{LONGFOR2b}, the Higgs mode. In particular, we focus on its dynamics at zero wave-vector, as shown in Fig. \ref{PLOTFW2}.
Near the critical point, we find that the Higgs mode is well approximated by a dispersion relation as in Eq.\eqref{LONGFOR2b}. We find numerically that:
\begin{equation}
    \Omega\sim(1-T/T_c)\,.
\end{equation}
This mode corresponds to the fluctuations of the amplitude of the order parameter. Its behavior is in perfect agreement with the expectation from GL theory, Eq.\eqref{hehe00}, i.e.,
\begin{equation}
    \Omega\,\,\leftrightarrow\,\,\frac{\omega_H^2}{2\gamma} \,,
\end{equation}
and also with the holographic results for superfluids in \cite{Amado:2009ts} and the analysis of \cite{Donos:2022xfd} (see also \cite{Flory:2022uzp,Cao:2022mep}).

Interestingly, by decreasing the temperature, this mode collides with a first non-hydrodynamic higher pole (indicated with blue color in Fig. \ref{PLOTFW2}). This collision produces a pair of complex modes, with a finite real part which are displayed in red color in Fig. \ref{PLOTFW2}. This behavior is, once more, well described qualitatively by GL theory, see Eq.\eqref{HGMO3}. To be precise, the complete dynamics is more complicated than an interactions between two modes as assumed in Eq.\eqref{HGMO3}. Indeed, the first non-hydrodynamic mode interacts as well with a second higher order non-hydrodynamic pole (green dots in Fig. \ref{PLOTFW2}) which is not included in Eq.\eqref{HGMO3}. Nevertheless, this mode does not strongly affect the low-energy dynamics.

We have also studied the behavior of the Higgs mode in Eq.\eqref{LONGFOR2b} and found that its dynamics is independent of the value of the EM coupling $\lambda$.
In other words, Fig. \ref{PLOTFW2} does not change with $\lambda$.\footnote{We have explicitly checked for $\lambda/T=0.1, 0.2, 0.3$ .} This is consistent with the properties of the Higgs mode as derived in the GL framework. In particular, the amplitude mode remains unaffected by Coulomb interactions. Importantly, this also implies that the position of the collision, and the temperature at which the Higgs mode acquires a real gap, do not depend on the value of the EM coupling $\lambda$.

\paragraph{Higgs mode at finite wave-vector.}
Considering the finite wave-vector case, we can try to push the comparison with GL theory further. For that purpose, we use the data in Fig. \ref{PLOTFW2} with the dispersion relation in Eq.\eqref{HGMO2} obtained from GL theory. Although, Eq.\eqref{HGMO2} cannot completely capture the dynamics of our Higgs mode near $T=T_c$, we can still use this approximation in the lower $T$ regime, i.e, for the red mode in Fig. \ref{PLOTFW2} up to near the collision point between the blue and black modes, $T/T_c\sim0.6$. 

From GL theory, Eq.\eqref{HGMO2}, we expect a dispersion relation of the form 
\begin{align}\label{FMDRH}
\begin{split}
\omega  = \pm \left(\sqrt{\tilde{\omega}_H^2 - \tilde{\gamma}^2} + \frac{\tilde{v}^2}{2\sqrt{\tilde{\omega}_H^2 - \tilde{\gamma}^2}}\,k^2\right) - i \tilde{\gamma}    \,,
\end{split}
\end{align}
where $\tilde{\omega}_H$ is the mass of the Higgs mode and $\tilde{\gamma}$ its attenuation constant. 
Here, as done for the transverse sector before, we use the tilde variables for the holographic quantities:
\begin{align}\label{}
\begin{split}
\tilde{\omega}_H \,\leftrightarrow\, \omega_H \,,\qquad \tilde{v} \,\leftrightarrow\, v \,, \qquad \tilde{\gamma} \,\leftrightarrow\, \gamma \,.
\end{split}
\end{align}
Performing this analysis, we find that the velocity $v$ coincides exactly with the velocity $\tilde v$, Eq.\eqref{vtilde2}, appearing in the dispersion of  the gauge fluctuations mode in Eq.\eqref{TRANSFIT1}. This is not surprising, and it is indeed expected from the GL theory (see Eq.\eqref{DISGA} and Eq.\eqref{HGMO2}). 

\begin{figure}[]
\centering
     \includegraphics[width=7.0cm]{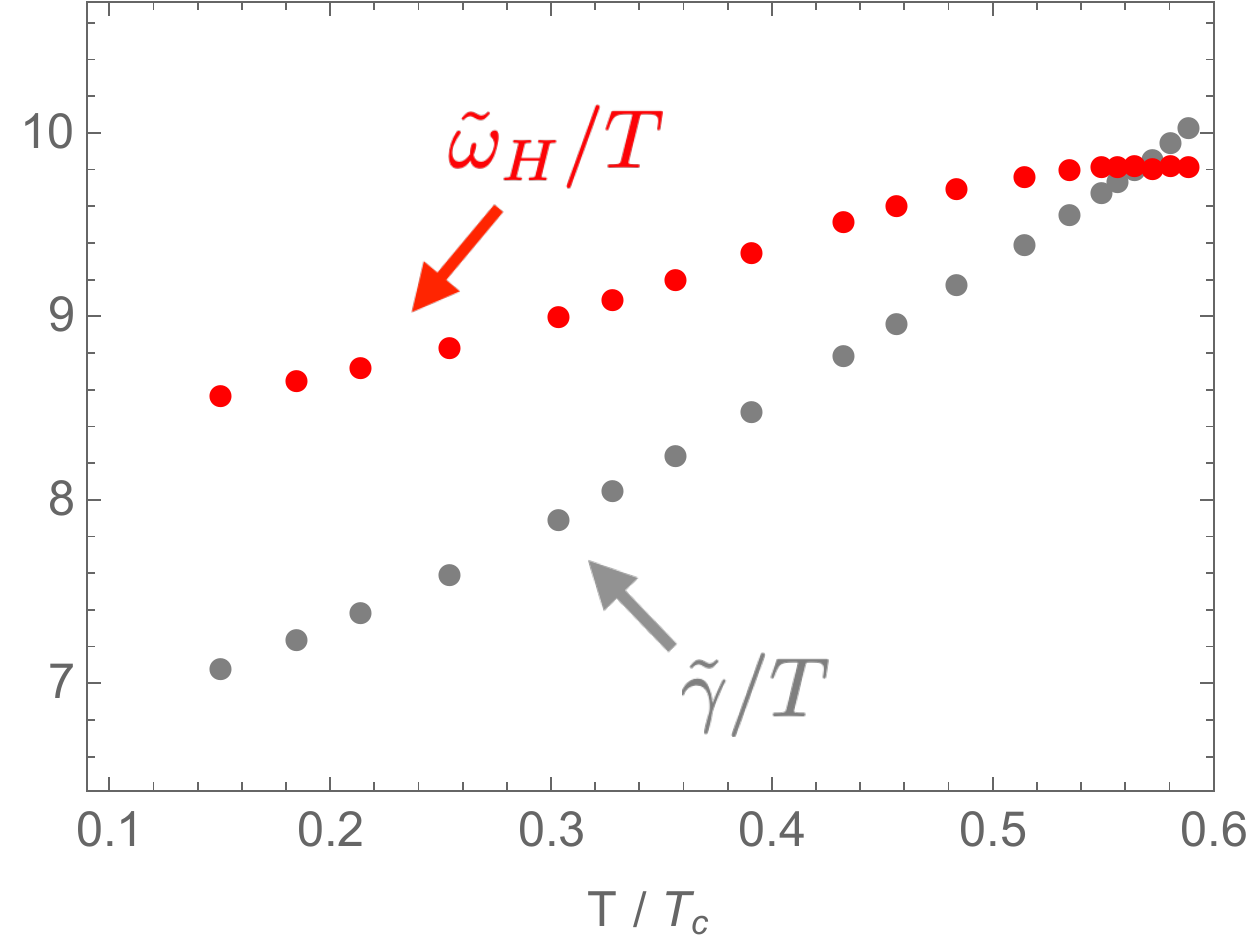} 
 \caption{The phenomenological parameters of GL theory, ($\tilde{\omega}_H, \tilde{\gamma}$). The Higgs frequency $\tilde{\omega}_H$ and the Higgs attenuation constant $\tilde{\gamma}$ at $\lambda/T=0.1$.}\label{NHPP1}
\end{figure}

Therefore, we have two fitting parameters ($\tilde{\omega}_H, \tilde{\gamma}$) which can be extracted from the zero wave-vector analysis.
Their temperature behavior for $\lambda/T=0.1$ is shown in  Fig. \ref{NHPP1}. At low temperature, we find that $\tilde{\omega}_H>\tilde{\gamma}$, which is consistent with the GL framework. Additionally, we find the two parameters are of the same order around  $T^*/T_c\approx 0.6$. This signals the crossover between the overdamped regime at large temperature and the underdamped one at low temperature and it is consistent with the results presented in Fig. \ref{PLOTFW2}. Finally, the dissipative parameter $\tilde{\gamma}$ does not seem to vanish towards zero temperature. This point deserves further investigation in the model with backreaction.

\begin{figure}[]
\centering
     \includegraphics[width=7.0cm]{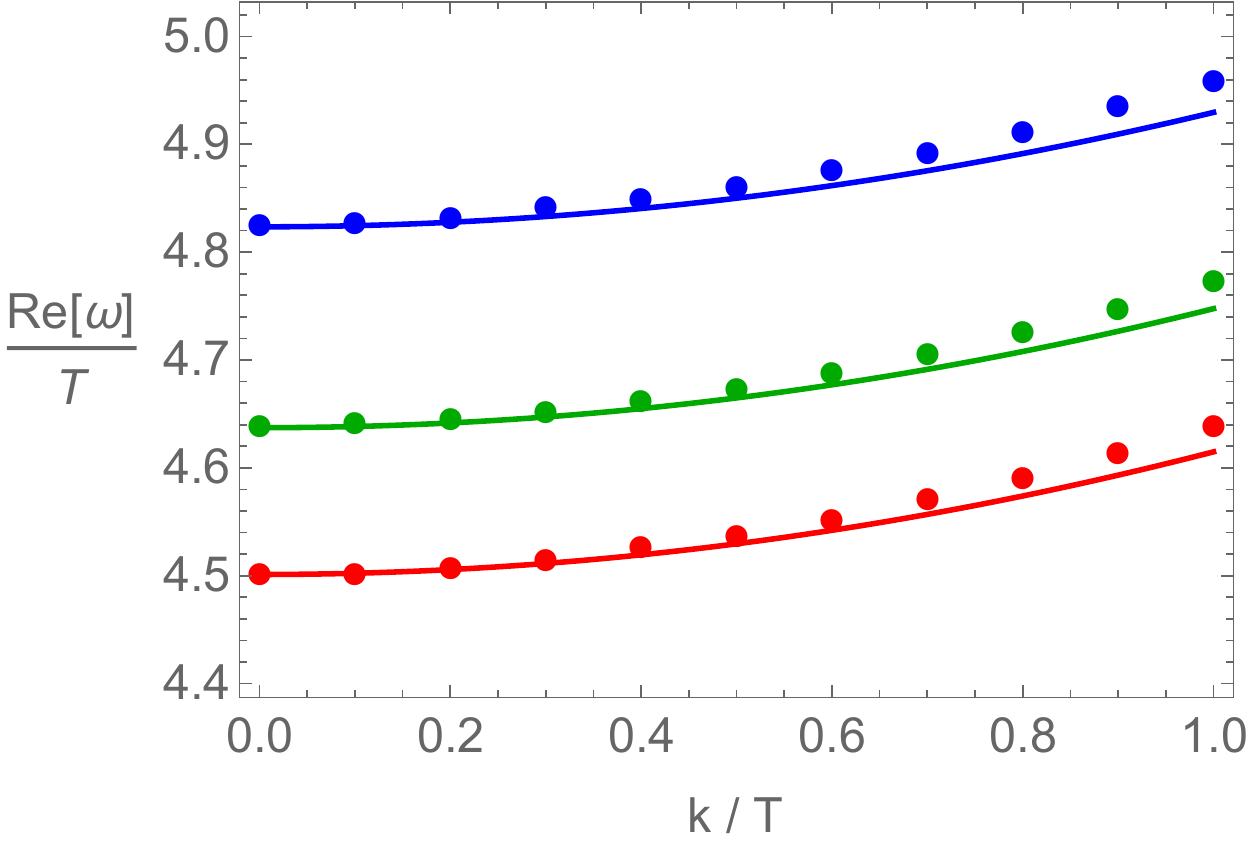} \quad
     \includegraphics[width=7.2cm]{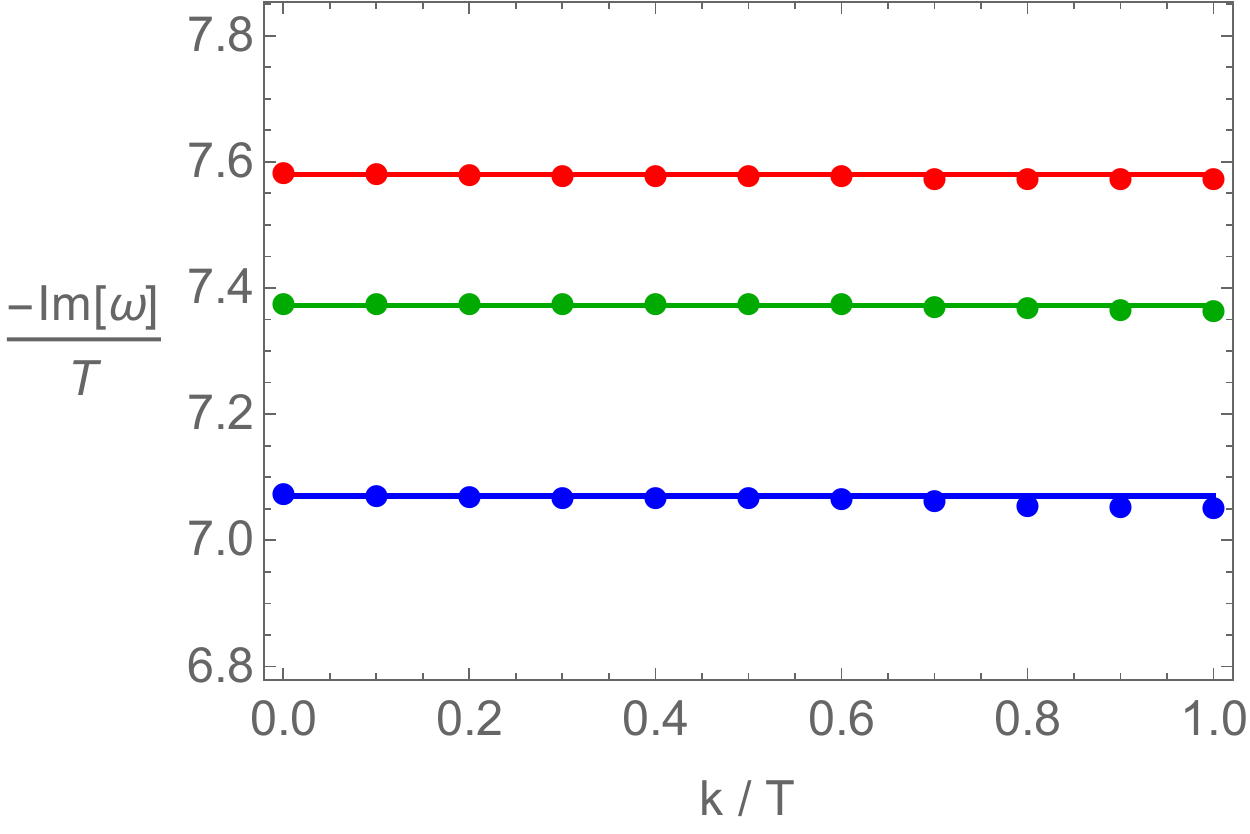} 
 \caption{The dispersion relation of the Higgs mode at low temperature: $T/T_c = (0.15, 0.21, 0.25)$ (blue, green, red). \textbf{Left:} $\mathrm{Re}[\omega]$ vs. $k$. \textbf{Right:} $\mathrm{Im}[\omega]$ vs. $k$. The solid lines are the predictions from GL theory, Eq.\eqref{FMDRH}.}\label{HIGKfig}
\end{figure}

After describing the dynamics of the Higgs mode at zero wave-vector, we can extend the analysis for $k\neq 0$, i.e., once we know ($\tilde{\omega}_H, \tilde{\gamma}$) together with $\tilde{v}^2=1-\lambda \,\chi_{BB}$, we can study the dispersion relation at finite wave-vector \eqref{FMDRH}. We show the real and imaginary parts of the dispersion relation of the Higgs mode at low temperature in Fig. \ref{HIGKfig}.

Interestingly, we see that the GL prediction fits very well the numerical data at low temperature.
This is yet another confirmation that the holographic results are in perfect agreement with the Ginzburg-Landau effective description.

\paragraph{Further comments on Higgs energy gap.}
Before closing this section, we discuss another feature related to the Higgs energy gap, $\tilde{\omega}_H$.
In (s-wave) BCS-type superconductors, under certain specific approximations, the Higgs mode energy gap $\tilde{\omega}_H$ obeys the following expression \cite{PhysRevB.26.4883}:
\begin{equation}\label{ex}
    \tilde{\omega}_{{H}} = 2\Delta \,,
\end{equation}
where $\Delta$ is the superconducting energy gap related to the order parameter as $2\Delta = \sqrt{\langle O_{2} \rangle}$ \cite{Hartnoll:2008vx}. Using our data in Fig. \ref{STSKF}, we estimate $\sqrt{\langle O_{2} \rangle} = 2\Delta \approx 8 \, T_c$ at $T/T_c = 0.15$, which implies $2\Delta/T \approx 53$ at $T/T_c = 0.15$. This result is not consistent with the value of the Higgs gap $\tilde{\omega}_H$ reported in Fig. \ref{NHPP1} which is $\tilde{\omega}_H/T \approx 8.6$ at approximately the same temperature $T/T_c = 0.15$. Combining these outcomes, we find:
\begin{equation}
 \frac{\tilde{\omega}_{H}}{2\Delta}\Big|_{T\approx 0.15 T_c} \approx 0.162  \,,
\end{equation}
which is much smaller than the expected value in Eq.\eqref{ex}.\\
We speculate about the origin of this discrepancy. First, from a practical perspective, working in the probe limit does not guarantee complete control on the low-temperature dynamics. Second, to the best of our knowledge, the result in Eq.\eqref{ex} is not of universal validity but rather limited to weakly coupled BCS-type superconductors. As explicitly shown recently in \cite{Schmalian:2022web,Inkof:2021ohk}, holographic superconductors do not fall into that simple class. It is tempting to attribute this novel outcome to the peculiar strongly-coupled and quantum critical nature of holographic superconductors. Further investigation is needed to ascertain the validity of such a statement.

\begin{figure}[]
\centering
     \includegraphics[width=7.3cm]{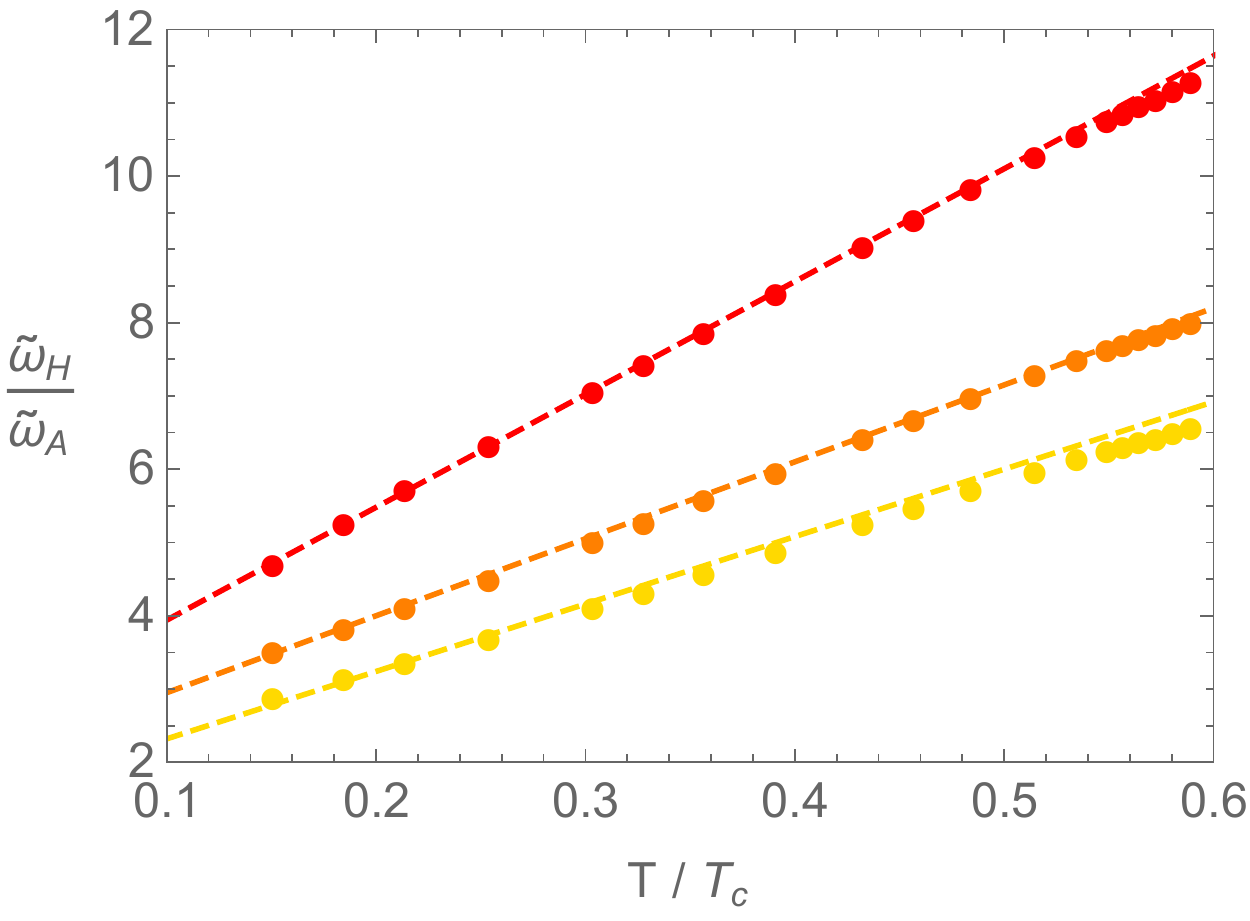}
 \caption{The ratio $\tilde{\omega}_H/\tilde{\omega}_A$ as a function of the reduced temperature for $\lambda/T=(0.1,0.2,0.3)$ (red, orange, yellow). The dashed line are the fitting to the function: $\zeta_1+\zeta_2 T/T_c$.}\label{NHPP2}
\end{figure}

Let us also discuss the GL parameter $\kappa_{GL}$ in \eqref{RATIOHA}, which is also associated with Higgs energy gap via:
\begin{align}\label{CCEFK}
\begin{split}
\frac{\tilde{\omega}_{H}}{\tilde{\omega}_A} = \sqrt{2} \, \kappa_{GL}    \,, \qquad \kappa_{GL}\approx\frac{1}{\sqrt{\lambda}} \,.
\end{split}
\end{align}
This parameter was studied recently in \cite{Natsuume:2022kic}. Using our numerical data ($\tilde{\omega}_H \,, \tilde{\omega}_A$), we can discuss the behavior of $\kappa_{GL}$.
In Fig. \ref{NHPP2}, we examine the ratio between $\tilde{\omega}_H$ and $\tilde{\omega}_A$ as a function of the temperature for different values of the EM coupling. We find a power law dependence of the type $\kappa_{GL}=\zeta_1+\zeta_2 T/T_c$. This has to be contrasted with the logarithmic behavior found in higher dimensions in \cite{Natsuume:2022kic}, which reflects the different nature of the EM coupling in different dimensions. Moreover, as expected from Eq.\eqref{CCEFK}, the ratio decreases at larger $\lambda$ (e.g., from red to yellow in Fig. \ref{NHPP2}). This is consistent with the fact that the $\lambda$-dependence in the GL parameter $\kappa_{GL}$ comes entirely from the propagation length $\lambda_{GL}\propto 1/\sqrt{\lambda}$. Finally, let us comment on the temperature dependence of the GL parameter $\kappa_{GL}$. In AdS$_5$ \cite{Natsuume:2022kic}, the GL parameter decreases with temperature. Here, it increases. The difference between the two scenarios is rooted in the dimension of the U(1) coupling $\lambda$ in 2D and 3D and could be possibly understood analytically by performing holographic perturbative computations near the critical point.

\begin{figure}[]
\centering
     \includegraphics[width=7.0cm]{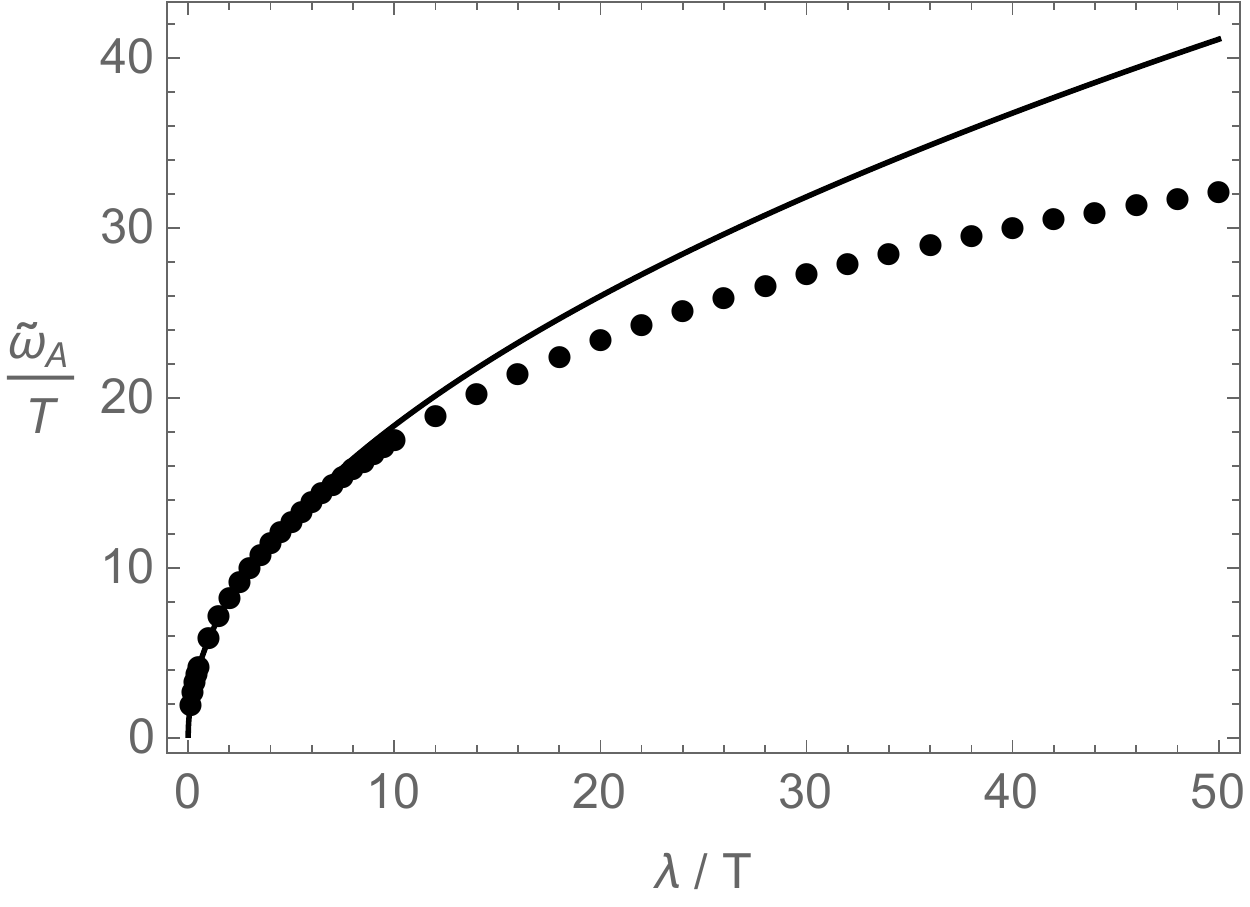} \quad
     \includegraphics[width=7.0cm]{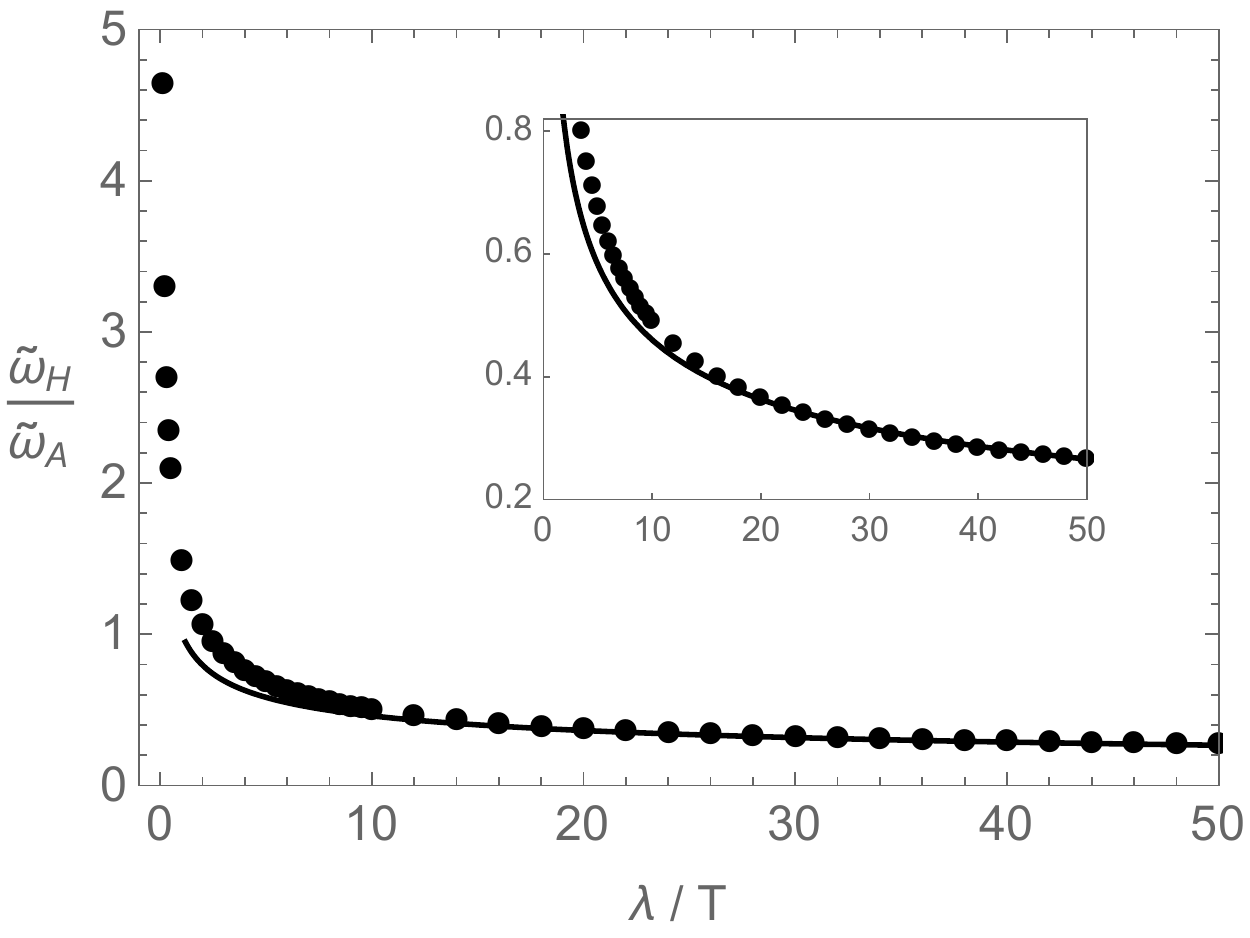} 
 \caption{The $\lambda$-dependence of the energy gaps ($\tilde{\omega}_A,\,\tilde{\omega}_H$) at $T/T_c=0.15$. \textbf{Left:} $\tilde{\omega}_A/T$ vs. $\lambda/T$. The solid line is the plasma frequency value, Eq.\eqref{TRANSFIT3}. \textbf{Right:} The ratio $\tilde{\omega}_H/\tilde{\omega}_A$ vs. $\lambda/T$. The solid line is the fitting curve at large $\lambda$: $\tilde{\omega}_H/\tilde{\omega}_A = \left(\lambda/T\right)^{-0.34}$.}\label{NHPP3}
\end{figure}

%%%
So far, we have focused our analysis on the weak-coupling regime, $\lambda/T \ll 1$. In Fig. \ref{NHPP3}, we also discuss the $\lambda$-dependence on the energy gaps ($\tilde{\omega}_A,\,\tilde{\omega}_H$) at fixed temperature and for larger values of the U(1) coupling $\lambda$. In the left panel, we display $\tilde{\omega}_A$ for different values of $\lambda$ at $T/T_c = 0.15$. We find that (I) $\tilde{\omega}_A$ is monotonically increasing as we enhance $\lambda$; (II) $\tilde{\omega}_A$ deviates from the plasma frequency value, Eq.\eqref{TRANSFIT3}, for large $\lambda$. Furthermore, dialing the value of $\lambda/T$ up to $\lambda/T=50$ at the same temperature $T/T_c = 0.15$, we also checked that the other energy gap, $\tilde{\omega}_H$, is independent of $\lambda/T$, which implies that the ratio between $\tilde{\omega}_H$ and $\tilde{\omega}_A$ is decreasing with $\lambda$ (see the right panel in Fig.\ref{NHPP3}). Our observation (II) also implies that $\kappa_{GL}$ does not follow $\approx1/\sqrt{\lambda}$ in the limit of $\lambda\rightarrow \infty$. On the contrary, we numerically find that $\tilde{\omega}_H/\tilde{\omega}_A = \left(\lambda/T\right)^{-0.34}$. This is another distinct feature from the higher dimensional case discussed in \cite{Natsuume:2022kic}, where $\kappa_{GL}$ remains finite even in the strong EM coupling limit $\lambda\rightarrow \infty$. It would be interesting to understand the large $\lambda$ limit better. We plan to revisit this question in the near future.
%%%
Finally, let us comment about the validity of the probe limit and the expectations in presence of backreaction. In general, we do not expect a qualitative difference in the nature of the low-energy modes, whose structure is mostly dictated by symmetries. Nevertheless, we do expect that the quantitative results, especially in the limit of small temperature, could radically change in presence of backreaction. This is also the reason why all our data are cut around $T\approx 0.2 T_c$, where we do expect such effects to become important. We leave the investigation of the backreacted model for the near future.

%%%%%%%%%%%%%%%%%%%%%%%%%%%%%%%%
\section{Outlook}
\label{secfin}
All previous studies on collective dynamics in holographic models with spontaneously broken U(1) symmetry (e.g., \cite{Amado:2009ts,Arean:2021tks}) have been focused on the case where the U(1) symmetry is global and the dual field theory describes a superfluid rather than a superconductor.
In this work, we have studied the low-energy collective dynamics of a \textit{bona fide} holographic superconductor model in which the gauge field in the boundary field theory is dynamical and the broken U(1) symmetry gauged. We have revealed the characteristic features  of the Anderson-Higgs mechanism and showed evidence for the presence of a Higgs mode presenting a real mass gap at low temperature. Interestingly, the pattern that gives rise to this mode seems to follow the GL logic. On the contrary, in holographic superfluids, the emergence of a pair of complex underdamped modes at low temperature has been observed to follow a very distinct dynamics \cite{Bhaseen:2012gg}.\\


Using a phenomenological attitude, and guided by the predictions of a simple Ginzburg-Landau approach, we have described the dispersion relations of the collective modes in both the transverse and longitudinal sector as a function of the temperature and the electromagnetic coupling. The agreement between the GL effective description and our holographic results is excellent. Our work proves that a holographic superconductor, following all the rules of superconductivity, including the characteristic dynamical excitations, can be constructed using mixed boundary conditions for the bulk gauge field.\\

There are several directions which are worth it investigating in the future. 
\begin{itemize}
    \item First and foremost, we have not presented a complete and formal effective description of the low-energy dynamics. This task can be performed using two slightly different approaches. From one side, one could try to gauge the model F of Hoenberg and Halperin \cite{RevModPhys.49.435} and perform an analysis similar to that of \cite{Donos:2022qao} for the case of holographic superfluids. An alternative approach would be to combine magnetohydrodynamics and superfluid hydrodynamics to construct a hydrodynamic framework for superconductors. This would need an extension of standard hydrodynamics in order to incorporate the dynamics of slowly relaxing non-hydrodynamic modes. Indeed, as emphasized above, ignoring the fluctuations of energy and momentum, a superconductor does not present any hydrodynamic gapless modes in the spectrum. This is very different from the case of superfluids which present a gapless propagating second sound mode easily described within ``standard'' hydrodynamics.
    \item It would be interesting to study in more detail the transport properties of a holographic superconductor. In particular, one would like to understand if any signature of the massive Higgs mode can be observed in the optical conductivity spectrum below the superconducting gap. Naively, one would expect that an underdamped Higgs mode with gap below the SC gap $\Delta$ should leave a clear signature in $\sigma(\omega)$. Here, one must deal with the subtleties regarding the electric response in presence of Coulomb interactions, see, e.g., \cite{Mauri:2018pzq,Romero-Bermudez:2018etn}.
    \item Quenches in our holographic superconductor model could be useful tools to explore the collective dynamics beyond linear approximation. In particular, one might think of extending the analysis of \cite{Flory:2022uzp,Cao:2022mep} to this case and use GL theory to interpret the numerical results. Nonlinear response is expected to be an excellent probe for the dynamics of the Higgs mode which is usually undetectable in the linear response regime \cite{Krull2016}.

    \item One could generalize our study to the case of multiband superconductors where a hydrodynamic mode, known as Leggett mode, should be present \cite{10.1143/PTP.36.901,RevModPhys.47.331}. It would be fascinating to study the dynamics of the Leggett mode using holography.
    \item A different way to promote the gauge field at the boundary as dynamical is by using the dual higher-form description in the bulk \cite{Grozdanov:2017kyl}. It would be interesting to construct a holographic superconductor model without advocating for any U(1) vector gauge field in the bulk.
 \item The dynamics and possible observation of the Higgs mode in superconductors has been topic of a long-standing debate in the condensed matter community \cite{Sherman2015,Endres2012,PhysRevB.59.14054,PhysRevLett.92.027203,PhysRevB.84.174522,PhysRevLett.109.010401,PhysRevB.86.054508}. In this work, we found very distinct features in the emergence of the Higgs mode in holographic superconductors with respect to the previous observations in holographic superfluids \cite{Bhaseen:2012gg}. In particular, we see that the Higgs mode arises, as expected from the Ginzburg Landau arguments, from the dynamics of the amplitude of the order parameter. On the contrary, in holographic superfluids, Ref.\cite{Bhaseen:2012gg} observed the emergence of an underdamped massive mode at low temperature from the spectrum of microscopic modes. It would be interesting to understand this difference further.
 \item It has been recently demonstrated that, in presence of a non-zero superflow, the fingerprints of the Higgs mode could be visible already in the linear response regime \cite{PhysRevLett.118.047001}. One could introduce a non-zero condensate flow (supercurrent) in the holographic model and investigate the dynamics of the amplitude mode therein. 
\end{itemize} 
We plan to return to some of these issues in the near future.

%%%%%%%%%%%%%%%%%%%%%%%%%%%%%%%%
%    Section: Acknowledgments
%%%%%%%%%%%%%%%%%%%%%%%%%%%%%%%%
\acknowledgments

We would like to thank Y. Ahn, C. Setty, S. Grieninger, A. Donos, P. Kailidis, J. Zaanen, M. Kaminski, G. Frangi, S. Grozdanov, K. Landsteiner, and A. Garcia-Garcia for valuable discussions and correspondence. We particularly thank M. Natsuume for several useful suggestions on a previous version of this manuscript.
This work was supported by the National Key R$\&$D Program of China (Grant No. 2018FYA0305800), Project 12035016 supported by National Natural Science Foundation of China, the Strategic Priority Research Program of Chinese Academy of Sciences, Grant No. XDB28000000, Basic Science Research Program through the National Research Foundation of Korea (NRF) funded by the Ministry of Science, ICT $\&$ Future Planning (NRF- 2021R1A2C1006791) and GIST Research Institute (GRI) grant funded by the GIST in 2022.
M.B. acknowledges the support of the Shanghai Municipal Science and Technology Major Project (Grant No.2019SHZDZX01) and the sponsorship from the Yangyang Development Fund. M.B. would like to thank IFT Madrid, NORDITA, GIST and Chulalongkorn University for the warm hospitality during the completion of this work and acknowledges the support of the NORDITA distinguished visitor program and GIST visitor program.
H.-S Jeong would like to thank GIST for the warm hospitality during the completion of this work. K.-Y Kim acknowledges the hospitality at APCTP where part of this work was done.

\appendix
%%%%%%%%%%%%%%%%%%%%%%%%%%%%%%%%
%    
%%%%%%%%%%%%%%%%%%%%%%%%%%%%%%%%
\section{The speed of transverse excitations in the normal phase}\label{appenaaa}
Let us remind the reader about the spectrum of transverse excitations in the normal phase. Since we are working in the probe limit, the dynamics of the transverse momentum is kept frozen. Because of this reason, in the normal phase, the shear diffusion mode will not appear and the whole low-energy dynamics will be controlled by the transverse fluctuations of the gauge field. Those follow the so-called telegrapher equation:
\begin{equation}
    \omega\,\left(\omega+i\,\frac{\sigma}{\epsilon_e}\right)\,=\,\frac{k^2}{\epsilon_e\,\mu_m}
\end{equation}
where $\epsilon_e,\mu_m$ are respectively the electric permittivity and the magnetic permeability. Because of the modified b.c.s. and the dynamical gauge field in the boundary description, the normal phase displays a transverse massless mode with diffusive dispersion, $\omega=-i \frac{k^2}{\sigma \mu_m}$, which can be thought as the diffusion of magnetic lines. A detailed check of this dynamics has been recently reported in \cite{Ahn:2022azl}.

Furthermore, from standard electrodynamics, we have that $\tilde v^2 =1/(\epsilon_e \mu_m)$, with $\epsilon_e,\mu_m$ respectively the electric permittivity and the magnetic permeability. In general, the latter is related to the electric and magnetic susceptibilities via
\begin{equation}\label{uss}
\chi_{EE}=\epsilon_e-\frac{1}{\lambda}\,,\qquad \chi_{BB}=\frac{1}{\lambda}-\frac{1}{\mu_m}\,.
\end{equation}
In \cite{Ahn:2022azl}, we found that, at least in the limit of small EM coupling $\lambda/T \ll 1$, $\chi_{EE}=0$ to a good approximation. Then, using Eq.\eqref{uss} in such a limit, we immediately find:
\begin{equation}
   \tilde {v}^2 = 1 - \lambda \, \chi_{BB} \quad \quad \text{for}\quad \lambda/T \ll 1\,.
\end{equation}
Moreover, as shown in \cite{Ahn:2022azl}, for our simple holographic model we have
\begin{equation}
    \chi_{BB}=-z_h=-3/(4\pi T)\,.
\end{equation}

%%%%%%%%%%%%%%%%%%%%%%%%%%%%%%%%
%    Section: END
%%%%%%%%%%%%%%%%%%%%%%%%%%%%%%%%

%\bibliography{HyunSikRefs}
\bibliographystyle{JHEP}

\providecommand{\href}[2]{#2}\begingroup\raggedright\endgroup

\end{document}